\documentclass[a4paper,12pt]{article}

\usepackage{amsmath,amssymb}
\usepackage[hypertex]{hyperref}
\usepackage{epsf}
\usepackage{cite}
\usepackage{color}

\makeatletter

\def\unit{{1\kern-.65ex {\rm l}}}
\def\1{{1\kern-.65ex {\rm l}}}

\def\Sym{\mathop{\mathrm{Sym}}\nolimits}
\def\Tr{\mathop{\mathrm{Tr}}\nolimits}

\def\CN{{\cal N}}
\def\CO{{\cal O}}

\def\bbR{{\mathbb{R}}}

\def\bbZ{{\mathbb{Z}}}

\newcount\hour \newcount\minute
\hour=\time \divide \hour by 60
\minute=\time
\def\now{%
\ifnum \hour<13
  \ifnum \hour=0 \advance \hour by 12 \number\hour:\else \number\hour:\fi%
     \ifnum \minute<10 0\fi%
     \number\minute%
\ A.M.%
\else \advance \hour by -12 \number\hour:%
  \ifnum \minute<10 0\fi%
  \number\minute%
  \ P.M.%
\fi%
}

\makeatother

\newcommand\be{\begin{equation}}
\newcommand\bea{\begin{eqnarray}}
\newcommand\ee{\end{equation}}
\newcommand\eea{\end{eqnarray}}
\newcommand\h{\frac{1}{2}}

\newcommand{\bdm}{\begin{displaymath}}
\newcommand{\edm}{\end{displaymath}}
\newcommand{\nn}{\nonumber \\}
\newcommand{\f}[2]{\frac{#1}{#2}}
\newcommand{\bref}[1]{(\ref{#1})}

\newcommand{\tM}{\tilde{M}}
\newcommand{\cJ}{\mathcal J}
\newcommand{\Dtube}{D_{\textrm{tube}}}
\newcommand{\Dbr}{D_{\textrm{BR}}}
\newcommand{\Stube}{S_{\textrm{tube}}}
\newcommand{\Sbr}{S_{\textrm{BR}}}

\begin{document}

\baselineskip=18pt  
\numberwithin{equation}{section}  
\allowdisplaybreaks  


%
\thispagestyle{empty}

\vspace*{-2cm} 
\begin{flushright}
IPhT-T11/164\\
ITFA11-11
\end{flushright}

\vspace*{2.5cm} 
\begin{center}
 {\LARGE  Moulting Black Holes}\\
 \vspace*{1.7cm}
 Iosif Bena$^1$,
 Borun D. Chowdhury$^2$,
 Jan de Boer$^2$,\\
 Sheer El-Showk$^1$, and
 Masaki Shigemori$^{3}$
\\
 \vspace*{1.0cm} 
 $^1$ Institut de Physique Th\'eorique,\\
CEA Saclay, CNRS URA 2306,
F-91191 Gif-sur-Yvette, France\\[1ex]
 $^2$ Institute for Theoretical Physics, University of Amsterdam,\\
Science Park 904, Postbus 94485, 1090 GL Amsterdam, The Netherlands\\[1ex]
 $^3$ Kobayashi-Maskawa Institute for the Origin of Particles and the Universe,\\
 Nagoya University, Nagoya 464-8602, Japan\\
\end{center}
\vspace*{1.5cm}

\noindent
We find a family of novel supersymmetric phases of the D1-D5
CFT, which in certain ranges of charges have more entropy than all known
ensembles. We also find bulk BPS configurations that exist in the same
range of parameters as these phases, and have more entropy than a BMPV
black hole; they can be thought of as coming from a BMPV black hole
shedding a ``hair'' condensate outside of the horizon.  The entropy of
the bulk configurations is smaller than that of the CFT phases, which
indicates that some of the CFT states are lifted at strong
coupling. Neither the bulk nor the boundary phases are captured by the
elliptic genus, which makes the coincidence of the phase boundaries
particularly remarkable.  Our configurations are supersymmetric, have
non-Cardy-like entropy, and are the first instance of a black hole
entropy enigma with a controlled CFT dual. Furthermore, contrary to
common lore, these objects exist in a region of parameter space (between the
``cosmic censorship bound'' and the ``unitarity bound'') where no black
holes were thought to exist.

\newpage
\setcounter{page}{1} 




\section{Introduction and summary}

The past few years have seen a great interest in the hair of black
holes in anti-de Sitter (AdS) spacetimes. In AdS gravity coupled to
other fields such as gauge fields and charged scalar fields, specifying
the mass and charge of the configuration does not necessarily determine
a unique black hole solution.  Instead, one sometimes finds infinitely
many solutions describing bound states of multiple black holes, or black
holes surrounded by a condensate of other fields which is often
referred to as ``hair''.\footnote{If one wants to reserve the word
``hair'' for genuine microstates of a black hole, then it is probably
better to call the condensate a ``halo'', because this configuration is
better thought of as a bound state of a black hole and the condensate
outside the horizon.  However, we will use the word ``hair'' because
this is a commonly used terminology in the literature.}
For non-extremal black holes the existence of condensates, or hair, can
be thought of as a thermodynamic instability for a charged black hole to
emit one or several of its charges; in certain regimes this can 
\emph{increase} the entropy of a black hole and thus it is entropically
favorable for the black hole to reduce its charge by shedding  charged
hair outside the horizon.

For example, \cite{Gubser:2008px, Hartnoll:2008vx, Hartnoll:2008kx,
Denef:2009tp, Gubser:2009qm, Gauntlett:2009dn} found that a
Reissner--Nordstrom black brane in AdS Maxwell gravity with a charged
scalar (in bottom-up settings or embedded in string theory) is
unstable against forming a charged scalar condensate outside its horizon
and breaking the $U(1)$ symmetry, and related this to the
superconducting phase transition in the boundary field theory.
As a different example embeddable in string theory,
\cite{Bhattacharyya:2010yg} studied a small $R$-charged black hole in
$AdS_5\times S^5$.  They found that the black hole is unstable against
forming an $R$-charged scalar condensate around it and constructed the
endpoint configuration perturbatively when the charge is small.

Another example of this instability is the so-called entropy enigma
\cite{Gauntlett:2004wh, Denef:2007vg}: certain two-center BPS black hole
configurations can have larger entropy than a single-center
solution with the same asymptotic charges. Since for some charge choice
one of these centers can uplift in five dimensions to a smooth geometry
with flux, these particular enigmas can be thought of as black holes
with hair around them.  In \cite{deBoer:2008fk}, the entropy enigma was
investigated in the context of the AdS/CFT correspondence by embedding
it in $AdS_3\times S^2$.  It was found that this phenomenon occurs in
the non-Cardy regime of the boundary CFT, where the entropy can deviate
from the one naively expected from the Cardy formula.  However, a
complete CFT understanding of the entropy enigma has not been reached
yet because of the limited knowledge on the dual MSW CFT
\cite{Maldacena:1997de, Minasian:1999qn}.

The purpose of this paper is to study the phase diagram\footnote{Unless 
stated otherwise, the word phase in this paper will refer to a 
microcanonical phase.} of the
three-charge BPS black hole in five dimensions, and to determine the
existence of new phases that contain black holes with hair that have
dominant entropy in certain regimes of parameters. For large angular
momentum the BPS states we find can be thought of as the endpoints of a
``thermodynamic'' instability of a rotating D1-D5-P BPS black hole in
$AdS_3\times S^3$, and we identify these endpoint configurations both in
the bulk and the boundary.  Unlike previous enigma examples, our system
has the advantage of being well-understood on both sides of holography.

More concretely, if $N_p$ is the momentum charge along the $S^1$
direction of $AdS_3$ and $J_L,J_R$ are the angular momenta\footnote{Our
conventions are such that $J_{L,R}$ are integers. } in $S^3$, then
the left-moving energy $L_0$ of the dual D1-D5 CFT is equal to $N_p$ (up
to a constant shift) and the CFT $R$-charges are $J_{L,R}$.
Now, let us consider a \emph{microcanonical} ensemble specified by 
given fixed values of $N_p>0$ and $J_L$ ($J_R$ is left unfixed), and ask
what is the entropy of the ensemble.  In the Cardy regime
\begin{align}
 N_p-J_L^2/4N\gg N,\label{Cardyregime}
\end{align}
where $c=6N$ is the central charge, the Cardy formula and the
spectral flow symmetry of the CFT give the entropy:
\begin{align}
 S_{\text{Cardy}}&=2\pi\sqrt{NN_p-J_L^2/4}.\label{SCardy}
\end{align}

\begin{figure}[tbh]
 \begin{center}
   \epsfxsize=10cm \epsfbox{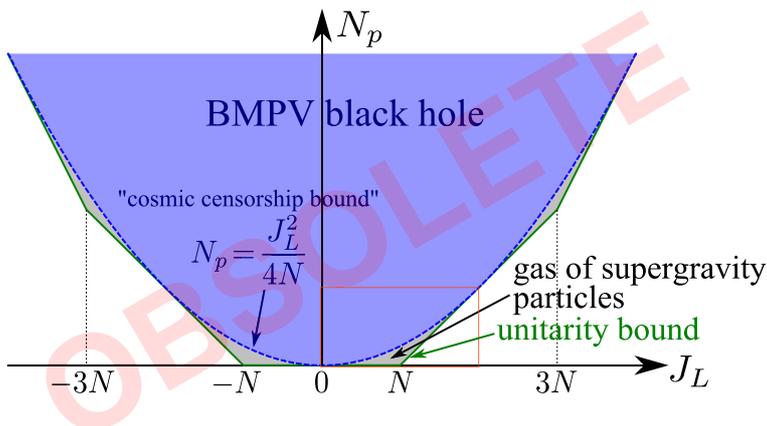}\vspace*{-.3cm} \caption{\sl The
 ``standard lore'' but incorrect phase diagram of the D1-D5 system. Above the blue dotted
  parabola $N_p=J_L^2/4N$ (the cosmic censorship bound) is the BMPV black
  hole phase (light blue), while below the parabola is the phase of a
  gas of supergravity particles (gray).  The range of $N_p,J_L$ is bounded from
  below by the unitarity bound (green solid polygon).
  \label{fig:old_phase_diag}}
\end{center}
\end{figure}
In the bulk this corresponds to a single-center BPS black hole -- the
BMPV black hole \cite{Breckenridge:1996is}, whose Bekenstein--Hawking
entropy nicely reproduces the Cardy entropy \eqref{SCardy}.  Although
the Cardy formula is valid only in the region \eqref{Cardyregime}, the
bulk BMPV black hole exists for any value of $N_p$ larger than the bound
$N_p=J_L^2/4N$.\footnote{This bound is oftentimes called the ``cosmic
censorship bound'' ({\it e.g.}, Ref.\ \cite{hep-th/0005003}), and we
follow this terminology.  Strictly speaking, this bound should instead
be called the ``chronological censorship bound'' because, below this
bound, the single-center black hole solution develops closed timelike
curves outside the horizon but not a naked singularity.}  Furthermore,
one can identify the CFT phase dual to the bulk BMPV black hole and show
that this CFT phase (known as the ``long string'' sector) also exists
all the way down to the cosmic censorship bound and that its entropy is
always equal to \eqref{SCardy} in the large $N$ limit.
Based on this, the phase diagram of the D1-D5 system has been thought
to be the one shown in Fig.~\ref{fig:old_phase_diag};
above the cosmic censorship bound, the system is in the BMPV black hole
phase while, below the bound, the system is in the phase of a gas of
supergravity particles.

However, in the parameter region outside \eqref{Cardyregime}, namely in
the \emph{non-Cardy regime}, the Cardy formula \eqref{SCardy} is no
longer valid and there is no guarantee that the BMPV black hole phase is
thermodynamically dominant.  We will analyze in detail the possible
phases both in the CFT and in the bulk, both analytically and numerically,
and find \emph{new phases} that for the same charges are thermodynamically dominant over
other known phases in the non-Cardy regime.  In the bulk, the new phase
corresponds to a black hole surrounded by a supertube, or to a black ring.  
We can interpret both bulk 
solutions as resulting from the moulting or hair-shedding of the BMPV black hole. In one 
configuration the hair is a supertube, and in the other one the hair is a Gibbons-Hawking or 
Taub-NUT center (corresponding to a D6 brane in four dimensions) whose shedding changes the topology of the black hole horizon and transforms it into a black ring. As a result,
the phase diagram shown in Fig.~\ref{fig:old_phase_diag} is
significantly modified in the non-Cardy regime.
\begin{figure}[tb]
 \begin{center}
\begin{tabular}{cc}
 \epsfxsize=8cm \epsfbox{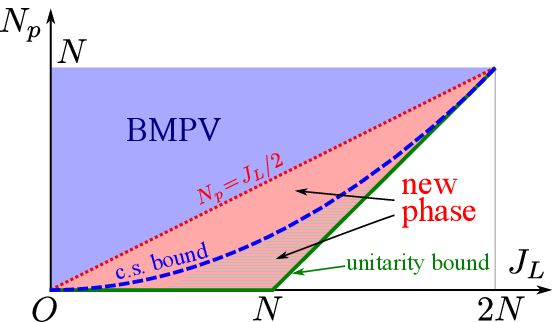} 
 &
 \epsfxsize=8cm \epsfbox{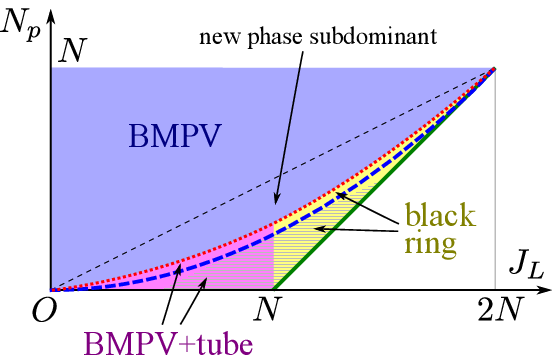} \\[-25.8cm] 
 \begin{minipage}{8cm}(a) CFT phase diagram in the RR sector\\
 at the orbifold point\end{minipage}
 &
 (b) Bulk phase diagram 
\end{tabular}
\caption{\sl The updated, correct phase diagram of the D1-D5 system for
the CFT and bulk (schematic, not to scale). The parameter range 
corresponds to the red rectangle in Fig.\ \ref{fig:old_phase_diag}. The abbreviation ``c.s.\
bound'' refers to the cosmic censorship bound $N_p=J_L^2/4N$.  For further
explanations, see the text.
  %
  \label{fig:new_phase_diag}}
\end{center}
\vspace*{-.5cm}
\end{figure}

The CFT phase diagram is shown in Fig.\ \ref{fig:new_phase_diag}a.  If
we start in the BMPV phase (light blue) with some large value of $N_p$
and decrease $N_p$, then at $N_p=J_L/2$ (red dotted line) a new phase
(light red region) becomes available before we reach the cosmic
censorship bound (thick blue dashed curve).  As soon as it becomes
available, this new phase entropically dominates over the BMPV\@ phase.
As we further decrease $N_p$, the BMPV phase disappears at the cosmic
censorship bound $N_p=J_L^2/4N$ (blue dashed line) while the new phase
continues to exist and is dominant all the way down to the unitarity bound (green solid
line).  Below the cosmic censorship bound, the phase of a gas of
supergravity particles is subdominant and not realized
thermodynamically.

The bulk phase diagram shown in Fig.\ \ref{fig:new_phase_diag}b is
somewhat similar, but there are some distinctive differences.  As we
start from the BMPV phase and lower $N_p$, a new phase appears at
$N_p=J_L/2$, but has less entropy than the BMPV black hole until we further decrease $N_p$
and reach the red dotted curve in Fig.\ \ref{fig:new_phase_diag}b.
After that, the new phase is dominant until the BMPV black hole disappears at the
cosmic censorship bound (thin blue dashed curve).  Below that, the new
phase is dominant all the way down to the unitarity bound.  Furthermore,
for $J_L<N$, the new phase is a BMPV black hole with a hair of smooth
geometry around it (light pink), while for $J_L>N$ it is a black ring
(light yellow).  On the $J_L=N$ line, these two configurations are
entropically degenerate but remain distinct configurations.

Although in Fig.\ \ref{fig:new_phase_diag} we have shown only a small
region of parameters $N_p$ and $J_L$, by the spectral flow symmetry of
the bulk and of the boundary, the new phase exists in all ``wedges''
below the cosmic censorship bound shown in Fig.\
\ref{fig:old_phase_diag} (see also Fig.
\ref{fig:spectralFlowedEnigmaticPhase}).

The entropy of the CFT new phase is \emph{larger} than that of the bulk
new phase.  Because the CFT computation was done in the free limit (at
the orbifold point), this implies that, as we increase the coupling,
some of the states that constitute the new phase in the CFT get lifted
and disappear by the time we reach the gravity point.  However, this
lifting is quite moderate, and does not change the power of $N$ that enters
in the entropy formula, but only its prefactor; the new phases both in
the CFT and the bulk are black hole states having an entropy of order
$\CO(N)$.

The fact that we have black holes below the cosmic censorship bound is
intriguing for the following reason.  In \cite{deBoer:1998us,
Maldacena:1999bp}, it was shown that the (modified) elliptic genus
computed in CFT and the one computed in supergravity agree exactly
for\,\footnote{In \cite{deBoer:1998us, Maldacena:1999bp} the relevant
inequality was given in terms of NS sector quantities as $L_0^{NS}\le
\f{N+1}{4}$.  Here this has been translated into the R sector.}
\begin{align}
 N_p\le {J_L\over 2}-{N-1\over 4}.
 \label{eq:Jan's_region}
\end{align}
This parameter range is shown in Fig.\ \ref{fig:new_phase_diag} as
horizontally hatched regions and is below the cosmic censorship bound.
One expects that, once one turns on coupling, all states that are not
protected will lift, and all that remain at strong coupling are the
states captured by the elliptic genus.  In \cite{deBoer:1998us,
Maldacena:1999bp}, the elliptic genus was correctly reproduced in
supergravity by counting particles, without including any black hole
states.  This appears to imply that in the region
\eqref{eq:Jan's_region} the only thing that exist in the bulk are
supergravity particles and there are no black hole states.  This was the
reason why the phase diagram was thought to be as shown in Fig.\
\ref{fig:old_phase_diag}.
On the contrary, in the current paper we find black hole (and ring)
states in supergravity even in the region \eqref{eq:Jan's_region}.  This
means that there are many states which are not protected and are thus
not captured by the elliptic genus but nevertheless do not
lift.\footnote{In $d=4,\CN=4$ theories, it has been argued
\cite{Dabholkar:2009dq} that multi-center solutions are not captured by
the supersymmetry index unless each center preserves 1/2 supersymmetry.
Our multi-center solution is made of a 1/4-BPS center and a 1/2-BPS
center and thus is not captured by the supersymmetry index by the
general argument of Ref.\ \cite{Dabholkar:2009dq}.  } This might be
suggesting the existence of a new index capturing these states.  It is
possible that such an index is related to the ``new moonshine''
\cite{Eguchi:2010ej} on the hidden underlying symmetry of K3 surfaces.

The original motivation for the current study was to find the
microscopic description of supersymmetric black rings
\cite{Elvang:2004rt, Bena:2004de, Elvang:2004ds, Gauntlett:2004qy} in
the D1-D5 CFT\@.\footnote{By a microscopic description we mean a
description in the UV CFT, corresponding to the asymptotic AdS$_3$
region at infinity.  Near the horizon of a supersymmetric black ring,
there is another AdS$_3$ region which corresponds to an IR CFT\@.  The
IR CFT description of supersymmetric black rings was discussed in
\cite{Bena:2004tk, Cyrier:2004hj}.  However, the IR CFT does not capture
many interesting dynamical features of the D1-D5 system, such as dipole
charges, multi-center solutions and the family of smooth geometries
\cite{Bena:2005va, Berglund:2005vb}, and thus is not of interest in
the present paper.} A CFT understanding of black rings and their dipole
charges \cite{Iizuka:2005uv, Dabholkar:2005qs, Alday:2005xj,
Dabholkar:2006za} is of much interest in its own right and may help us
identify the boundary description of the family of smooth supergravity
solutions found in \cite{Bena:2005va, Berglund:2005vb}.  In
\cite{Bena:2004tk}, a possible microscopic description of supersymmetric
black rings in the D1-D5 CFT was proposed but it was based on a
phenomenological assumption, and hence not entirely satisfactory.  Here,
we made attempts to make progress in this direction by asking what is
the most entropic configuration for given charges $N_p,J_L$.  The new
phase on the CFT side has already been reported in \cite{MasakiMITTalk},
and in the current paper we are reporting progress on the bulk side
based on recent developments.  It is interesting that the most entropic
configuration is indeed a black ring in a certain parameter region.  We
hope to come back to the microscopics of black rings in near future.

It was noted in \cite{Gauntlett:2004wh} that certain configurations of
multi-center black rings can have entropy larger than a single-center
black hole with the same values of charges and angular momenta.
However, to our knowledge, no systematic search for the maximum entropy
configuration of multi-center black holes/rings has been done, and such
configurations have never been investigated in the context of the
AdS/CFT correspondence.\footnote{The configurations found in
\cite{Gauntlett:2004wh} are not in the regime of parameters discussed in
the current paper.  Their configurations have $N_p\sim \sqrt{N}$ while
we are interested in $N_p\sim N$.}

The plan of the rest of the paper is as follows.  In section
\ref{sec:CFTAnalysis}, after reviewing some necessary background material, 
we study the
phase diagram of the D1-D5 CFT\@.  We find a new phase that has more
entropy than the BMPV phase, and give a physically-intuitive picture for this.
Then, we confirm the existence of the new phase more rigorously by
numerically evaluating the CFT partition function.  In section
\ref{sec:sugraAnalysis}, we explore the phase diagram of the D1-D5
system in the dual supergravity description.  We perform a thorough analysis of
two-center solutions and find black hole and ring configurations that
have more entropy than a single-center BMPV black hole in a certain
region of parameters.  Section \ref{sec:discussion} is devoted to the
discussion of the results and future directions.  In the Appendices, we
present technical details and further clarifications on the subjects
discussed in the main text.

\section{CFT analysis}
\label{sec:CFTAnalysis}

In this section we study the possible phases of the D1-D5 CFT, for given
momentum and angular momentum charges.  For large values of these
charges (in the Cardy regime), the Cardy formula predicts the entropy of
the system which is known to be reproduced by the entropy of the BMPV
black hole in the bulk.  However, outside the Cardy regime, there is no
formula for the entropy of general CFTs.  In the D1-D5 CFT, however, the
explicit orbifold construction of the CFT allows us to make an educated
guess on the phase outside the Cardy regime and its entropy formula,
which we will confirm by computer analysis.  We will find that, in
a certain regime of parameter space, a new phase appears and entropically
dominates over the BMPV phase.

\subsection{D1-D5 CFT}

In this subsection we give a quick review of the D1-D5 CFT\@. For a more
detailed review, see for example \cite{David:2002wn}.

Consider type IIB string theory on $S^1 \times M^4$ with $N_1$ D1-branes
wrapping $S^1$ and $N_5$ D5-branes wrapping $S^1 \times M^4$, where
$M^4=T^4$ or $K3$.  We take the size of $M^4$ to be string scale. The
Higgs branch of this system flows in the IR to an $\mathcal N=(4,4)$
SCFT whose target space is a resolution of the symmetric product
orbifold $\mathcal M=(M^4)^N/S_N\equiv {\rm Sym}^N(M^4)$, where $S_N$ is
the permutation group of order $N$ and $N=N_1 N_5$ ($N=N_1N_5+1$) for
$M^4=T^4$ (for $M^4=K3$). The orbifold $\mathcal M$ is called the
``orbifold point'' in the space of CFTs and the theory is 
easy to analyze
at that point.

The CFT is dual to type IIB string theory on $AdS_3 \times S^3 \times
M^4$. To have a large weakly-coupled $AdS_3$, $N$ must be large and the
CFT must be deformed far from the orbifold point by certain marginal
deformations (for recent work see \cite{Avery:2010er,Avery:2010hs,Avery:2010vk}). In this work we will
consider a new phase at the orbifold point and look for it at the
supergravity point.

For presentation purposes, we will henceforth take $M^4=T^4$, but much
of the discussion goes through also for $M^4=K3$.  In particular, the
existence of the new phase does not depend on whether $M^4=T^4$ or $K3$
because it is constructed using structures common to both.

The theory has an $SU(2)_L \times SU(2)_R$ $R$-symmetry which originates
from the $SO(4)$ rotational symmetry transverse to the D1-D5 system.
There is another $SU(2)_1 \times SU(2)_2$ global symmetry which is
broken by the toroidal compactification but can be used to classify
states. We label the charges under these symmetries as $\alpha,\dot
\alpha$ and $A,\dot A$ respectively. At the orbifold point each copy of
the CFT has four left-moving fermions $\psi^{\alpha A}$, four left-moving bosons $\partial X^{A \dot A}$, four right-moving fermions
$\psi^{\dot \alpha A}$ and four right-moving bosons $\bar \partial X^{A
\dot A}$. In addition the CFT has twist fields $\sigma_n$ which
cyclically permute $n \le N$ copies of the CFT on a single $T^4$. One
can think of these twist fields as creating winding sectors in the D1-D5
worldsheet with winding over different copies of the $T^4$.

The D1-D5 CFT is in the  Ramond-Ramond sector because of asymptotic flatness and supersymmetry. Elementary bosonic twist fields (without any
bosonic or fermionic excitations) are charged under $SU(2)_L \times
SU(2)_R$ {\it viz.}\ $\sigma_n^{\alpha \dot \alpha}$ or under $SU(2)_1 \times
SU(2)_1$ {\it viz.}\ $\sigma_n^{A B}$ while elementary fermionic twist
fields are charged under $SU(2)_L \times SU(2)_1$ {\it viz.}\ $\sigma_n^{\alpha
A}$ or $SU(2)_R \times SU(2)_1$ {\it viz.}\ $\sigma_n^{\dot \alpha A}$. A
general Ramond sector ground state is made up of these bosonic and
fermionic twist fields with the total twist $\sum n=N$ as
\begin{gather}
 |gr,gr \rangle = \prod_{n,\alpha,\dot \alpha, A,\dot A} (\sigma_n^{\alpha \dot \alpha})^{N_{n,\alpha\dot \alpha}} (\sigma_n^{A B})^{N_{n, AB}} (\sigma_n^{\alpha A})^{N_{n ,\alpha A}} (\sigma_n^{\dot \alpha A})^{N_{n,\dot \alpha A}}, \nonumber \\
 \sum_{n,\alpha,\dot \alpha, A,\dot A}  n( N_{n,\alpha\dot \alpha} + N_{n,AB} + N_{n,\alpha A}+N_{n,\dot \alpha A})=N, \nonumber \\
\qquad N_{n,\alpha\dot \alpha}=N_{n,AB}=0,1,2,\dots, \quad N_{n,\alpha A}=N_{n,\dot \alpha A}=0,1.
\end{gather}

\begin{figure}
\begin{center}
\begin{tabular}{ccc}
   \epsfxsize=5cm \epsfbox{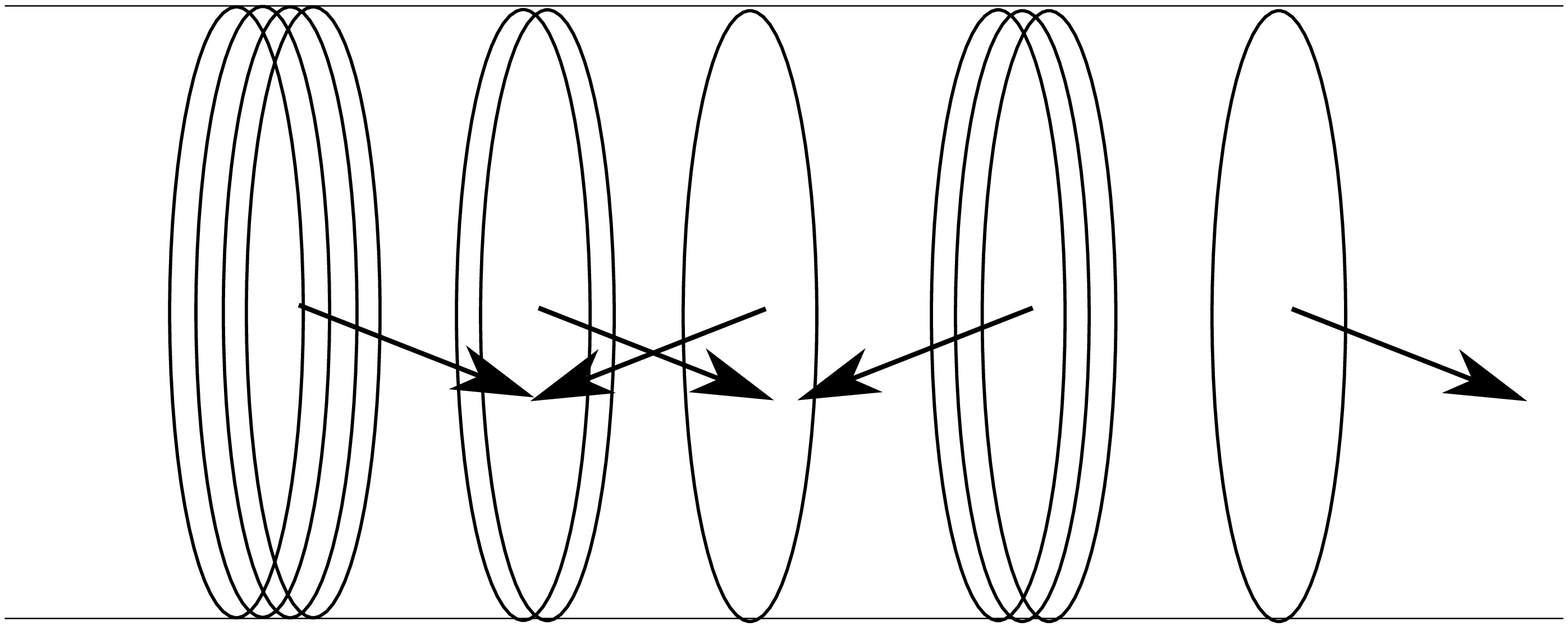} &
   \epsfxsize=5cm \epsfbox{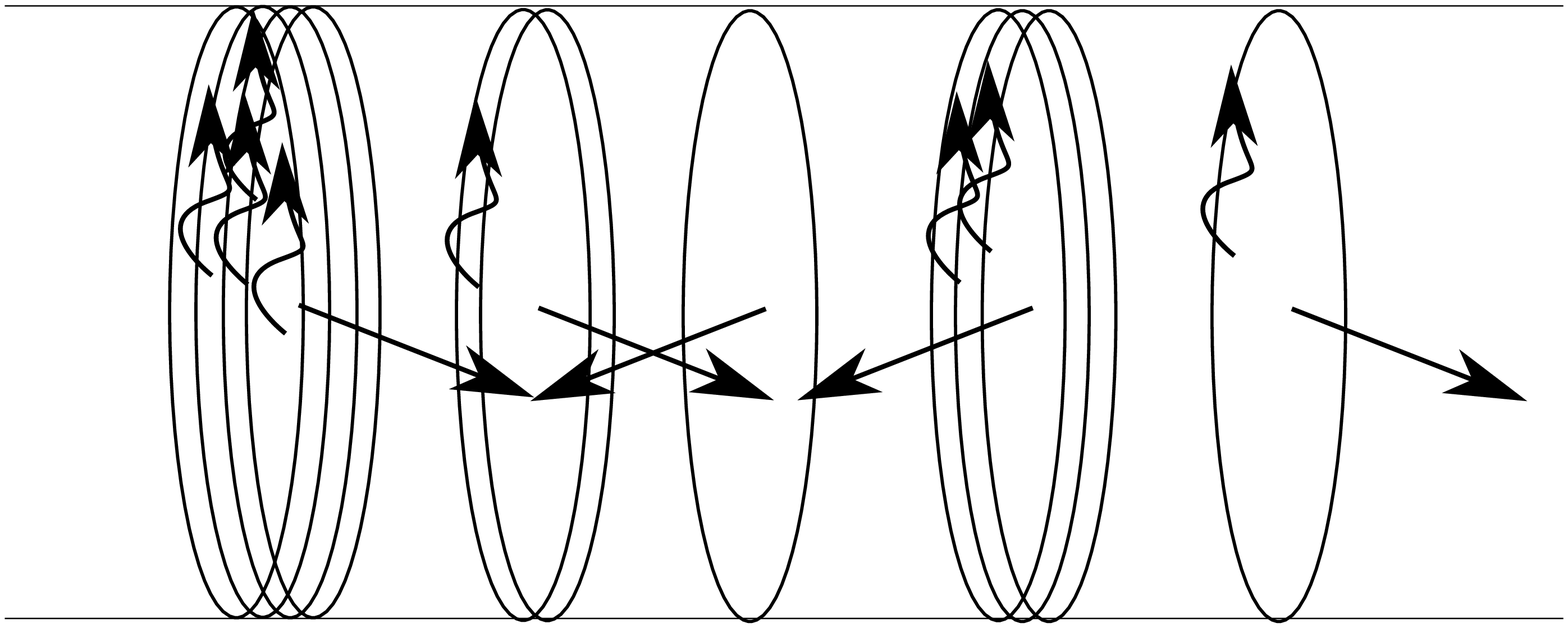} \\
(a) A ground state &
(b) A state with left excitations \\[3ex]
   \epsfxsize=5cm \epsfbox{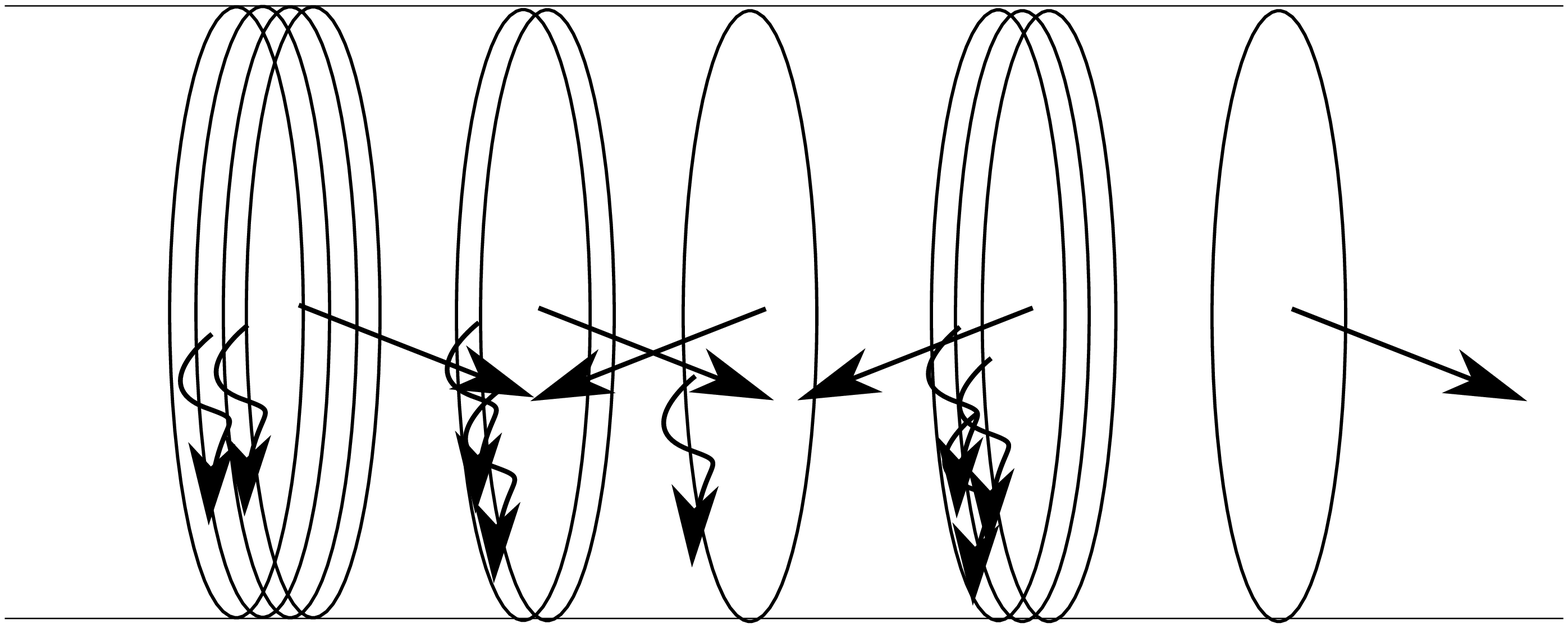} &
   \epsfxsize=5cm \epsfbox{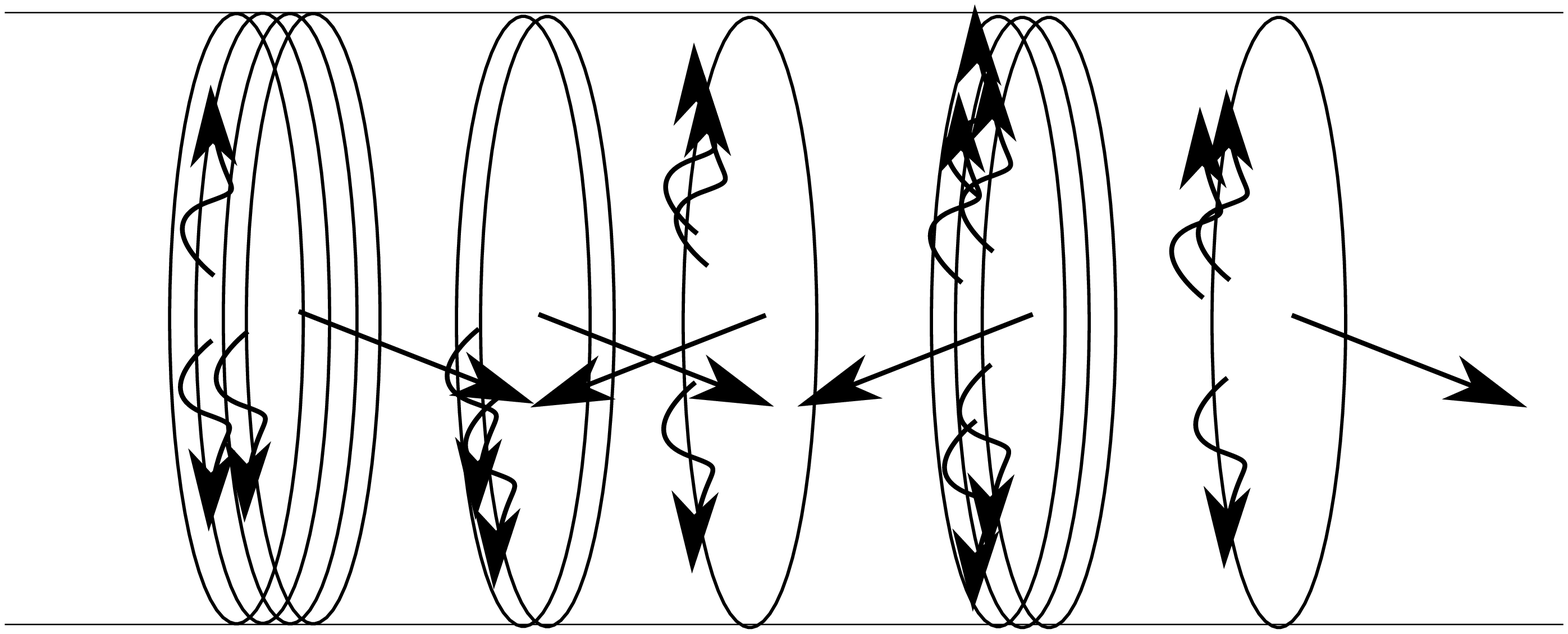} \\
(c) A state with right excitations &
(d) A state with left and right excitations
\end{tabular}\\[2ex]
 \caption{\sl Various states in the Ramond sector of the D1-D5 CFT.\label{fig:D1-D5states}}
\end{center}
\end{figure}

A general Ramond sector state is made of left- and right-moving excitations on the Ramond ground states
\begin{equation}
 |ex,gr\rangle, \qquad |gr,ex\rangle, \qquad |ex,ex \rangle
\end{equation}
where ``$ex$'' means acting on Ramond ground states ``$gr$'' by the bosonic and
fermionic modes. In Fig.~\ref{fig:D1-D5states} we diagrammatically
represent a Ramond ground state with no excitations, left excitations only, right
excitations only, and both. The arrows represent different $R$-charges of
elementary twists.
The states of the CFT are characterized by their left and right
dimension ($L_0$ and $\bar L_0$) and $R$-charges ($J_L$ and $J_R$). 
In our conventions, $J_{L,R}$, the third components of the $SU(2)_{L,R}$ generators $\vec
J_{L,R}$ are integers.  The Ramond
sector ground states all have the same dimension $L_0 = \bar
L_0=\f{N}{4}$. An excited state has dimension greater than that of the
ground state and any additional dimension is related to the left- and
right-moving momentum along the branes by
\be
N_p = L_0 - \f{N}{4}, \qquad \bar N_p = \bar L_0 - \f{N}{4} \label{DimAndMomentum}
\ee
The relation between the momentum and dimension is not so straightforward in the NS sector as different twist sectors have different dimensions.

The CFT also has an outer automorphism called ``spectral flow''
\cite{Schwimmer:1986mf}. Spectral flow by odd units maps states from NS
to R sector and vice versa whereas spectral flow by even units maps
states to states in the same sector. Under spectral flow by $\alpha$
units we have
\be
L_0'=L_0+ \h \alpha J_L + \f{1}{4} \alpha^2 N , \qquad J_L'=J_L+\alpha N.  \label{Eq:SpectralFlow}
\ee

\subsection{The enigmatic phase}

In this subsection, we will first describe two phases in CFT at the
orbifold point which are dual in the bulk to the BMPV black hole and to the
maximally-spinning smooth solution found by
Balasubramanian, Keski-Vakkuri, Ross and one of the authors, and by
Maldacena and Maoz \cite{Balasubramanian:2000rt, Maldacena:2000dr}.  We
will then explicitly construct the new phase, which we will call the
enigmatic phase, in CFT by combining properties of the BMPV and the
maximally-spinning phases. This will become clear as we proceed.  We
will then put this on a more rigorous footing by identifying the
enigmatic phase in the BPS partition function of the CFT\@. We will also
show that the elliptic genus fails to capture the enigmatic phase.

Our construction will be at the orbifold point of the CFT\@. Since the
elliptic genus fails to capture the enigmatic phase, it is logically
possible that this phase gets lifted once we move away from the orbifold
point of the CFT moduli space by turning on deformation and go to the
supergravity point.  We will explore the possibility of an enigmatic
phase on the gravity side in the next section.

\subsubsection{The BMPV phase}

\begin{figure}
 \begin{center}
 \begin{tabular}{ccc}
   \epsfxsize=5cm \epsfbox{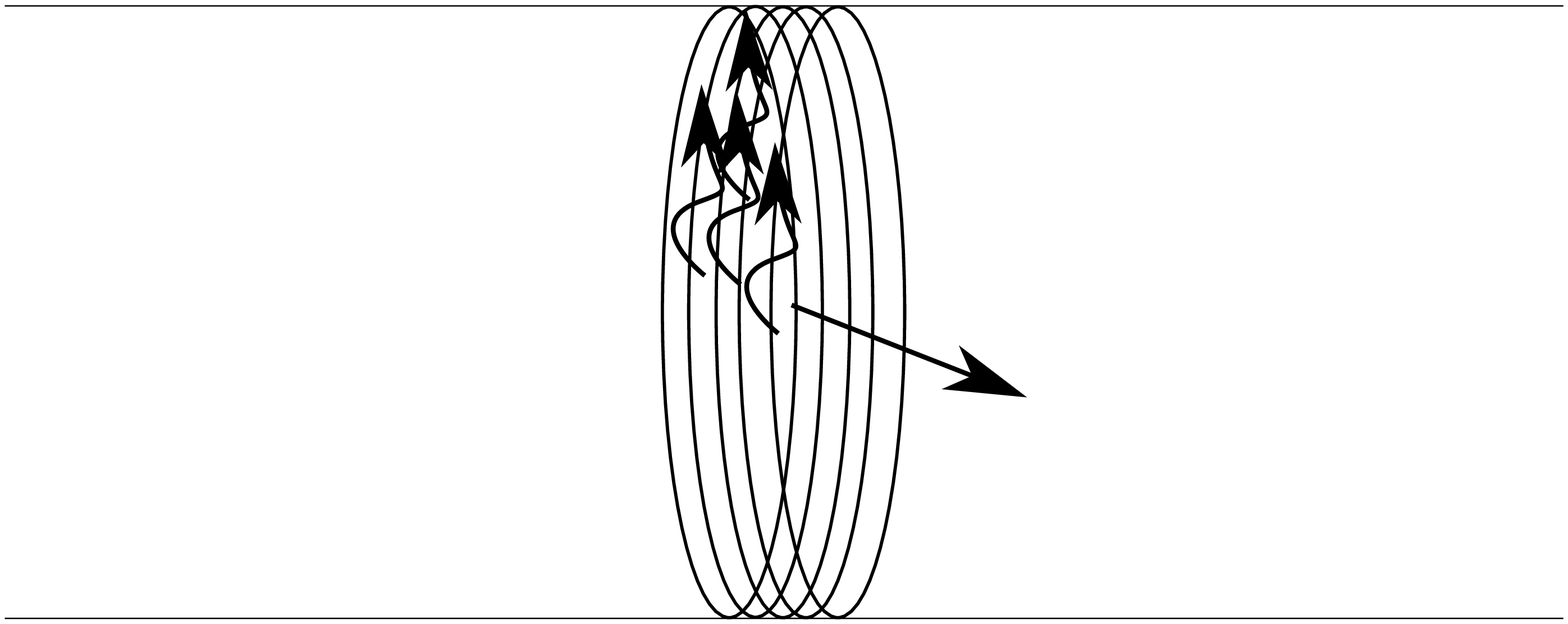} &
   \epsfxsize=5cm \epsfbox{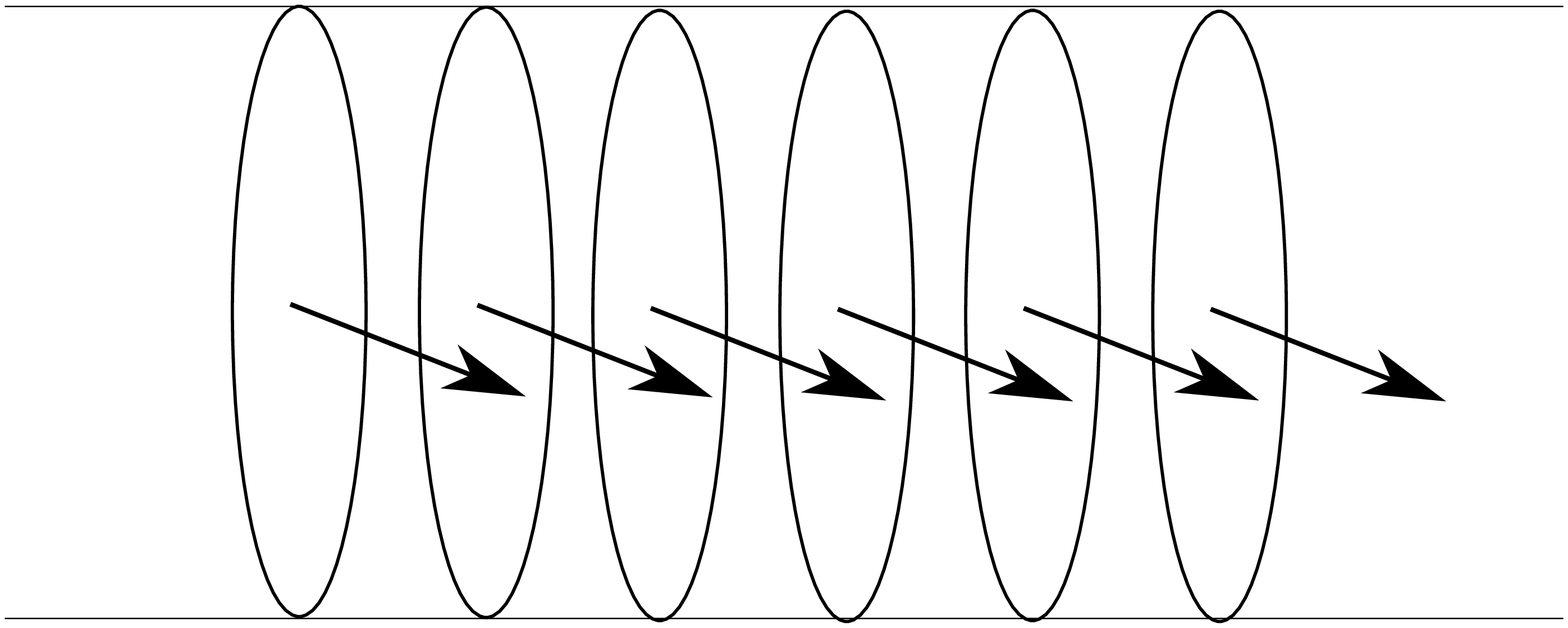} &
   \epsfxsize=5cm \epsfbox{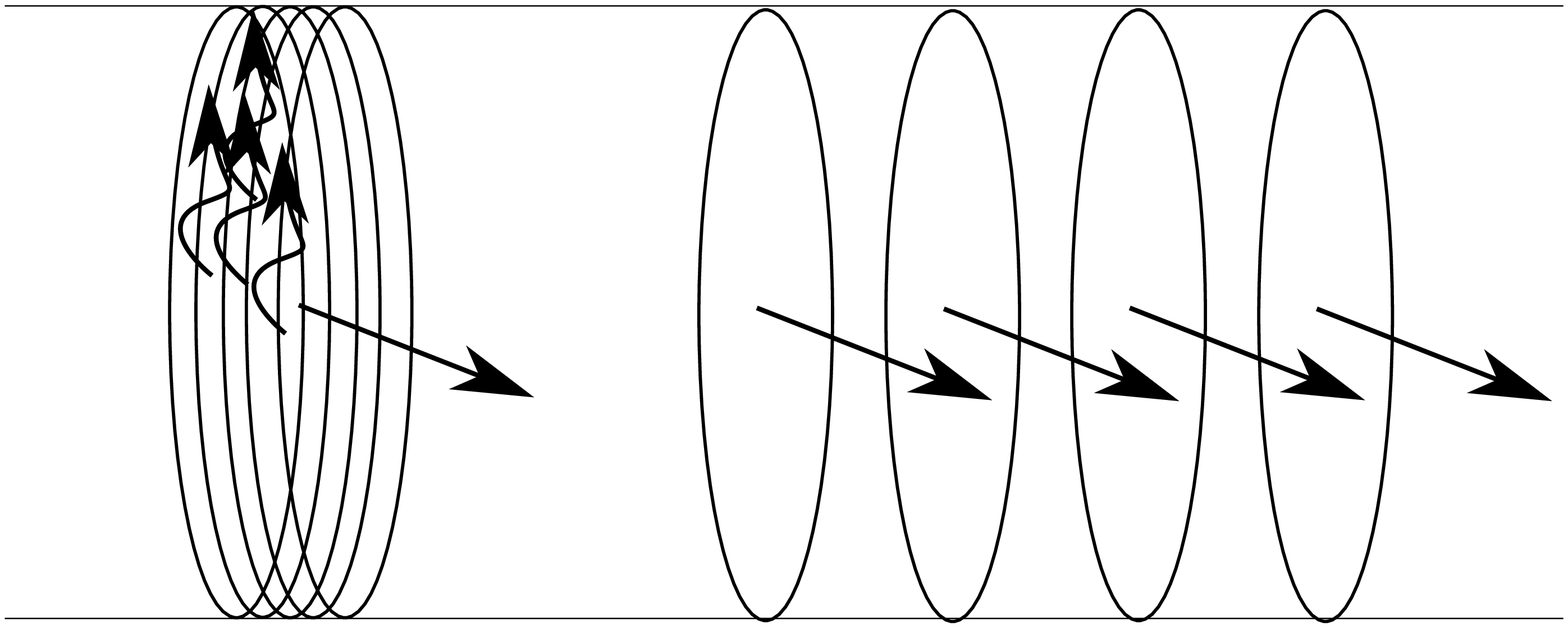}\\
 (a) BMPV&
 (b) Maximally Spinning&
 (c) Enigmatic Phase
 \end{tabular}\\[2ex]
 \caption{\sl Three phases at the orbifold point of the D1-D5 CFT.\label{fig:3phases}}
 \end{center}
\end{figure}

The BMPV black hole \cite{Breckenridge:1996is} has $U(1)_L \times
SU(2)_R$ symmetry and has an entropy \be S_{\text{BMPV}} = 2\pi \sqrt{N
N_p - J_L^2/4}.  \label{SBMPV} \ee Black holes have entropy and thus
their CFT duals are ensembles of states. The dual to BMPV black holes
consists of an ensemble of thermal excitations on the left-moving sector
on a long string \be (ex_L) \sigma_N^{++}.  \ee The $SU(2)_L$ charge is
carried by left-moving fermions. This phase is shown in a diagrammatic
way in Fig.~\ref{fig:3phases}(a).
The subleading corrections to the above picture come from $O(1)$
winding in short strings.

When the charges are large so that we are in the Cardy regime
$N_p-J_L^2/4N\gg N$, the Cardy formula (and the spectral flow symmetry) yields the same entropy as the Bekenstein-Hawking entropy of the black hole \eqref{SBMPV}.  Thus, in the Cardy
regime, we have a nice matching of the CFT and the bulk. 

\subsubsection{The maximally-spinning state}

Refs. \cite{Balasubramanian:2000rt, Maldacena:2000dr} found a family of smooth
solutions with $U(1)_L \times U(1)_R$ symmetry that have no horizon and thus no entropy. 
Their CFT dual states can be uniquely determined: they have all the
winding in single twists and their R-charges are in the largest
multiplet:
\be
(\sigma_1^{++})^N.
\ee
The phase is shown diagrammatically in Fig.~\ref{fig:3phases}(b).

This state has the largest possible value of $J_L$ among the ground
states, namely $J_L=N$.  Among other possible ground states with $J_L=N$
are
\begin{align}
 (\sigma_1^{++})^{N-j} (\sigma_1^{-+})^{j}, \qquad j=0,1,\dots, N.\label{SU(2)R_multiplet}
\end{align}
These form an $SU(2)_R$ multiplet with $|\vec J_R|=N$.

\subsubsection{The enigmatic phase}\label{subsec_enig_phase}

In the above, we discussed the BMPV phase which dominates at large momenta and the maximally-spinning state which has no momentum but very large angular momentum.
Now let us consider combining these two, namely an ensemble of
states where there is one long string and a condensate of short
strings, and ask what is the entropy maximizing ensemble with given $N_p,J_L$ (we assume $J_L>0$ without loss of generality).

All the excitations are carried by the long string (fractionation
ensures this is dominant \cite{Mathur:2005ai}). Let $l$ be the number of short
strings. Thus the long string has winding $N-l$. The short strings are
aligned with the left-moving angular momentum of the long strings so
have $J_L=l$, and symmetrization ensure that the short strings form an
$SU(2)_R$ multiplet with $|\vec J_R|=l$ just as in
\eqref{SU(2)R_multiplet}. Thus this phase has R-symmetry broken down to
$U(1)_L \times U(1)_R$. This phase is shown in Fig.~\ref{fig:3phases}(c).

\begin{figure}[htb]
\vspace*{-1cm}
\begin{center}
   \epsfxsize=12cm \epsfbox{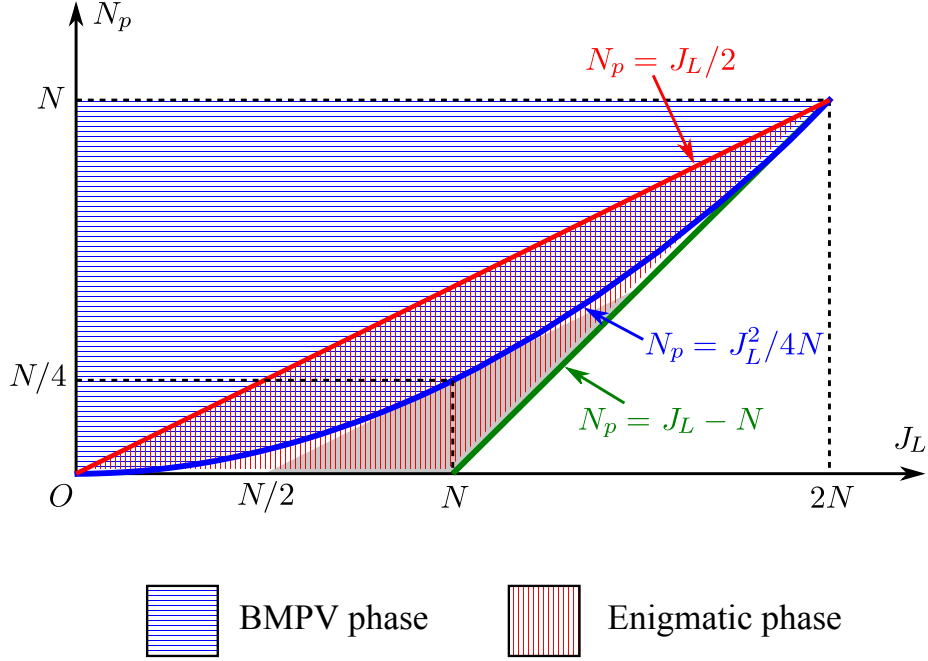}
\end{center}
\vspace*{-114cm}
\caption{\sl Phase diagram of D1-D5 CFT at the orbifold point.}  \label{fig:CFTphasediag}
\end{figure}

The entropy of this ``enigmatic'' phase comes from the long string
sector which is the same as that in the BMPV phase albeit with different
winding number and angular momentum
\be
S_{\text{enigma},l} = 2\pi \sqrt{(N-l)N_p - \f{1}{4} (J_L-l)^2}.
\ee
Maximizing this entropy with respect to $l$, the winding in the short
strings, we get the optimal number of short strings to
be\footnote{Splitting the system into two parts and choosing the way of
splitting so that the entropy is maximized is reminiscent of the
procedure taken in \cite{Bhattacharyya:2010yg} where the system is split
into a ``non-interacting mix'' of a black hole and a charged
condensate.}
\be
l=J_L-2 N_p
\label{maxmizingl}
\ee
and the entropy for this is
\be
S_{\text{enigma}}=2\pi \sqrt{N_p ( N_p+N-J_L)}.
\label{Senigma}
\ee
The enigmatic phase exists in the region where the square of the entropy is positive and the number of short strings is greater than zero, namely
\be
N_p>0, \qquad  N_p+N-J_L>0,\qquad J_L-2 N_p>0. \label{Eq:BoundariesNewPhase}
\ee
This means that $N_p\sim J_L\sim N$ and therefore this phase exists
\emph{outside} the Cardy regime.  In addition the new phase is charged under
$SU(2)_R$ with
\be\label{Eq:CFT_JL_JR_rel}
|\vec J_R|=J_L-2N_p.
\ee
The $N_p$-$J_L$ diagram showing the BMPV and the enigmatic phases
are plotted in Fig.~\ref{fig:CFTphasediag}.
It is straightforward to see that
\be
S_{\text{enigma}}^2 - S_{\text{BMPV}}^2 = \left(\f{J_L}{2}-N_p\right)^2\ge 0 
\ee
and thus the enigmatic phase is dominant
over the BMPV phase and smoothly merges into it at the upper boundary of the
``wedge'' in Fig.~\ref{fig:CFTphasediag}.  We emphasize that in the region
where  the enigmatic phase and the BMPV phase coexist the former dominates in
entropy.

\subsubsection*{Spectral-flowed enigmatic phase}
\vspace{-1ex}

The enigmatic phase was constructed by splitting the CFT effective string into
two parts. The long string carried all the excitations and thus the
entropy and the short strings carried part of the angular momentum but no
excitations. There is another configuration where the short strings
carry part of the angular momentum but no entropy and that is gotten by making each
short-string excitation of the form
\be
\psi^{+1}_{-1} \psi^{+2}_{-1} \sigma_1^{++}.
\ee
In fact we can fill the fermions on the short strings up to a higher level $\eta$ this way. For example, the short string for $\eta=3$ corresponds to
\be
\psi^{+1}_{-3} \psi^{+2}_{-3}\psi^{+1}_{-2} \psi^{+2}_{-2} \psi^{+1}_{-1} \psi^{+2}_{-1} \sigma_1^{++}.
\ee
Such configurations are obtained from the original configuration by spectral flow \bref{Eq:SpectralFlow} by $2\eta$ units. Using \bref{DimAndMomentum} to rewrite the enigmatic phase in terms of the dimension rather than the momentum: 
\be
S=2 \pi \sqrt{ \Bigl(L_0 - \f{N}{4}\Bigr)\Bigl(L_0 - J_L +\f{3N}{4}\Bigr)}
\ee
one obtains the entropy of these spectral-flowed states:
\be
\label{eq:flowedSnew}
S=2\pi \sqrt{ 
\Bigl[L_0 - \eta  J_L +\Bigl( \eta^2 - \f{1}{4}\Bigr) N\Bigr]
\Bigl[L_0 - (\eta+1)  J_L +\Bigl( (\eta+1)^2 - \f{1}{4}\Bigr) N\Bigr]  },
\ee
which is just a spectral-flowed version of the entropy formula by $-2 \eta$ units.
This expression is valid in both NS and R sectors. We can then express this result in the Ramond sector in terms of the momentum using \bref{DimAndMomentum} 
\be
S=2\pi \sqrt{[N_p-\eta J_L+ \eta^2 N][N_p-(\eta+1)J_L+(\eta+1)^2 N]}.
\ee
As a simple example of this formula we see that we get the expression for the mirror image wedge ($J_L \to -J_L$) by taking $\eta=-1$.
\begin{figure}[htb]
\begin{center}
   \epsfxsize=12cm \epsfbox{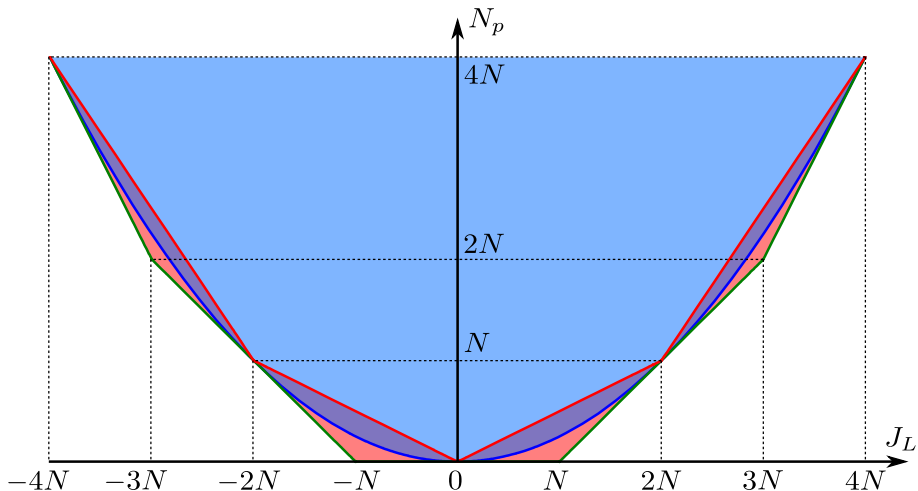}
\end{center}
\caption{\sl Spectral-flowed enigmatic phases}  \label{fig:spectralFlowedEnigmaticPhase}
\end{figure}
The region in which the spectral-flowed new phase exists is found by mapping the boundaries of the non-spectral-flowed new phase \bref{Eq:BoundariesNewPhase}:
\be
N_p-\eta J_L+ \eta^2 N>0, 
\quad N_p-(\eta+1)J_L+(\eta+1)^2 N>0, 
\quad J_L(1+2\eta)-2 N_p-2\eta(1+\eta)N>0. \label{Eq:BoundariesNewPhaseSF}
\ee
In Fig.~\ref{fig:spectralFlowedEnigmaticPhase} we show four such enigmatic phases for $\eta=-2,-1,0,1$.

Note that, although the arguments above are for $M^4=T^4$, the new phase
should exist also for $M^4=K3$ with the same entropy formula
\eqref{Senigma}.  This is because the structures we used above, such as
effective strings and operators $\sigma^{\pm +}$, are common to both
$T^4$ and $K3$.

\subsection{Numerical evaluation of partition function}
\label{subsec:numericalEvaluation}



The analysis of the previous section showing a new ``enigmatic'' phase
can be put on a firmer footing by looking at the partition function
which counts all the states of the system with given charges. We will
evaluate the BPS partition function at the orbifold point of the CFT for both
$T^4$ and $K3$ compactifications,
and find that it indeed shows the growth expected from
the entropy of the enigmatic phase. 
In the non-Cardy regime where the enigmatic phase exists, the BPS
partition function is not easy to evaluate because we cannot use its
modular properties.  We overcome this problem by evaluating it numerically.

The BPS partition function computes the absolute degeneracy but is not
protected under marginal deformations unlike the elliptic genus. We will
also look at the (modified) elliptic genus on $K3$ ($T^4$) in the
non-Cardy regime where the enigmatic phase exists to see if we find any
trace of the enigmatic phase.  The result is that these elliptic genera
do not capture the enigmatic phase.  This is as it should be, because
the new phase exists in a region where the supergravity elliptic genus
was found to match that of the CFT \cite{deBoer:1998us,
Maldacena:1999bp}. Thus finding a new black object phase in the elliptic
genera would have been a contradiction.  In Appendix
\ref{sec:newphasedoesntcontributetoindex}, we give an argument why the
particular states of the form of a long string with excitations on it
plus multiple short strings of length one do not contribute to the
elliptic genus.

The readers who are not interested in the details of the computation can directly
jump to \S\ref{sss:num_part_func} where the final results are presented.


\bigskip
We begin by first defining the quantities we compute. The BPS partition function is defined as
\be
\chi_{PF}^{} = \Tr_{RR;any,gnd}[q^{L_0-\f{c}{24}} y^{J_L}],
\qquad q=e^{2\pi i \sigma},\qquad y=e^{2\pi i\upsilon}
\ee
where the trace is taken over all states in the left-moving Ramond
sector and ground states in the right-moving Ramond sector.  Namely,
$\chi_{PF}^{}$ counts BPS states only.  The elliptic genus is defined as
\be
\chi_{EG}^{} = \Tr_{RR}[(-1)^{J_L-J_R} q^{L_0-\f{c}{24}}  y^{J_L}]
\label{def-EG}
\ee
and the 
modified elliptic genus as
\be
\chi_{MEG}^{} = \Tr_{RR}[(-1)^{J_L-J_R}(J_R)^2 q^{L_0-\f{c}{24}}  y^{J_L}].
\label{def-MEG}
\ee
where the traces are taken over all states in the left and right Ramond
sectors.  Even though the trace is taken over all states, it is easy to see that the elliptic genera only count states in the right-moving sector.


\subsubsection{BPS partition function and elliptic genera on single copy of $K3$ and $T^4$}

Let us first discuss the BPS partition function and elliptic genera on a
single copy of $T^4$ and $K3$. Based on this, we will compute the BPS partition
function and elliptic genera for symmetric products $\Sym^N(K3)$ and
$\Sym^N(T^4)$.

\begin{itemize}
\item{\bf Elliptic genus on $K3$}\newline
The elliptic genus on K3 was found in \cite{Eguchi:1988vra,Kawai:1993jk} and is given by
\be
\chi_{EG}(q,y;K3)= 8 \left[ \left( \f{\vartheta_2(\upsilon,\sigma)}{\vartheta_2(0,\sigma)} \right)^2+\left( \f{\vartheta_3(\upsilon,\sigma)}{\vartheta_3(0,\sigma)} \right)^2+\left( \f{\vartheta_4(\upsilon,\sigma)}{\vartheta_4(0,\sigma)} \right)^2 \right].
\ee
The elliptic genus is protected and is the same everywhere in the moduli space of $K3$ surfaces.
From the definition of elliptic genus we see that the coefficient of
     $q^m y^l$ counts the difference between the number of bosonic and
     fermionic states with $L_0=m+\f{c}{24}$ and $J_L=l$. Thus we have
\be
\chi_{EG}(q,y;K3)= \sum_{m\ge 0,l} (c^B_{K3}(m,l) - c^F_{K3}(m,l)) q^m y^l.
\ee
\item{\bf BPS partition function on $K3$}\newline The BPS partition
     function is not invariant under changes in moduli and thus depends
     on the point in the $K3$ moduli space where it is evaluated.  So,
     in principle we should evaluate it at all points in the moduli space
     in order to show that it points toward the existence of the
     enigmatic phase.  However, the BPS partition function can be
     computed only at special points in the $K3$ moduli space, and that
     is what we will content ourselves with.

     The partition function for $K3$ can be computed
     \cite{Eguchi:1988vra} at the orbifold points\footnote{These
     orbifold points in the moduli space of $K3$ surfaces are not to be
     confused with the orbifold points in the moduli space of D1-D5 CFT
     where the target space is a symmetric orbifold of the $K3$
     surface.} in the $K3$ moduli space, where $K3$ can be written as
     $T^4/\bbZ_l$ ($l=2,3,4,6$), and the BPS partition function can be
     extracted from it.
     For illustrative purposes, we present the BPS partition function at
     the orbifold point where $K3=T^4/\bbZ_2$. This can be evaluated in a
     straightforward way using the results of \cite{Eguchi:1988vra} and
     is found to be\footnote{We ignored zero modes because they do not contribute to the BPS partition function for generic moduli of the parent $T^4$.}
\begin{align}
 \chi_{PF}(q,y;K3=T^4/\mathbb Z_2) 
 =2 \f{\vartheta_2(\upsilon,\sigma)^2}{\eta(\sigma)^6} +
 16 \left(\f{\vartheta_4(\upsilon,\sigma)}{\vartheta_3(0,\sigma)} \right)^2 
 +8
 \left(\f{\vartheta_2(0,\sigma)}{\vartheta_4(0,\sigma)} \right)^2
 \left(\f{\vartheta_2(\upsilon,\sigma)}{\vartheta_3(0,\sigma)} \right)^2.
 \nn
\end{align}
From the definition of the partition function we can see that the coefficient of $q^my^l$ counts the total number of states, both bosonic and fermionic, with $L_0=m+\f{c}{24}$ and $J_L=l$. Thus we have
\be
\chi_{PF}(q,y;K3=T^4/\mathbb Z_2)= \sum_{m\ge 0,l} (c^B_{K3}(m,l) + c^F_{K3}(m,l)) q^m y^l.
\ee
\item{\bf The modified elliptic genus on $T^4$}\newline
The usual elliptic genus on $T^4$ vanishes identically because of extra
     fermion zero modes. On the other hand, the modified elliptic genus,
     which soaks up the extra fermion zero modes,
     is nonvanishing and given by \cite{Maldacena:1999bp}
\be
\chi_{MEG}^{}(q,y;T^4)= -2 \left[\f{\theta_1(\upsilon,\sigma)}{\eta(\sigma)^3} \right]^2.
\ee
The coefficient of $q^m y^l$ again counts the difference between number
of bosons and fermions of with $L_0=m+ \f{c}{24}$ and $J_L=l$. However
because half the fermion zero modes are soaked up, the coefficient only
counts the states built on the other half of the fermion zero modes
     \cite{Maldacena:1999bp}.
To find the modified elliptic genus for the symmetric product, we will only
need the total coefficient
\be
\chi_{MEG}(q,y;T^4)=  \sum_{m\ge 0,l} c_{MEG;T^4}(m,l) q^m y^l.
\ee
\item{\bf BPS partition function on $T^4$}\newline
The BPS partition function for $T^4$ is straightforward to evaluate
     because it is a free theory.  The result is found to
     be\footnote{Again, we ignored zero modes because they do not
     contribute for the generic moduli of
     $T^4$.}
\be
\chi_{PF}(q,y;T^4)= 4 \left[\f{\vartheta_2(q,y)}{\eta(q)^3} \right]^2.
\ee
The coefficient of $q^m y^l$ counts the total number of states, both bosonic and fermionic, with $L_0 = m+ \f{c}{24}$ and $J_L=l$. However from the vanishing of the elliptic genus for $T^4$ we know that the number of bosonic and fermionic states are equal and so we have
\be
\chi_{PF}(q,y;T^4)= \sum_{m\ge 0,l} c_{PF;T^4} (m,l) q^m y^l,
\ee
where
\be
c^B_{T^4}(m,l) = c^F_{T^4}(m,l) = \h c_{PF;T^4}(m,l).
\label{cB=cF_T4}
\ee
\end{itemize}

\subsubsection{BPS partition function and elliptic genera on  $\Sym^N(K3)$ and $\Sym^N(T^4)$}

Next we discuss the elliptic genus and BPS partition function on $\Sym^N
(K3)$ and the modified elliptic genus and BPS partition function on
$\Sym^N(T^4)$.
\begin{itemize}
\item{\bf Elliptic genus on $\Sym^N(K3)$}\newline In
\cite{Dijkgraaf:1996xw} the generating function for the elliptic genus
on the symmetric product $\Sym^N(K3)$ was found to be
\be
\sum_{N\ge 0} p^N \chi_{EG}(q,y;\Sym^N(K3))= \prod_{n\ge 1,m\ge 0,l} \f{1}{(1-p^n q^m y^l)^{c^B_{K3}(mn,l) - c^F_{K3}(mn,l) }}.
\ee
We can expand the elliptic genus for $\Sym^N(K3)$ as
\be
\chi_{EG}(q,y;\Sym^N(K3))=\sum_{M \ge 1,L} C_{EG;K3}(N,M,L) q^M y^L,
\ee
where $C_{EG;K3}(N,M,L)$ counts the difference in bosonic and fermionic
states with $L_0=M+ \f{c}{24}$ and $J_L=L$ on $\Sym^N(K3)$. Let us define
\be
S_{EG;K3}(N,M,L)=\log |C_{EG;K3}(N,M,L)|. \label{Eq:EntropyEGK3}
\ee
     By a slight abuse of terminology, we will refer to  the logarithm of
     (modified) elliptic genus, such as $S_{EG;K3}(N,M,L)$ above,  as ``entropy''.

\item{\bf BPS partition function on $\Sym^N(K3=T^4/\bbZ_2)$}\newline
The generating function for the BPS partition function on the symmetric
product $\Sym^N (K3=T^4/\bbZ_2)$ can be easily found using the techniques
of \cite{Dijkgraaf:1996xw} to be
\be
\sum_{N\ge 0} p^N \chi_{PF}(q,y;\Sym^N (K3=T^4/Z_2))= \prod_{n\ge 1,m\ge 0,l} \f{(1+p^n q^m y^l)^{c^F_{K3}(mn,l)}}{(1-p^n q^m y^l)^{c^B_{K3}(mn,l)}}.
\ee
We can expand the BPS partition function for $\Sym^N(K3=T^4/\bbZ_2)$ as
\be
\chi_{PF}(q,y;\Sym^N (K3=T^4/\bbZ_2))=\sum_{M \ge 1,L} C_{PF;K3}(N,M,L) q^M y^L,
\ee
where $C_{PF;K3}(N,M,L)$ counts the total number of states, both bosonic
and fermionic, with $L_0=M+ \f{c}{24}$ and $J_L=L$ on
$\Sym^N(K3=T^4/\bbZ_2)$. We denote the associated entropy by
\be
S_{PF;K3}(N,M,L)=\log C_{PF;K3}(N,M,L) \label{Eq:EntropyPFK3}.
\ee
\item{\bf Modified elliptic genus on $\Sym^N(T^4)$}\newline
In \cite{Maldacena:1999bp} the generating function for the modified elliptic genus on $\Sym^N(T^4)$ was found to be
\be
\sum_{N\ge 0} p^N \chi_{MEG}(q,y;\Sym^N(T^4)) =\sum s~(p^nq^m y^l)^s\,  c_{MEG}( mn ,l).
\ee
The modified elliptic genus for $\Sym^N(T^4)$ can be expanded as
\be
\chi_{MEG}(q,y;\Sym^N(T^4))=\sum_{M \ge 1,L} C_{MEG;T^4}(N,M,L) q^M y^L,
\ee
where $C_{MEG;T^4}(N,M,L)$ counts the difference in bosonic and
fermionic states with $L_0=M+ \f{c}{24}$ and $J_L=L$ on
$\Sym^N(T^4)$. However because it soaks up half the zero modes it counts
only the states built on the other half of the zero modes. We denote the associated ``entropy'' by
\be
S_{MEG;T^4}(N,M,L)=\log |C_{MEG;T^4}(N,M,L)|. \label{Eq:EntropyMEGT4} 
\ee
\item{\bf BPS partition function on $\Sym^N(T^4)$}\newline
The generating function for the BPS partition function on the symmetric product $\Sym^N (T^4)$ can also be easily found using the techniques of \cite{Dijkgraaf:1996xw}  to be
\be
\sum_{N\ge 0} p^N \chi_{PF}(q,y;\Sym^N (T^4))= \prod_{n\ge 1,m\ge 0,l} \left(\f{1+p^n q^m y^l}{1-p^n q^m y^l} \right)^{\h c_{PF;T^4}(m n ,l)}
\ee
where we used the relation \eqref{cB=cF_T4}.
We can expand the partition function for $\Sym^N(T^4)$ as
\be
\chi_{PF}(q,y;\Sym^N (T^4))=\sum_{M \ge 1,L} C_{PF;T^4}(N,M,L) q^M y^L,
\ee
where $C_{PF;K3}(N,M,L)$ counts the total number of states, both bosonic and fermionic, with $L_0=M+ \f{c}{24}$ and $J_L=L$ on $\Sym^N(T^4)$. We denote the associated entropy by
\be
S_{PF;T^4}(N,M,L)=\log C_{PF;T^4}(N,M,L). \label{Eq:EntropyPFT4}
\ee

\end{itemize}

\subsubsection{Numerical evaluation of partition functions and elliptic genera}
\label{sss:num_part_func}

Here we give the results of the numerical evaluation of the various
entropies
\bref{Eq:EntropyEGK3}, \bref{Eq:EntropyPFK3}, \bref{Eq:EntropyMEGT4} and
\bref{Eq:EntropyPFT4}.  We present the results by plotting
$S(N,N_p,J_L)$ (blue dots) against $J_L$ for different values of $N$ with
$\f{N}{N_p}=5$ along with the BMPV entropy (thin interior blue line) given by the
Cardy formula \bref{SBMPV} and the enigmatic phase entropy \bref{Senigma} (thin exterior red line).
\begin{itemize}
\item{\bf Elliptic genus on $\Sym^N(K3)$}\newline In Fig.~\ref{fig:egK3}
     we plot $S_{EG;K3}(N,N_p,J_L)$ against $J_L$. We see that for small
     values of $J_L$ the elliptic genus matches the Cardy formula but
     not the new phase.  At some value of $J_L$ ``shoulders'' appear in
     the elliptic genus and it deviates from the Cardy formula but still
     does not match the enigmatic phase entropy.  Further plots for
     larger values of $N,N_p$ keeping $N_p/N=1/5$ fixed show us that the
     shoulders appear at larger values of $J_L$ and are smaller. In
     fact, a numerical analysis hints at the bump coming from
     logarithmic corrections to the Cardy formula that vanish as
     $N,N_p\to\infty$. We thus conclude that the elliptic genus does not
     capture the enigmatic phase and asymptotes to the BMPV
     entropy.\footnote{This is consistent with \cite{Castro:2008ys}
     where it was shown that the $K3$ elliptic genus goes as
     \eqref{SCardy} as long as all charges $N,N_p,J_L$ are large, both
  in the Cardy  and non-Cardy regimes.}
\begin{figure}
 \begin{center}
 \begin{tabular}{ccc}
   \epsfxsize=5cm \epsfbox{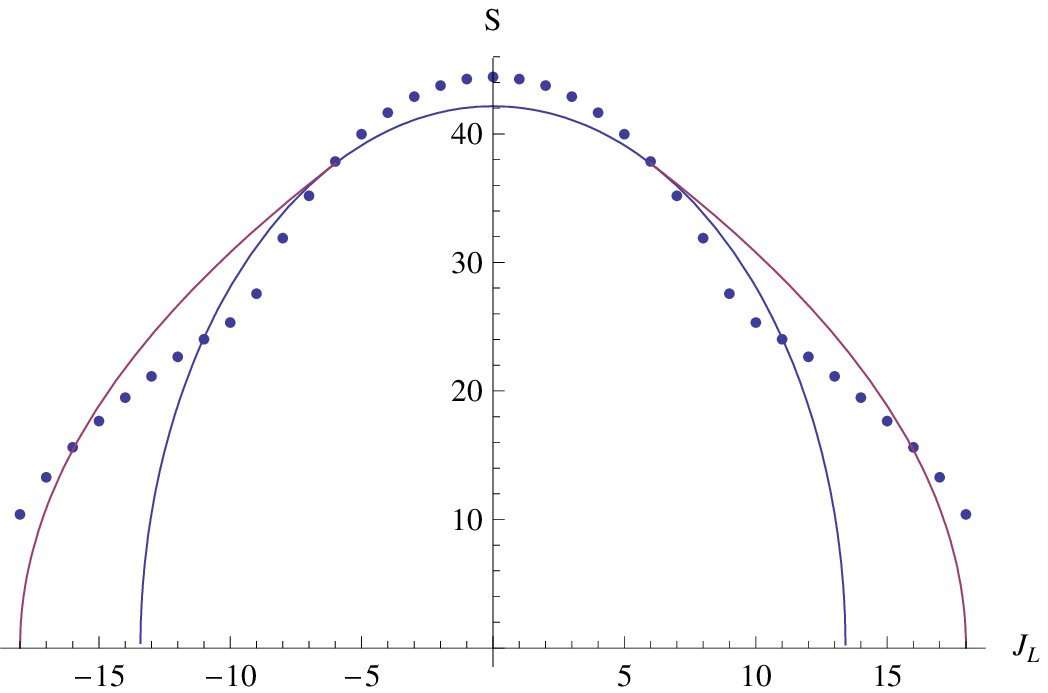} &
   \epsfxsize=5cm \epsfbox{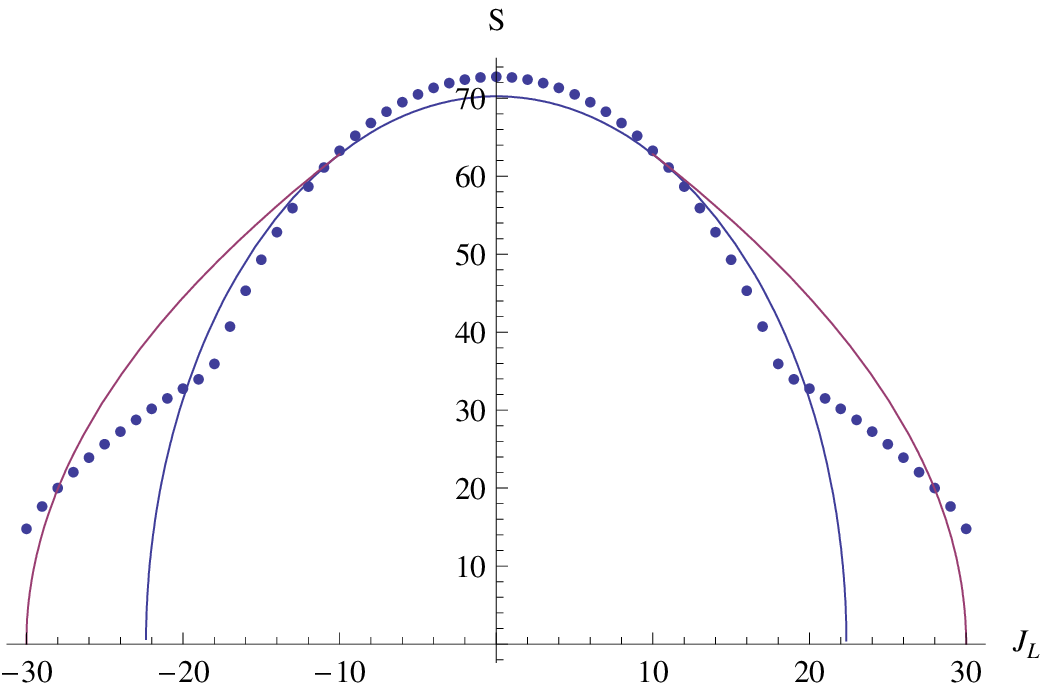} &
   \epsfxsize=5cm \epsfbox{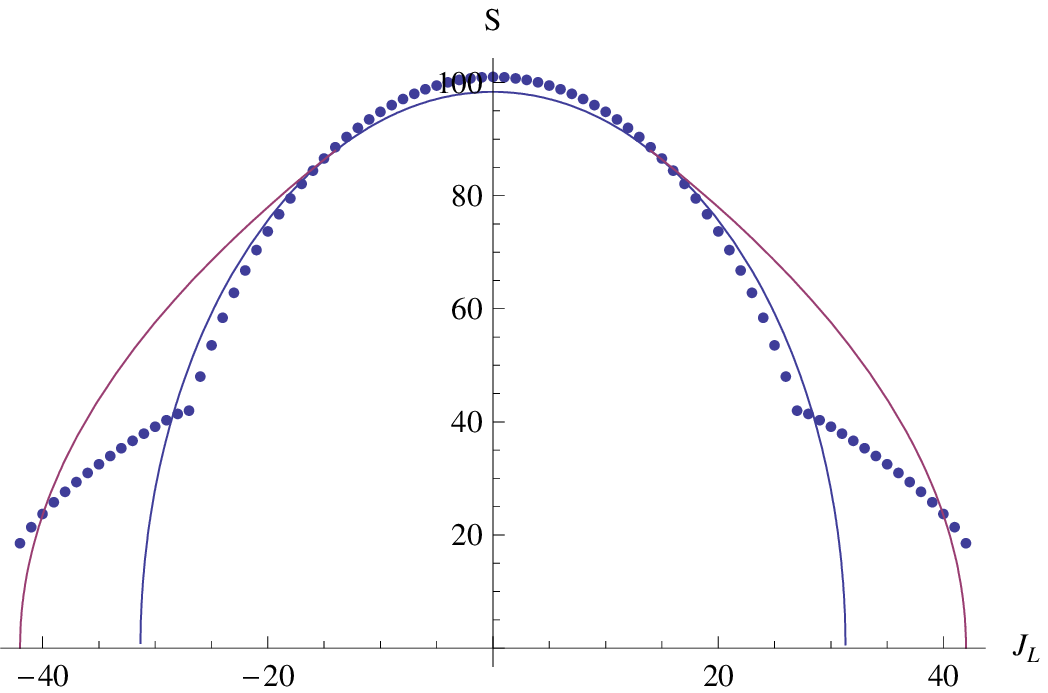}\\
 $N=15,N_p=3$&
 $N=25,N_p=5$&
 $N=35,N_p=7$
 \end{tabular}
 \caption{\sl The logarithm of the elliptic genus for $\Sym^N (K3)$.
  \label{fig:egK3}}
 \end{center}
\end{figure}
\item{\bf BPS partition function on $\Sym^N(K3=T^4/\bbZ_2)$}\newline In
Fig.~\ref{fig:pfK3} we plot $S_{PF;K3}(N,N_p,J_L)$ against $J_L$. We
see that the partition function for $\Sym^N(K3)$ indeed captures the new
phase.  With larger values of $N$ the match of the partition function to
the enigmatic phase entropy \bref{SBMPV} calculated in the large $N$ limit in
the previous subsection seems to get better but we were limited in our
analysis by computational power.
\begin{figure}
 \begin{center}
 \begin{tabular}{ccc}
   \epsfxsize=5cm \epsfbox{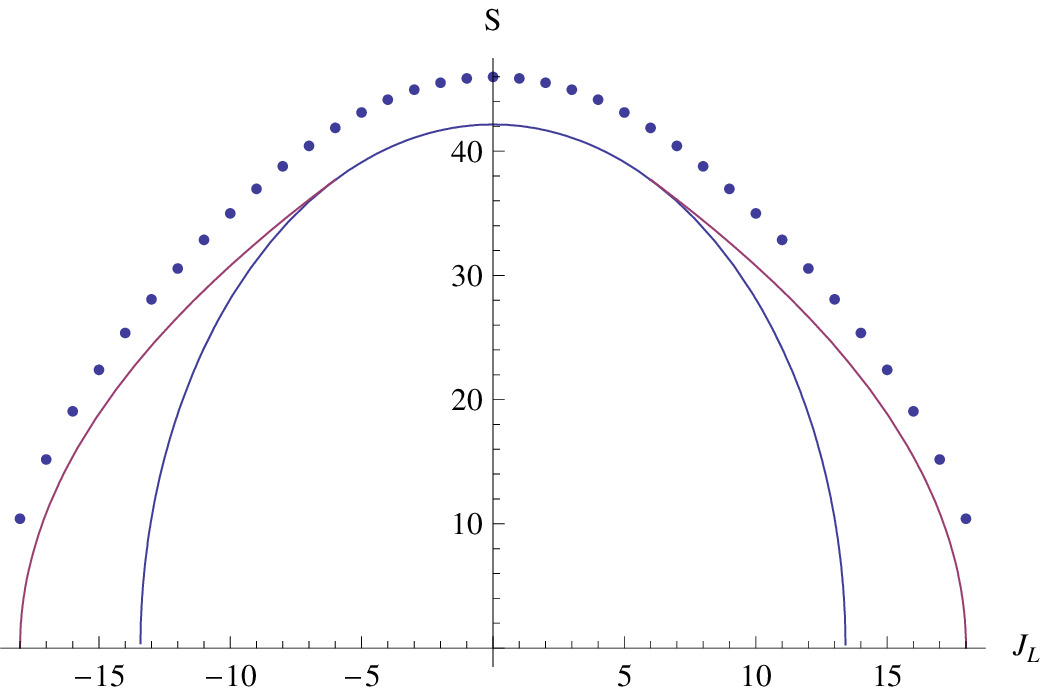} &
   \epsfxsize=5cm \epsfbox{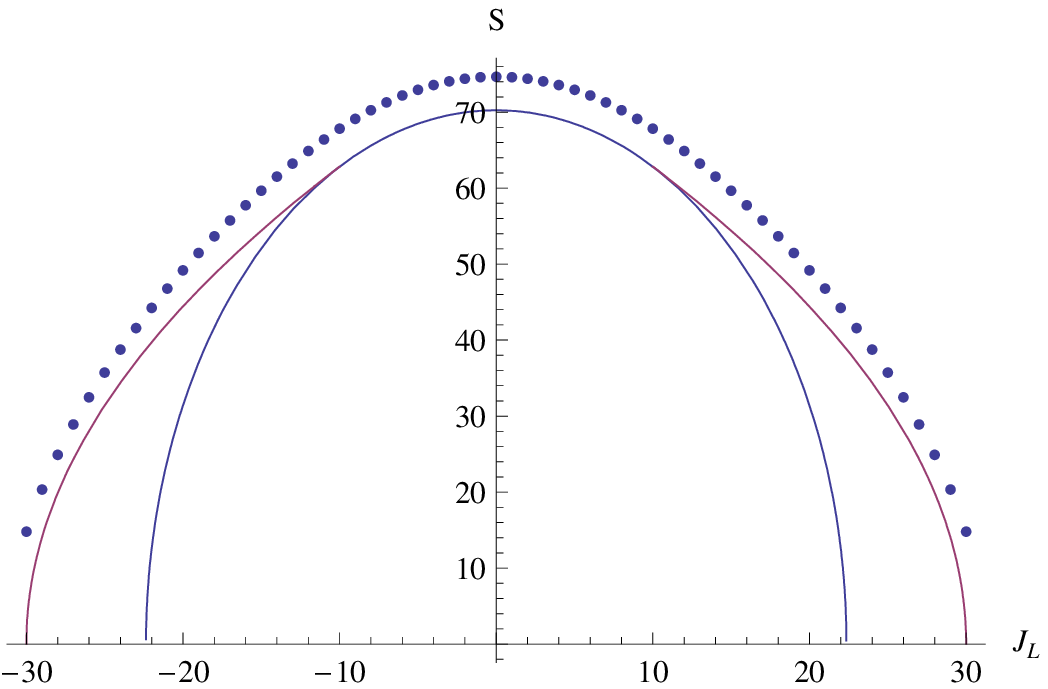} &
   \epsfxsize=5cm \epsfbox{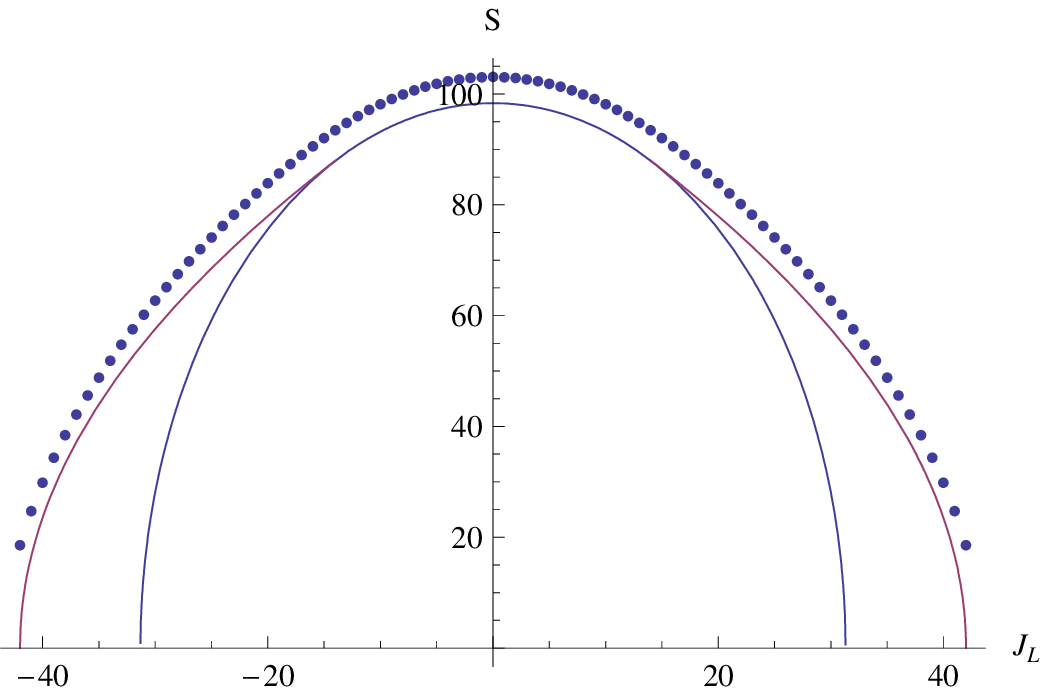}\\
 $N=15,N_p=3$&
 $N=25,N_p=5$&
 $N=35,N_p=7$
 \end{tabular}
 \caption{\sl The logarithm of the BPS partition function for $\Sym^N(K3=T^4/\mathbb Z_2)$.
  \label{fig:pfK3}}
 \end{center}
\end{figure}
Although here we presented the result for $\Sym^N(K3=T^4/\bbZ_l)$ with $l=2$,
     we worked out the other cases $l=3,4,6$ as well and obtained similar
     behavior.

\item{\bf Modified elliptic genus on $\Sym^N(T^4)$}\newline
In Fig.~\ref{fig:megT4} we plot $S_{MEG;T^4}(N,M,L)$ against $J_L$. Just as the elliptic genus for $\Sym^N(K3)$, the modified elliptic genus approaches the BMPV entropy for large $N$ but fails to capture the enigmatic phase.
\begin{figure}
 \begin{center}
 \begin{tabular}{ccc}
   \epsfxsize=5cm \epsfbox{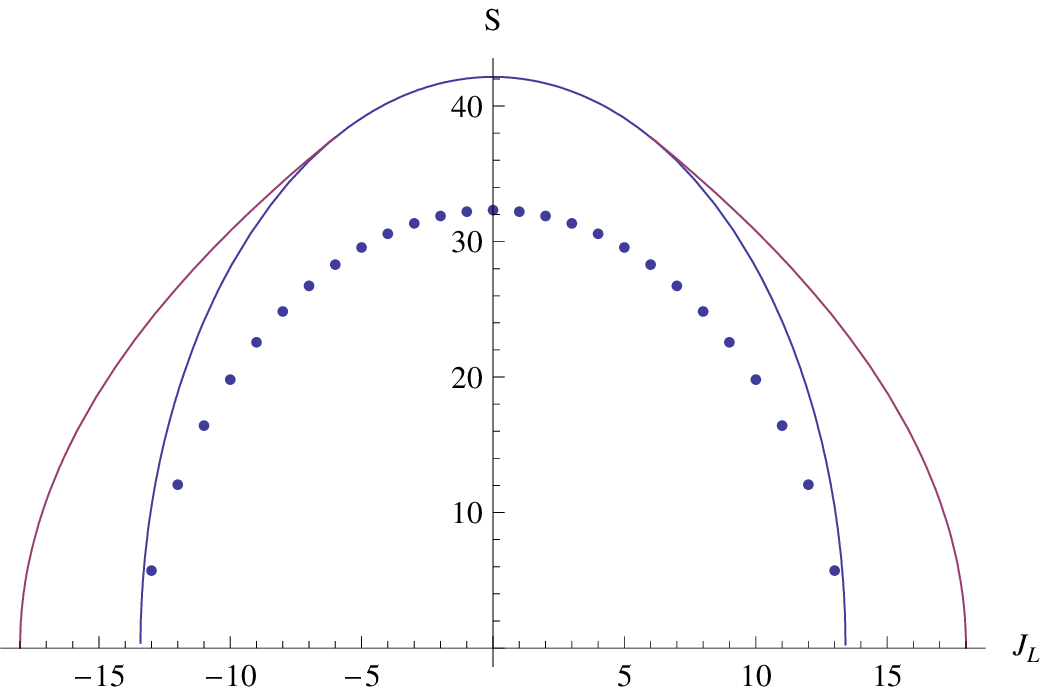} &
   \epsfxsize=5cm \epsfbox{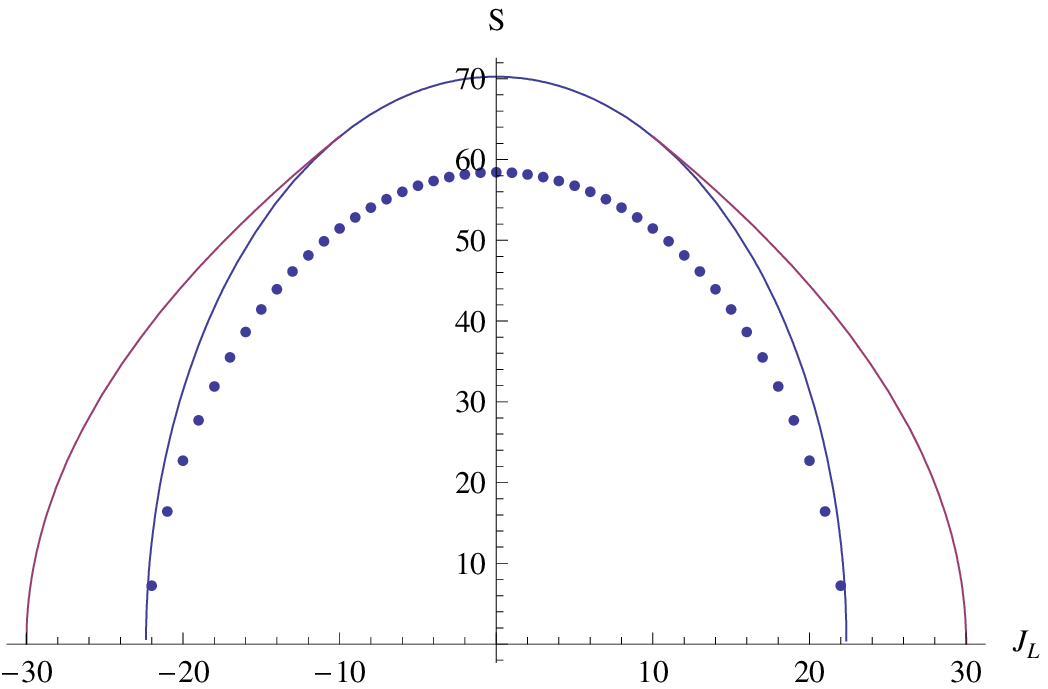} &
   \epsfxsize=5cm \epsfbox{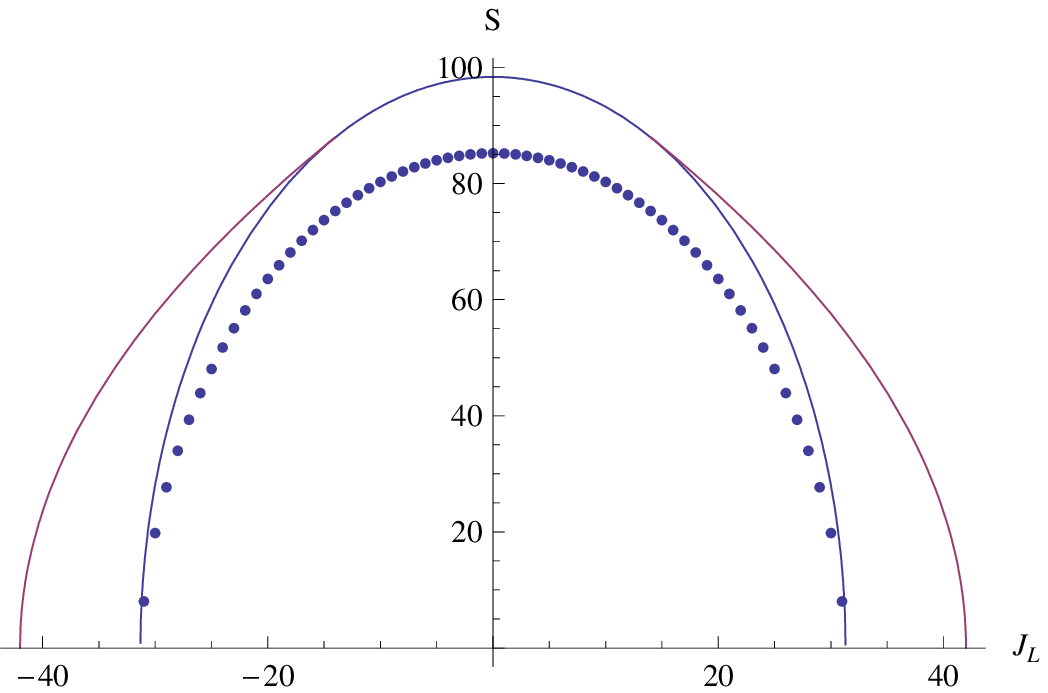}\\
 $N=15,N_p=3$&
 $N=25,N_p=5$&
 $N=35,N_p=7$
 \end{tabular}
 \caption{\sl The logarithm of the modified elliptic genus for $\Sym^N(T^4)$.
  \label{fig:megT4}}
 \end{center}
\end{figure}
\item{\bf BPS partition function on $\Sym^N(T^4)$}\newline
In Fig.~\ref{fig:pfT4} we plot $S_{PF;T^4}(N,N_p,J_L)$ against
$J_L$. We see that the BPS partition function for $\Sym^N(T^4)$ 
captures the enigmatic phase,
just as that for $\Sym^N(K3)$.
\begin{figure}
 \begin{center}
 \begin{tabular}{ccc}
   \epsfxsize=5cm \epsfbox{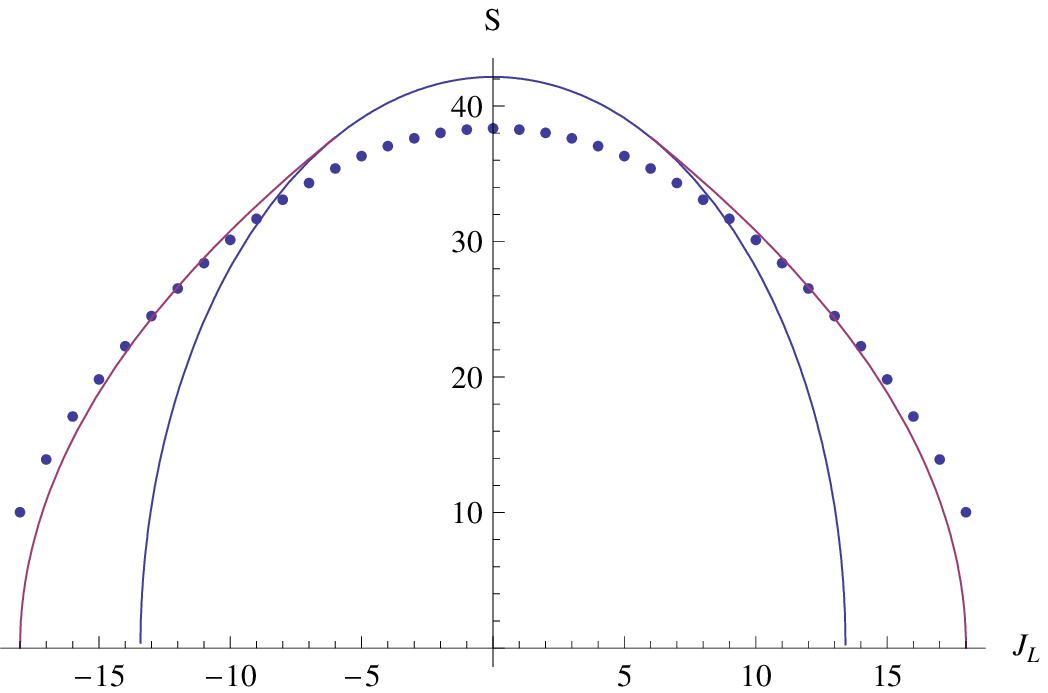} &
   \epsfxsize=5cm \epsfbox{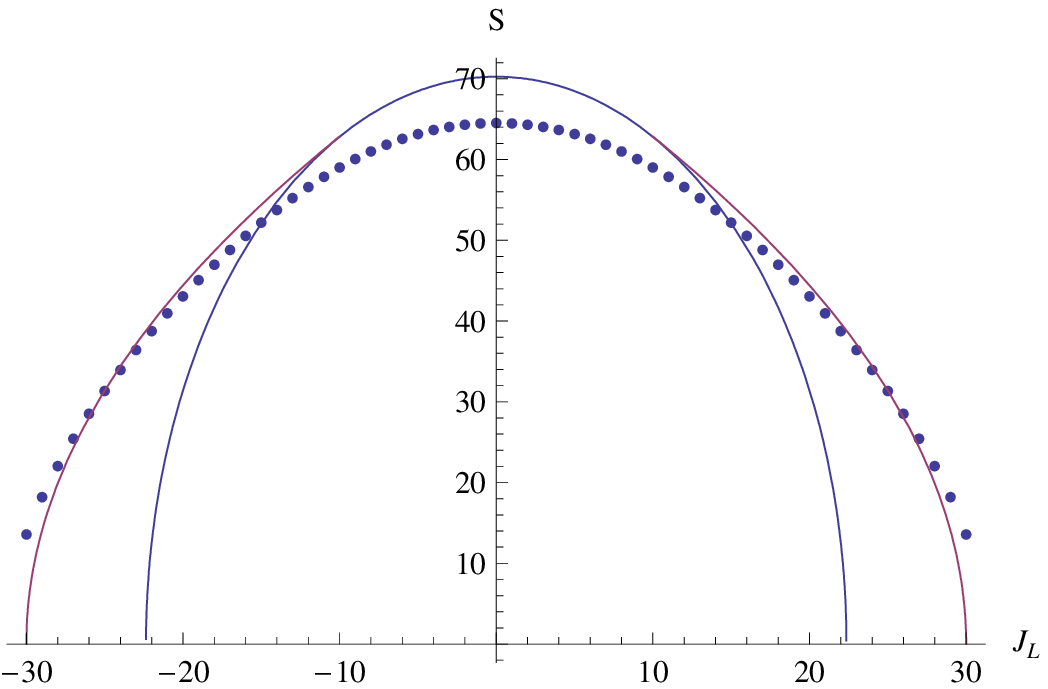} &
   \epsfxsize=5cm \epsfbox{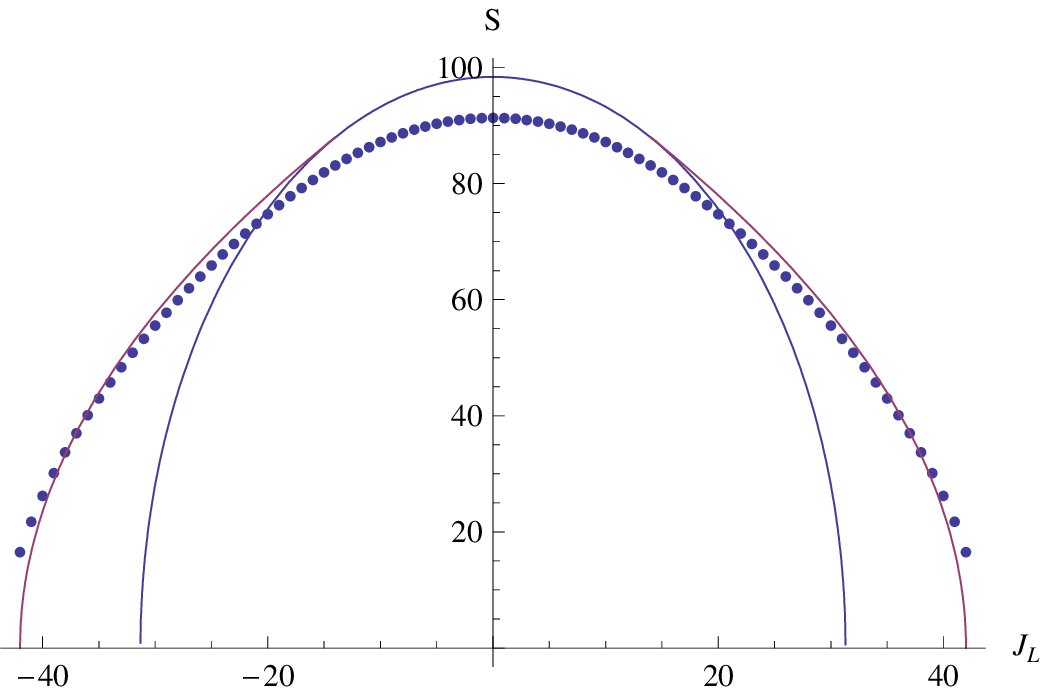}\\
 $N=15,N_p=3$&
 $N=25,N_p=5$&
 $N=35,N_p=7$
 \end{tabular}\\[2ex]
 \caption{\sl The logarithm of the BPS partition function for $\Sym^N(T^4)$.
  \label{fig:pfT4}}
 \end{center}
\end{figure}

\end{itemize}
%

In conclusion, the numerical analysis of the BPS partition function,
which counts the absolute degeneracy, confirms the existence of the new
enigmatic phase with entropy \eqref{Senigma} at the orbifold point of
the D1-D5 CFT, both for $T^4$ and $K3$.  On the other hand, the
(modified) elliptic genus, which is an index, does not capture the new
enigmatic phase.  One might naively take this as indicating that, once
we depart the orbifold point of the CFT, the enigmatic phase gets lifted
and is nowhere to be found in the supergravity regime.  However, we will
see that this is \emph{not} so.

\section{Supergravity analysis}
\label{sec:sugraAnalysis}

Having established the existence of a new ensemble in the CFT let us now
consider the possible bulk dual.  From the structure of the boundary theory we
might imagine that the dual configuration is a BMPV black hole surrounded by a
maximally spinning supertube \cite{Mateos:2001qs} (which can be thought as sourcing the geometries dual to the maximally-spinning state)
carrying \emph{some} of its $J_L$.  To systematically analyze possible bulk
configurations we will first dualize to a IIA or M-theory frame where the full
set of $U(1)\times U(1)$ symmetric configurations were classified in
\cite{Denef:2000nb, Bena:2005va, Berglund:2005vb}.  We will then argue that the
putative duals are necessarily two-centered, and then use the bulk version of
spectral flow symmetry to scan through all possible two center duals.  More
specifically, we ``flow'' any given two-centered configuration to a particular,
tractable class of configurations where we can search for entropy-maximizing
configurations.

\subsection{Multi-centered Solutions in IIA/M-theory}

Let us consider M-theory compactified on a $T^6$ (spanning $x_5, \dots,
x_{10}=z$) to five dimensions\footnote{We could equally well consider
M-theory on K3$\times T^2$ and the five dimensional part of the
discussion would go through unaltered.}.  
In \cite{Bena:2004de, Gutowski:2004yv} the most general class of solutions
preserving the same supersymmetries as three stacks of M2 branes
were written down, and in \cite{Gauntlett:2004qy} the most general class of solutions
preserving a U(1) isometry were classified (see also \cite{Bena:2005ni, Bena:2005va,
Berglund:2005vb, Denef:2000nb}).

We will review the form of these solutions, using notation mostly\footnote{Our
notation will differ in that our harmonic $\tM$ is twice the $M$ appearing in
\cite{Bena:2008dw}, $M_{\textrm{there}} = \frac{\tM_{\textrm{here}}}{2}$. To make
this distinction clear we use $\tM$ for $M$ with our normalization.} following
\cite{Bena:2008dw}.  Our treatment will be somewhat concise; the reader is
referred to \cite{Bena:2005va, Bena:2008dw} for more details.  The metric of the solutions is 
\begin{equation}
	ds_{11}^2 = - Z^{-2/3} (dt + k)^2 + Z^{1/3} ds_{\rm HK}^2 + Z^{1/3} \left(Z_1^{-1}
	dx_{56}^2 + Z_2^{-1} dx_{78}^2 + Z_3^{-1} dx_{9z}^2\right)\,,
\end{equation}
with $Z=Z_1 Z_2 Z_3$.  We mention in passing that this solution also requires a
five dimensional gauge field but this will not be relevant for our analysis.
This is the most general metric preserving the same supersymmetries as the
three-charge black hole \cite{Bena:2004de}.

The 4-d metric $ds^2_{\rm HK}$ is hyperk\"ahler and if we take it to be tri-holomorphic
(possessing a translational U(1) isometry) then the most general solution is a Gibbons--Hawking space \cite{Gibbons:1987sp}
\begin{equation}
	ds_{\rm HK}^2 = V^{-1} (d\psi + A)^2 + V (dy_1^2 + dy_2^2 + dy_3^2), \qquad dA
	= *dV
\end{equation}
with $\psi \cong \psi + 4\pi$ the periodic coordinate.  Here, and in what
follows, $*$ denotes the Hodge dual with respect to the flat $\bbR^3$ 
base space with coordinates $y_{1,2,3}$.

The parameters entering into the above metric are
\begin{equation}\label{Zmu_def}
	\begin{split}
	Z_I &= L_I + \frac{C_{IJK} K^J K^K}{2V}, \qquad k = \mu (d\psi+A) + \omega \\
	\mu &= \frac{\tM}{2} + \frac{K^I L_I}{2 V} + \frac{C_{IJK} K^I K^J K^K}{6 V^2}
	\end{split}
\end{equation}
with $I=1,2,3$ and $C_{IJK}=|\epsilon_{IJK}|$. The one-form $\omega$ satisfies
\begin{equation}
 *d\omega = \frac{1}{2} \left(V d \tM - \tM d V + K^I d L_I - L_I d K^I	\right).
\end{equation}
The solution is entirely specified by eight harmonic functions in $\bbR^3$: $V$, $K^I$, $L_I$ and $\tM$
\begin{equation}
	V = \sum_p \frac{n_p}{r_p} + n_0, \qquad K^I = \sum_p \frac{k^I_p}{r_p} +
	k^I_0, \qquad L_I= \sum_p \frac{l_I^p}{r_p} + l_I^0, \qquad \tM = \sum_p
	\frac{m_p}{r_p} + m_0.
\end{equation}
The labels $p=1,\dots, N$ run over the number of centers with $r_p = |\vec{x} -
\vec{x}_p|$ the distance from each center in the flat $\bbR^3$ metric. The choice of constants in the harmonic functions 
\begin{equation}
	h = \{n_0, k^I_0, l_I^0, m_0\}
\end{equation}
fixes the asymptotic structure of the spacetime.  The charges 
\begin{equation}
	\Gamma_p = \{n_p, k^I_p, l_I^p, m_p\} \label{eqn:CenterCharges}
\end{equation}
at a given center $p$ correspond, in the M-theory frame, to KK-monopole, M5,
M2, and KK-momentum charge, respectively, where the monopole and momentum charge are 
along the $\psi$ circle.  As usual, when we reduce M-theory to IIA along $\psi$
these respectively become D6, D4, D2, and D0 charges, and we will mostly use this
language\footnote{In the conventions used in this paper the numbers appearing in \bref{eqn:CenterCharges} are the integer charges of KK-monopole, M5, M2 and KK-momentum. For details see appendix \ref{app:Conventions}.}.

The charges that appear in the harmonic functions above are dimensionful
quantities that characterize a supergravity solution, and can be related
to the quantized brane charges of the solution (that give the CFT
charges) via proportionality constants that depend on the moduli and
coupling constants of the solution. These relations depend on the
duality frame, and can be straightforwardly derived or found in many
references (see for example eq.\ (2.3) in \cite{Bena:2004tk} or appendix
D of \cite{Bena:2008dw}). To un-clutter notation, in the rest of the
paper we pick a particular set of values for the moduli such the
supergravity charges are always equal to the quantized charges.

\subsubsection{Entropy, Angular momentum and CTCs}

The solutions given above generically carry angular momentum in
${\mathbb R}^3$ coming from crossed electric and magnetic fields (recall
that in four dimensions D4 branes and D2 branes are electromagnetic
duals of each other, and so are D6 branes and D0 branes).  This can be
read off from the asymptotic value of $\omega$ as (see {\it e.g.}\
\cite{Denef:2000nb})
\begin{equation}\label{def_j3}
	\vec{J}^{(3)} = \sum_{p < q} \frac{\langle \Gamma_p, \Gamma_q \rangle}{r_{pq}}
	\vec{r}_{pq}
\end{equation}
where $\vec{r}_{pq}\equiv \vec x_p-\vec x_q$ and
\begin{equation}
	\langle \Gamma_p, \Gamma_q \rangle =  n_p m_q + k^I_p l_I^q - l_I^p k^I_q 
	-m_p n_q
\end{equation}
gives the electromagnetic pairing.

A given center $\Gamma_p$ may correspond to a black object with a horizon at
$r_p = 0$; the area can be computed by evaluating
\begin{equation}\label{entropyR}
	S(r) = 2 \pi \sqrt{D(r)} = 2 \pi \sqrt{ Z_1 Z_2 Z_3 V - \mu^2 V^2}
\end{equation}
at the horizon giving (in terms of the charges) the E$_{7(7)}$ invariant
\begin{equation}\label{e77inv}
	\begin{split}
D(\Gamma_p) = 
- \frac{1}{4} m_p^2 n_p^2 - \frac{1}{6} m_p C_{IJK} k_p^I k_p^J k_p^K - \frac{1}{2} m_p n_p k_p^I l^p_I - \frac{1}{4} (k_p^I l^p_I
)^2 \\
+ \frac{1}{6} n_p\, C^{IJK} l^p_I l^p_J l^p_K + \frac{1}{4} C^{IJK} C_{IMN} l^p_J l^p_K
k_p^M k_p^N .
	\end{split}
\end{equation}

The function $D(r)$ is proportional to
the $g_{\psi\psi}$ component of the metric, and so its positivity effects the causal
structure of the spacetime; if $D(r) < 0$ in some region the metric will
contain closed timelike curves (CTCs) and must be discarded as unphysical.  This
constraint will play an important role in what follows.

Another necessary but not sufficient set of conditions for the absence of CTCs
are the $N-1$ so-called integrability (or ``bubble'') equations
\begin{equation}\label{bubble}
	\sum_{q=1, q \neq p}^N \frac{\langle \Gamma_p, \Gamma_q\rangle}{r_{pq}} =
	\langle h, \Gamma_p\rangle
\end{equation}
which, for two centers, fix the inter-center separation.

The condition (\ref{bubble}) is a no-CTC condition in the neighborhood of the
centers but a more general no-CTC condition is the global
positivity of 
\begin{equation}\label{noctc}
Z_1 Z_2 Z_3 V - \mu^2 V^2 - |\omega|^2 \geq 0,
\end{equation}
which ensures the existence of a time function \cite{Berglund:2005vb}.
This is, in general, a difficult condition to check but it will play a role in
simplifying our analysis.  We will often employ the weaker (necessary but not
sufficient) condition $D(r) > 0$ to constrain our choice of solutions.

\subsubsection{Gauge symmetries and ``Spectral Flow''}

The solutions above are invariant \cite{Bena:2005ni} under a family of ``gauge
transformations''\footnote{These are generated by gauge transformations of
the 10-dimensional B-field in the IIA frame.} parametrized by three constants, $g^I$:
\begin{equation}\label{gauge_trans}
	\begin{split}
	&V \rightarrow V, \qquad K^I \rightarrow K^I + g^I V,\\
	&L_I \rightarrow L_I - C_{IJK} g^J K^K  - \frac{1}{2} C_{IJK} g^J g^K V, \\
	&\tM \rightarrow \tM -  g^I L_I + \frac{1}{2} C_{IJK} g^I g^J K^K +
	\frac{1}{6} C_{IJK} g^I g^J g^K V.
	\end{split}
\end{equation}
Another set of transformations\footnote{These transformations were called ``Spectral Flow'' transformations in \cite{Bena:2008wt} because when the solutions are dualized to asymptotically $AdS_3 \times S^3$ IIB solutions (as explained below) one of them corresponds to a spectral flow of the dual CFT\@.} parametrized by $\gamma_I$, are \cite{Bena:2008wt}
\begin{equation}\label{flow_trans}
	\begin{split}
	&\tM \rightarrow \tM, \qquad L_I \rightarrow L_I - \gamma_I \tM, \\
	&K^I \rightarrow K^I - C^{IJK} \gamma_J L_K  + \frac{1}{2} C^{IJK} \gamma_J
	\gamma_K \tM, \\
	&V \rightarrow V +  \gamma_I K^I - \frac{1}{2} C^{IJK} \gamma_I \gamma_J L_K +
	\frac{1}{6} C^{IJK} \gamma_I \gamma_J \gamma_K \tM.
\end{split}
\end{equation}
While the latter are \emph{not} symmetries of the solutions they are clearly
related to the transformation of (\ref{gauge_trans}) via electric-magnetic
duality.  These transformations can, in fact, be generated by $U$-dualities and,
in the IIB frame, by diffeomorphisms (see \cite{Bena:2008wt} for more details).
As the entropy function (\ref{e77inv}) is, by construction, $U$-duality invariant
these transformations preserve the entropy of the centers.  The charges at each
center, however, are not invariant so we can use these transformations to
transform the charges to a convenient form.  Thus these symmetries will greatly
simplify the task of scanning through putative bulk dual solutions.  We will
generally refer to these as $g$- and $\gamma$-transformations, respectively.

\subsubsection{T-dualizing to the IIB frame}

The metrics given above correspond to solutions of IIA string theory
compactified on $T^6$ with D6, D4, D2, D0 charges but, as we are interested in
the D1-D5 system in IIB, we must dualize to this frame.  An appropriate set of
dualities consists of a KK reduction to IIA on $x_9$ (rather than
$\psi$), followed by three $T$-dualities along $x_5$, $x_6$ and $z=x_{10}$ yielding the
following charges.
\begin{align}
\left\{ 
\begin{array}{l l}
  M2(56) \\
  M2(78) \\ 
  M2(9z) \\ 
  \end{array} 
  \right\} 
  \xrightarrow{\textrm{KK on $x_9$ to IIA}}
\left\{ 
\begin{array}{l l}
  D2(56) \\
  D2(78) \\ 
  F1(z) \\ 
  \end{array} 
  \right\} 
  \xrightarrow{\textrm{$T_{56z}$}}
\left\{ 
\begin{array}{l l}
  D1(z) \\
  D5(5678z) \\ 
  P(z) \\ 
  \end{array} 
  \right\} 
\end{align}
while the M5 charges become dipole charges in IIB
\begin{align}
\left\{ 
\begin{array}{l l}
  m5(\psi 789z) \\
  m5(\psi 569z) \\ 
  m5(\psi 5678) \\ 
  \end{array} 
  \right\} 
  \xrightarrow{\textrm{KK on $x_9$ to IIA}}
\left\{ 
\begin{array}{l l}
  d4(\psi 78z) \\
  d4(\psi 56z) \\ 
  ns5(\psi 5678) \\ 
  \end{array} 
  \right\} 
  \xrightarrow{\textrm{$T_{56z}$}}
\left\{ 
\begin{array}{l l}
  d5(\psi 5678) \\
  d1(\psi) \\ 
  kk(\psi 5678;z) \\ 
  \end{array} 
  \right\} 
\end{align}
The final solution has a KK-monopole dipole charge along the $\psi 5678$
directions with its special transverse circle in the $z$ direction,
denoted by $kk(\psi 5678;z)$. The original M-theory KK and momentum
modes along the Gibbons--Hawking isometry direction $\psi$ (corresponding
upon $\psi$-reduction to D6 and D0 in the IIA) are relatively inert
under these transformations and go over to $kk(56789z;\psi)$ and
$P(\psi)$ in IIB\@.

The resultant NSNS fields are \cite{Bena:2008dw}
\begin{align}\label{IIB_sol}
	ds_{\rm IIB}^2 &= -\frac{1}{Z_3\sqrt{Z_1 Z_2}} (dt + k)^2 + \sqrt{Z_1Z_2} \,
	ds_{\rm HK}^2 + \frac{Z_3}{\sqrt{Z_1 Z_2}} (dz + A^3)^2 + \sqrt{\frac{Z_1}{Z_2}}\,
	dx_{5678}^2\\
	e^\Phi &= \frac{Z_1}{Z_5}, \qquad B_{\mu\nu} = 0
\end{align}
and there is also an RR potential, $C^{(2)}$, corresponding to the D1 and D5
charge.  In order to determine when this metric is asymptotically-AdS (as we will explain in detail in Appendix
\ref{app_decoupling}) we will need
\begin{equation}
	A^3 = \frac{K^3}{V} (d\psi + A) + \xi^3 - \frac{1}{Z_3} (dt + k), \qquad *d
	\xi^3 = - d K^3.
\end{equation}
When there are no dipole charges, ($K^I = 0$), we see from (\ref{Zmu_def})
that the $Z_I$ reduce to simple harmonic functions and the metric above is the
usual D1-D5-P black hole metric.

\subsubsection{AdS$_3\times$S$^3$ and the AdS/CFT Dictionary}

If we consider a system with $n \neq 0$ (net D6 charge in IIA) then we
can take a decoupling limit such that the solution is asymptotically
AdS$_3\times$S$^3/{\mathbb Z}_n$.  We review the AdS/CFT dictionary for
these solutions as we will need it in what follows; for details of the
decoupling limit the reader is referred to Appendix
\ref{app_decoupling}\footnote{See also the Appendix of
\cite{Bena:2008wt}.}.

We consider a total charge 
\begin{align}
	\Gamma = \{n, k^I, l_I, m\}
\end{align}
and we set all the constants $h = \{n_0, k^I_0, l_I^0, m_0\}$ to zero except $l_3^0
= 1$ and 
\begin{equation}\label{ads_consts}
	m_0 = - \frac{k^3}{n}.
\end{equation}
The choice to set $l_3^0\neq 0$ yields AdS asymptotics and the
requirement\footnote{Which simply follows from also summing over the index $p$
on both sides of eqn.\ (\ref{bubble}) and using the anti-symmetry of the pairing
$\langle \cdot, \cdot \rangle$.} that $\langle \Gamma, h\rangle= 0$ then fixes
the choice of $m_0$.  Of course if $k^3=0$ then even $m_0 = 0$ (but we do not
allow $n=0$ as this does not generate an AdS$_3\times$S$^3$ geometry).  Note
that as a consequence of these asymptotics the integrability equations
(\ref{bubble}) imply that centers that do not have either D6 charge or the D4 charge $k_p^3$ cannot 
form bound states inside AdS$_3$ (otherwise the right hand side of (\ref{bubble}) is zero).

Most CFT quantum numbers can be read off from the asymptotic values of the
charges as follows
\begin{equation}\label{cft_from_asym}
	\begin{split}
		N_1 &\sim Z_1^{(1)} \sim l_1 + \frac{k^2 k^3}{n}, \qquad
N_p \sim Z_3^{(1)} \sim l_3 + \frac{k^1 k^2}{n}, \\
N_5 &\sim Z_2^{(1)} \sim l_2 + \frac{k^1 k^3}{n}, \qquad
\cJ_L \sim 2 \mu^{(1)} \sim m + \frac{k^I l_I}{ n} + \frac{2 k^1 k^2 k^3}{n^2}
	\end{split}
\end{equation}
where the superscript ``$(1)$'' on a quantity $f$ means to pick out the
coefficient of the order $1/r$ term from the large $r$ expansion of
$f$.  For $Z_1$, $Z_2$ this is the leading term but for $Z_3$ the
leading piece is the constant $l_3^0$.  Note that the constraint
(\ref{ads_consts}) guarantees that $\mu$ has no leading constant piece. 

As we explained above, the quantized charges that characterize the CFT
are related to the ``supergravity'' charges that one obtains from the
asymptotics of the warp factors via proportionality constants that
depend on the moduli and coupling constants of the solution. However,
given that our phase is a hybrid between the BMPV phase and the
maximally spinning phase, we can always use the known relation between
these CFT phases and their dual bulk solutions to relate the
supergravity and CFT charges. Alternatively, we can work at some values
of the moduli where the supergravity and quantized charges are always
equal, which is what we will do through the rest of this paper.

In particular, the charge $\cJ_L$ is to be identified with $J_L$ of the
CFT up to a sign that we will discuss below. Furthermore, as with $J_L$,
we will define a bulk charge $\cJ_R$ related to the CFT charge $J_R$ up
to a sign.  The charge $\cJ_R$ comes from the SO(3)$\,\cong\,$SU(2)
angular momentum, $\vec{J}^{(3)}$, of the ${\mathbb R}^3$ base of the
solutions (which becomes one of the SU(2)'s in the SO(4) isometry group
of S$^3$ in the near horizon geometry) so can be read off from the
asymptotic value of $\omega$
\begin{equation}
	\cJ_R \sim 2 \omega^{(1)}.
\end{equation}
Unlike the other CFT charges $\cJ_R$ depends not on the total bulk charge
but on the distribution of charges between the centers.  For two centers
(to which we will turn presently) this reduces to 
\begin{equation}
\cJ_R = \langle \Gamma_1, \Gamma_2\rangle.
\end{equation}
Note that we could just as well have chosen $\cJ_R = \langle \Gamma_2, \Gamma_1
\rangle$ which would differ from the definition above by a minus sign.  To fix
conventions however we will \emph{define} $\cJ_R$ as above and then relate it to
the $J_R$ in the CFT (which we have defined to be positive) via $J_R = \pm
\cJ_R$.

\subsubsection{A Note on Signs}

When identifying the CFT quantum numbers with those of the bulk we must
be careful to incorporate potential physically-meaningful sign
differences.  The sign of $N_1 N_5$ is fixed by the requirement of
giving a positive AdS$_3$ central charge and the sign of $N_p$ is fixed
with respect to this.\footnote{When $M^4=T^4$, neither sign of
$N_p$ will break supersymmetry.  However, in the $\CN=2$ formalism
we are working in, only one sign is manifestly supersymmetric and
allowed.}

The angular momenta $\cJ_L$ and $\cJ_R$ are related to those in the CFT but
there is no canonical way to fix the signs.  In the CFT we have taken, without
loss of generality, $J_L, J_R > 0$ and we would like to do the same in the bulk.
From the expression for $\cJ_L$ we see that we can flip its sign simply by
sending $k^I, m$ to $-k^I, -m$ which is a symmetry of the solution.  This also
flips the sign of $\cJ_R$ as the intersection product
$\langle\Gamma_1,\Gamma_2\rangle$ is odd under this symmetry.  As mentioned
above we also have the further freedom to change the sign of $\cJ_R$ by
switching the order $\Gamma_1 \leftrightarrow \Gamma_2$ 
but for notational clarity let us fix the definition of the scalar quantity
$\cJ_R := \langle \Gamma_1, \Gamma_2\rangle$ and then define $\vec{J}_R = \cJ_R
\, \hat{x}_{12}$ with $\vec{x}_{12}$ a unit vector between the two centers.  In
order to match conventions with the CFT we will measure the angular momentum at
infinity along an axis aligned with $\vec{J}_R$ so that $J_R > 0$.  In terms of
$\cJ_R$ this gives $J_R = \pm \cJ_R$.  Thus we can always arrange for $J_L, J_R
> 0$ in the bulk  but, as we will see below, once we have fixed charge conventions such that
$J_L > 0$, we have to check the sign of $\cJ_R$ to determine if $J_R = \cJ_R$
or $-\cJ_R$.

\subsection{Two-centered Solutions in AdS$_3\times$S$^3$}

While a general bulk solution in the class considered above may have many
centers we will now argue that the putative duals to the new CFT phase must be 
two-center configurations.  This follows from the observation that an $N$-center
solution has $3 N - 3$ parameters ($3 N$ given by $\vec{r}_p$ minus the three center of mass
parameters) constrained by $N-1$ equations giving a $2N-2$ dimensional
solution space.  Generically, each point in this space corresponds to a different
value of $\cJ_R$ via (\ref{def_j3}).  As the leading entropy comes from summing
the entropy of each center and does not depend on the locations of the centers,
an $N$-center configuration generically has a fixed (leading) entropy but a
range of $\cJ_R$.  

The new phase in the CFT, however, is characterized by a fixed value of $J_R$
that maximizes the entropy.  We thus expect a bulk configuration with fixed
$\cJ_R$. It is not hard to see that this corresponds to $N=2$; for two centers
$J^{(3)}$ is fixed and only its orientation is unfixed (yielding two parameters
$\{\theta, \phi\}$).  Thus we can restrict our analysis to two-center
configurations.

\subsubsection{Stability and Smoothness}

To further constrain the problem, let us consider the partition of a
fixed total charge into two centers $\Gamma = \Gamma_1 + \Gamma_2$ and
the entropy of the associated configuration
\begin{equation}
S_{\text{2-center}} = S(\Gamma_1) + S(\Gamma_2).\label{fvyk6Jul11}
\end{equation}
One might naively imagine, based on the intuition that black holes are thermodynamic
ensembles, that such a partition is always entropically disfavorable
as combining two ensembles generally increases the total entropy:
\begin{equation}
S_{\text{1-center}}(\Gamma_1 + \Gamma_2) 
 \stackrel{\text{?}}{>}
S_{\text{2-center}}=
S(\Gamma_1) + S(\Gamma_2).
\end{equation}
It is clear, however, from examples such as the entropy enigma of
\cite{Denef:2007vg} that such intuition is misguided and there are examples when a two-center configuration has larger entropy than a
single-center configuration.  Because of this, and because of the
non-vanishing constant value of $\cJ_R$ observed in the CFT, we restrict
ourselves to two-center configurations and look for the ones with the most entropy.

To find them, one would need to do a
stability analysis based on maximizing the total entropy of a partition into two charges:
\begin{equation}
S_{\text{2-center}} = S(\Gamma-\Gamma_2) + S(\Gamma_2),
\end{equation}
If the configuration is stable (locally entropy-maximizing), the Hessian of
$S_{\text{2-center}}$ with respect to $\Gamma_2$ should have only
negative eigenvalues.  If there are some positive eigenvalues, the
configuration is entropically unstable against shedding charge from one center to the other.

Although the analysis of the Hessian for the general partition
$\Gamma_2$ is technically rather difficult, there is one situation where one
might expect stability: when one center is smooth and carries no
macroscopic or microscopic entropy -- such a center can no longer shed
charge to the other center without producing closed timelike curves.
Such smooth centers, first discussed in \cite{Bena:2005va,
Berglund:2005vb}, correspond in four dimensions to D6 branes with Abelian worldvolume fluxes  \cite{Balasubramanian:2006gi}, and have also appeared in the ${\mathcal N}=2$ entropy enigma \cite{deBoer:2008fk}. In Appendix \ref{app_convexity} we will demonstrate 
local entropic stability for two-center configurations where all the entropy is carried by one center.

While other two center configurations \emph{might} be entropically
stable, they are probably non-generic and thus will impose many
additional charge constraints.  Although we cannot entirely rule out
stable configurations with two horizons, motivated by the entropy enigma
of \cite{deBoer:2008fk} and the fact that configurations with a smooth
center live on the boundary of charge space and are isolated (in the
sense of not being continuously connected to other charge
configurations), we will restrict our analysis to configurations with
one smooth center.

Requiring $S(\Gamma_2) = 0$ and smoothness\footnote{If there is a singularity at
$r=r_2$ this is usually associated with a microscopic horizon and subleading
entropy so we can re-apply the entropy maximization argument
above including subleading corrections.} at $r_2$ fixes the charge
$\Gamma_2$ to satisfy \cite{Bena:2005va,Berglund:2005vb,Bena:2008dw}
\begin{equation}\label{smoothness}
	l_I = - \frac{C_{IJK} k^J k^K}{2 n}, \qquad  m = \frac{k^1 k^2 k^3}{n^2} 
\end{equation}
from which it follows  \cite{Balasubramanian:2006gi} that
center ``2'' carries no microscopic entropy as it is gauge-equivalent (by choosing
the appropriate $g^I$ in eqn.  (\ref{gauge_trans})) to $n$ D6-branes in IIA or
to a ${\mathbb Z}_n$ quotient singularity in M-theory.

Thus we can reduce our problem to considering solutions specified by the
following charges and asymptotics
\begin{equation}\label{init_charge}
	\begin{split}
\Gamma_1 &=
\left\{1-\alpha,\left\{k^1,k^2,k^3\right\},\left\{l_1,l_2,l_3\right\},m\right\},\\
\Gamma_2 &= \left\{\alpha,\left\{\alpha\, p^1,\alpha \,p^2,\alpha\, p^3\right\},\left\{-\alpha \,p^2
p^3,-\alpha\, p^1
p^3,-\alpha \,p^1 p^2\right\},\alpha \,p^1 p^2 p^3\right\},\\
h &= \left\{0,\{0,0,0\},\{0,0,1\},-k^3-\alpha p^3\right\}
	\end{split}
\end{equation}
where $h$ denotes the ``vector'' of constants in the harmonic functions.
One can check that $\Gamma_2$ satisfies (\ref{smoothness}) and so
corresponds to a smooth center with $S(\Gamma_2) = 0$ for any choices of
$\alpha$, $p_i$.  By charge quantization, all the entries of
$\Gamma_{2}$, such as $\alpha$, $\alpha p^1$, and $\alpha p^2p^3$, are
assumed to be integers.  Note we have taken the total KKM/D6 charge to
be 1.  When this charge is $n$ the decoupling limit discussed in
Appendix \ref{app_decoupling} gives an AdS$_3\times$S$^3/{\mathbb Z}_n$
space. Nevertheless, since the CFT phase we found exists in the standard
unquotiented orbifold theory, we are only interested in asymptotically
AdS$_3\times$S$^3$ solutions so we restrict to $n=n_1 + n_2 = 1$.

The smoothness condition \eqref{smoothness} insures that a certain
center is smooth in all duality frame. Nevertheless, in the D1-D5-P
duality frame in which we are working it is also possible to have smooth
centers that correspond in the IIA frame not to fluxed D6 branes but to
fluxed D4 branes that have a nonzero $k^3$. These are the supertubes
dual to the maximally spinning phase, and can be thought of as coming
from \eqref{init_charge} by taking the limit $\alpha \rightarrow 0$
keeping $\alpha p^3$ fixed (note that this limit is rather formal,
because we take $\alpha$ to be an integer). When the smooth center is a
supertube, the KKM charge of the first center is one.

\subsubsection{Spectral Flow in the Bulk}

While our analysis of the last subsection has reduced the problem to considering
all two-center configurations with one smooth center, this is still a rather daunting problem.  On the other hand
these solutions enjoy a great deal of symmetry arising from
(\ref{gauge_trans})--(\ref{flow_trans}) and we can use this to simplify our
analysis.

From (\ref{flow_trans}) we see that a general $\gamma$-flow modifies the
asymptotics and may not preserve an asymptotically AdS$_3\times$S$^3$ form of
the metric.  Recall that the latter requires that only $m_0$ and $l_3^0$ are
non-vanishing and that they satisfy (\ref{ads_consts}).  Moreover, as mentioned
above, we want to keep the total D6-charge equal to one.

Let us see how these constraints restrict the transformations we can perform.
First since gauge transformations ($g^I$-flows) preserve the solution we are
free to perform them with impunity.  On the other hand, a general $\gamma_I$-flow
modifies $V$ in a way dependent on all the other harmonics, and hence
will generically modify the asymptotics. Since we are interested in keeping the $AdS_3 \times S^3$ asymptotics, it is not hard to see from \eqref{init_charge} that we can only use a  $\gamma$-flow with a nonzero $\gamma_3$. In the IIB duality frame we are in, this is a geometrized component of the $U$-duality group, and is
the bulk dual to spectral flow in the CFT \cite{Bena:2008wt}.

To ensure that $\gamma_3$ does not modify $V$, we first have to use our
gauge-freedom to set $K^3$ to vanish asymptotically.  This is simply
accomplished by $g^3 = -k^3-p^3 \alpha$.  We are then free to flow by $\gamma_3$
and find this affects the solution asymptotically as follows:
\begin{align}
	\tilde{Z}_1 &\sim Z_1, \qquad \tilde{Z}_2 \sim Z_2, \\
	\tilde{Z}_3 &\sim Z_3 - 2 \gamma_3 \mu + (\gamma_3)^2 Z_1 Z_2 \\
	\tilde{\mu} &\sim \mu - \gamma_3 Z_1 Z_2
\end{align}
with $\sim$ meaning that the leading asymptotic terms (as well as the subleading term in $Z_3$)
Recalling (\ref{cft_from_asym}) and comparing this with (\ref{Eq:SpectralFlow})
we see the above maps to spectral flow in the CFT by $\eta = \mp \gamma_3$ for
$\cJ_L = \pm J_L$.  As we will fix conventions such that $\cJ_L = J_L$ this
gives $\eta = - \gamma_3$.

\subsubsection{Spectral Flowing to BMPV plus Supertube}\label{flow_to_bmpv}

We will now take advantage of the above transformations to flow
arbitrary charges of the form (\ref{init_charge}) to a more tractable
form.  We first spectral flow using the following transformations
\begin{equation}\label{flow_param}
	g^3 =  -k^3- p^3 \alpha \quad \textrm{  followed by }\quad  \eta = -\gamma_3  = -\frac{1}{k^3+p^3
	\alpha -p^3}.
\end{equation}
This has the effect of removing the D6 charge from $\Gamma_2$ and turning the
latter into a supertube.  After this $\Gamma_1$ still generically has
non-vanishing D4 charges but we can use a gauge transformation (which has no effect
on the CFT quantum numbers) to set $k'^I$ (these flowed charges are generally
inequivalent to those of (\ref{init_charge})) to zero via
\begin{equation}\label{gauge_param}
(g^1, g^2, g^3) = (-k'^1, -k'^2, -k'^3).
\end{equation}
The resultant charge vectors after these transformations are
\begin{equation}\label{bmpvtube_charge}
	\begin{split}
		\Gamma_{\textrm{bmpv}} &=
\left\{1,\left\{0,0,0\right\},\left\{Q_1,Q_2,Q_3\right\}, m\right\},\\
\Gamma_{\textrm{tube}} &= \left\{0,\left\{0,0,d\right\},\left\{q_1, q_2,
0\right\}, \frac{q_1 q_2}{d} \right\},\\
h &= \left\{0,\{0,0,0\},\{0,0,1\},-d \right\}.
	\end{split}
\end{equation}
The relation between these charges and those of \eqref{init_charge} can
be found in Appendix \ref{app_spectral}\@. 
As indicated in the
labeling in \eqref{bmpvtube_charge}, the first center is nothing but a
BMPV black hole while the second is a maximally spinning supertube.
As we explain below we have chosen this choice of spectral flow to
simplify our analysis.


One may wonder if a spectral flow by a fractional flow parameter
\eqref{flow_param} is allowed.  Actually, the spectral flow is a
transformation which maps a legitimate configuration into another
legitimate configuration in both supergravity and the CFT, and is
defined in principle for \emph{any} flow parameter.  Therefore, such
flows are indeed allowed.

It is true that on the CFT side the flow parameter must be integer
quantized if one wants to map a state in a sector to another state
\emph{in the same sector} with the same periodicity of fermions.
Non-integral spectral flows, on the other hand, change the fermion
periodicity both in the boundary and bulk.  However, they also modify
the VEV of the asymptotic gauge field in a compensating way so that the
bulk geometry remains regular in a suitable sense; {\it e.g.},
supersymmetry stays preserved due to the modified VEV of the gauge field
(see \cite{Balasubramanian:2000rt} and references therein for more
details).  By assumption our solution (\ref{init_charge}) is dual to the
original D1-D5 CFT which is in the Ramond sector.  After flowing by
$\eta$ units the solution (\ref{bmpvtube_charge}) will not be in the
Ramond sector if $\eta$ is non-integral but this is of no consequence as
we ultimately flow this solution back by $-\eta$ units once we have
found the maximal entropy configuration.  Thus our final configuration
will once more be in the Ramond sector (in fact we will see in our
analysis that the entropy-maximizing value of $\eta$ turns out to be
integral so such concerns are moot).


The astute reader may also notice that the spectral flowed charges
\eqref{bmpvtube_charge}, whose explicit expressions can be found in
Appendix \ref{app_spectral}, are not integers in general and wonder if
they are allowed.  However, note that we initially started with integral
charges \eqref{init_charge} and thus a manifestly regular geometry.
Spectral flow merely gives different frames to look at the same physical
situation, and hence a regular configuration is mapped into a regular
configuration again, no matter how it may look.  In the present case,
the fractional charges are allowed because of the gauge field VEV
mentioned above, which modifies the charge quantization.  This point is
perhaps easier to understand in the IIB frame \eqref{IIB_sol}, where the
spectral flow transformation is nothing but a coordinate change of the
$\psi$-$z$ torus \cite{Bena:2008wt}.  A fractional flow mixes $\psi$ and
$z$ in a non-standard way such that they are not independently periodic.
However, a coordinate transformation does not change the physical torus,
which remains regular.  The fractional charges just reflect the
non-standard periodicity of the torus coordinates and pose no problem at
all.



\subsection{The BMPV plus Supertube System}

\begin{figure}[htbp]
 \begin{center}
  \epsfxsize=15cm \epsfbox{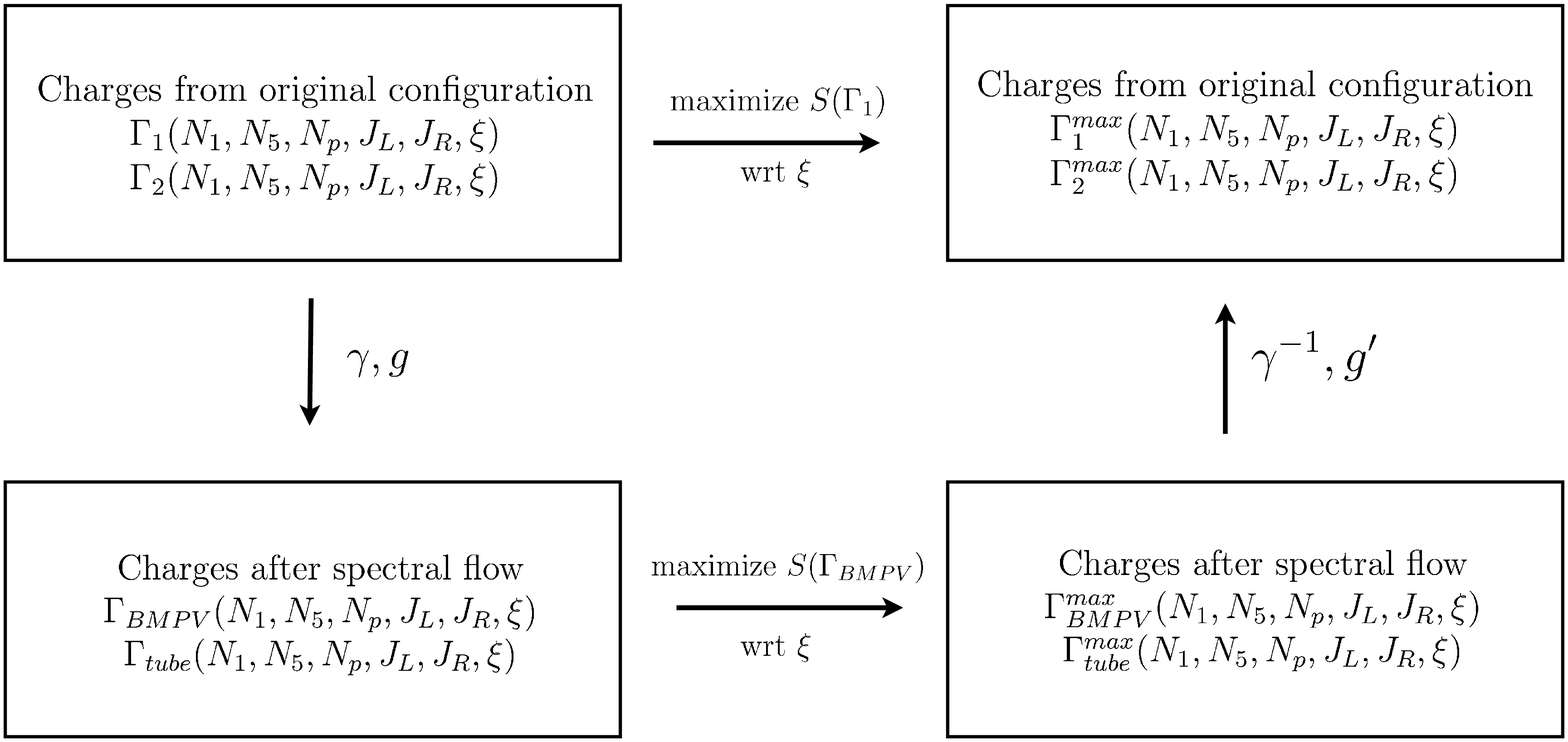}
   \label{fig:flowchart}
 \end{center}
  \vspace*{-4cm}
 \caption{\sl We want to maximize the bulk entropy with charges given in
 \bref{init_charge} with respect to parameters (collectively denoted
 by $\xi$) describing the
 distribution of charges between the centers;  this procedure is described by the top horizontal arrow, and is quite non-trivial. Alternatively we can spectral flow and gauge transform to the
 BMPV plus tube configuration with charges given in \bref{bmpvtube_charge} (left vertical arrow) and then maximize the entropy (lower horizontal arrow). We can then spectral flow the configuration back (right vertical arrow) to get the required maximum entropy solution dual to the CFT\@.}
\end{figure}

Thus far we have argued that it is possible to use the bulk analog of spectral
flow (combined with gauge transformations that do not affect the CFT) to
transform a two-center solution where one center carries no entropy and the other has an arbitrary set of charges to a BMPV black hole surrounded by a supertube (\ref{bmpvtube_charge}).  The spectral flows and gauge transformations do not alter the entropy of the bulk configuration, and the particular $\gamma_3$ spectral flow also leaves the smooth center smooth (although it may change it from a D6 center to a supertube).  On the other hand the charges
(\ref{bmpvtube_charge}) are rather simple and maximizing the entropy of the
total system with fixed CFT charges is relatively straightforward. This process is described in Fig.~\ref{fig:flowchart}.

The reason to take this somewhat indirect approach is that it is rather
non-trivial to maximize $S(\Gamma_1)$ (with $\Gamma_1$ arbitrary) with
respect to fixed CFT charges while making sure that the bulk solution
stays regular and free of CTC's. On the other hand, it is very easy to
understand the origin of CTC's in the BMPV+supertube system (they appear
when the charge of the supertube and those of the BMPV black hole are
opposite, or when either object has too much angular momentum) and hence
it is much more straightforward to maximize $S(\Gamma_{\textrm{bmpv}})$
and to relate the CFT charges to the parameters appearing in
(\ref{bmpvtube_charge}).  In fact the relation is simply given by
\begin{equation}\label{bulk2cft}
	\begin{split}
N_1 &= Q_1 + q_1, \qquad
N_5 = Q_2 + q_2, \qquad
N'_p = Q_3  \\
\cJ'_L &= d\, N'_P+\frac{q_1 q_2}{d}+m, \qquad \cJ_R = \frac{q_1 q_2}{d}-d\, N'_P \\
r_{12} &= \frac{q_1 q_2}{d^2}-N'_P = \frac{\cJ_R}{d}.
	\end{split}
\end{equation}
Note that the $N'_P$ that enters in the formula for the inter-center separation $r_{12}$ is always positive, so the radius always becomes small when the magnitude of $N'_P$ grows. This reflects the fact that a supertube near a black hole can be merged into the black hole when the horizon radius of the latter becomes large enough \cite{Bena:2004wt,Marolf:2005cx,Bena:2005zy}. 

Thus we can immediately fix the $Q_I$ and consider the $q_I$ as
parameters.  We simplify the analysis by assuming $N_1 = N_5$ and take
$q_1 = q_2 = q$.  This reduces our parameter space by one dimension and
corresponds physically to restricting our attention to a system with an
equal number of $N_1$ and $N_5$ branes.

Note that we have used the CFT charges $N'_P$ and $\cJ'_L$ above, as these are related
by spectral flow to the enigmatic phase discussed in section
\ref{subsec_enig_phase} (the other charges do not flow).  A very useful
constraint coming from the CFT is eqn.\ (\ref{Eq:CFT_JL_JR_rel}) which holds only for $\eta = 0$ (but
can be flowed to any frame).  In terms of the bulk charges it is
\begin{equation}\label{bulk_JLRrel}
|\cJ_L| - |\cJ_R| = 2 N_p
\end{equation}
where we use the unflowed $N_p$ and $\cJ_L$.  Recall that we can fix conventions
so that $\cJ_L > 0$, but once such conventions are chosen we still need to check
the sign of $\cJ_R$.  It follows from $\cJ_R = d \, r_{12}$ that the sign of
$\cJ_R$ is the same as that of $d$, so we cannot fix the sign of $\cJ_R$, as
defined in (\ref{bulk2cft}), as this would over-constrain the bulk charges.  However, as
mentioned above, we do have the freedom to choose the axis along which we
measure $J_R$ at infinity so we can always choose an axis such that the latter
is positive\footnote{Equivalently we can always flip the orientations of
the centers in the bulk so that $\langle \Gamma_1, \Gamma_2\rangle > 0$ by
interchanging $\Gamma_1 \leftrightarrow \Gamma_2$.} giving $J_R = \pm \cJ_R$
with the $\pm$ corresponding to the sign of $d$. Thus in the bulk we have
\begin{align} 
J'_L = d\, N'_P+\frac{q^2}{d}+m,& \qquad J_R = 
\left\{ 
\begin{array}{l l}
\frac{q^2}{d}-d\, N'_P  \quad (d > 0)\\
d\, N'_P -\frac{q^2}{d} \quad (d < 0)\\
  \end{array} \right. ,
\label{bulkJLJR}\\
  J_L - J_R = 2 N_P.
\end{align}
Hence we need to consider separately the supertubes with $d < 0$ and $d > 0$.  Let us also
recall that the spectral-flowed charges are
\begin{equation}
J'_L = J_L + 2 \eta N, \qquad N'_p = N_p + \eta J_L + \eta^2 N.
\end{equation}

\subsubsection{Spectral Window}

Let us examine the configuration above and attempt to constrain it as best we
can by the various no-CTC conditions.  We first note that if $Z_I = 0$ for
any $I$ then the condition (\ref{noctc}) is violated and we generate CTC's.
This necessarily happens if $q$ and $Q$ have different signs since then $Z_1$
and $Z_2$ will become zero at some point (this argument is insensitive to having
$N_1 = N_5$).  Thus we must have
\begin{equation}
	0 \leq q < N_5 = \sqrt{N}	.
\end{equation}
Likewise, positivity of $r_{12}$ requires $q^2 \geq d^2 N'_p$, and hence
\begin{equation}
d^2 (N_p + \eta J_L + \eta^2 N) \leq q^2 \leq N
\end{equation}
and we recall from the CFT discussion that we are interested in the range of
charges
\begin{equation}
2 N_p \leq J_L \leq N + N_p.
\end{equation}
From the condition $d^2 N_P' < N$ it is not hard to see that the spectral flow
parameter, $\eta$, is constrained to be between $0$ and $ -1$.  As noted before
$\eta$ need not be integral so in principle any value $-1 \le \eta \le 0$ is
allowed but it is possible to check, numerically, that the entropy is always
maximized for $\eta=0,-1$ so from now on we will allow only these two
possibilities.

Let us review this logic.  In section \ref{flow_to_bmpv} we showed that an
arbitrary two-center configuration with one smooth center can be flowed to a
BMPV plus a supertube.  By construction this flow does not change the character
of either center (the black hole remains a black hole or a black ring, and the
smooth D6 center remains a smooth D6 center or is transformed into a smooth
``supertube'' center). In particular, the value of the black hole or black ring
entropy remains the same, although its dependence on the charges changes.

Since it is always possible to flow to a BMPV plus supertube, there exists some
value of $\eta$ which flows the putative bulk dual of the CFT phase to a
supertube plus BMPV configuration.  We analyze the CFT  constraints on the
charges in the flowed frame as a function of $\eta$ and find they can only be
satisfied for $-1 \le \eta \le 0$.  A further numerical scan shows that the
entropy is always maximized on the boundaries of this region ($\eta = 0, -1$).
Since, by assumption, we started with a well-defined two-center configuration
and spectral flow does not generate CTCs this shows that the BMPV plus
supertube configuration must correspond to a spectral flow of $\eta = 0$ or
$\eta = -1$ of the bulk configuration maximizing the entropy. 

Note this argument did not involve any sort of entropy maximization over
the set of all two-center solutions, which is notoriously difficult
because of the absence of intuition about the relation between the
charges of the centers and the appearance of CTC's. Rather, we maximize
the entropy of a BMPV black hole surrounded by a supertube (where it is
well-understood where CTC's come from) and use the charge relations
given in section \ref{subsec_enig_phase} to argue that the only possible
dual bulk configuration can be a BMPV plus supertube, or its spectral
flow by $\eta=-1$.

\subsubsection{Entropy Maximization of a BMPV Black Hole Surrounded by a Supertube}

The entropy of this system, comes from the BMPV center, and is 
\be
S(\Gamma_1)
= 2\pi \sqrt{D(\Gamma_1)}\,,
\ee
 where
\begin{equation}
	D(\Gamma_1) = (\sqrt{N} - q)^2 N'_p - \frac{m^2}{4} \,.
\end{equation}
It is clear that the entropy is maximized by minimizing $q$ and $m$. These can be expressed in terms of the original CFT charges via 
\begin{equation}
	\begin{split}
	q^2 &= d^2 (r_{12} + N_p + \eta J_L + \eta^2 N)\\
		&= d (\pm J_R + d (N_p + \eta J_L + \eta^2 N)),\\
	m &= J'_L \mp J_R - 2 d\, N'_p 
	\end{split}
\end{equation}
where the $\pm$ sign in the second lines corresponds to the cases $d > 0$ and $d < 0$,
respectively, (and likewise the $\mp$ in the third line) as follows from
eqn.\ (\ref{bulkJLJR}).  To facilitate the analysis let us simplify our notation
and use variables $J_L = j N$ and $N_p = p N$ with $2p \leq j \leq 1 +p$
(and $p >0$).  We also use $J_R = J_L - 2 N_P$ to arrive at
\begin{equation}
	\begin{split}
	q^2 &= N d ( \pm (j - 2 p) + d (p + \eta j + \eta^2)),\\
	m &= N( (1\mp1) j \pm 2 p + 2 \eta - 2 d (p + \eta j + \eta^2).\\ 
	\end{split}
\end{equation}
It is not hard to see that $q^2$ is monotonic in $|d|$ in the regime $2 p \leq j
\leq 1+p$ for any $\eta$ while for $m^2$ this also happens for $\eta = 0,
-1$.  Thus it is always entropically favorable to take $|d|=1$.

Let us combine these constraints to compute $D(\Gamma_1)$ in terms of the
CFT parameters.
We can restrict the four cases $\eta = 0, -1$
and $d=\pm 1$ and we find the two dominant combinations
\begin{equation}
	\begin{split}
D_a &= N^2 \left(1-\sqrt{j-p}\right)^2 p, \qquad (\eta = 0, d=1) \\
D_b &= N^2 \left(1-\sqrt{1-p}\right)^2 (1 + p -j), \qquad (\eta = -1, d=-1) \\
	\end{split}
\end{equation}
The cross-over between the two entropies seems to occur at $j=1$ and this
will be borne out from the numerical evaluations below.

Although $D_a$ and $D_b$ are positive and real for $p \leq j \leq
1+p$ (or $0 \leq j \leq 1+p$ for $D_b$) this is misleading as we know the
relation $J_L - J_R = 2 N_p$ restricts $J_L$ from below.  In the bulk this
relation simply follows from $r_{12} = J_R = J_L - 2 N_P$.
Thus these configurations exist only for $j > 2p$.

Let us consider the new maximal-entropy configurations we have found.  For $j <
1$ the maximal entropy bulk configuration has $\eta=0$ and is thus a BMPV plus
supertube (as we did not have to flow). The charges for this configuration after maximization are found to be
\bea
\Gamma_{\text{bmpv}} &=& \{1, \{0,0,0 \}, \{\sqrt{N}- \sqrt{J_L-N_p} , \sqrt{N}- \sqrt{J_L-N_p}, N_p \}, 0 \}, \\
\Gamma_{\text{tube}} &=& \{0, \{0,0,1\},\{ \sqrt{J_L-N_p}, \sqrt{J_L-N_p},0 \}, J_L-N_p \}.
\eea
For $j > 1$ the maximal entropy
phase corresponds to a configuration which must be flowed by $\eta =-1$ to give a
BMPV plus tube.  

Recall that spectral flow is accomplished by a $\gamma$-transformation, but only
after a $g$-transformation whose coefficient is fixed by the constraint that the total D6
charge after the flow be equal to 1 (\ref{flow_param}):
\begin{equation}
	g^3 = -d  \quad \textrm{followed by} \quad \gamma_3 = -1.
\end{equation}
The resulting solution is a black ring in a background with non-trivial Wilson lines\footnote{These Wilson lines correspond in four dimensions to Abelian flux on the D6 center}, which we can undo by a further  $g$-transformation: 
\begin{equation}
	\{g^1, g^2, g^3\} = \{-q, -q, 1\}\,.
\end{equation}
The final solution is thus an asymptotically $AdS_3 \times S^3$ black ring:
\begin{equation}
	\begin{split}
 \Gamma_1 &= \{ 0, \{\sqrt{N} - \sqrt{N-N_p},\sqrt{N} - \sqrt{N-N_p},1 \}, \nn
 & ~~~~~~ \{ \sqrt{N-N_p},\sqrt{N-N_p}, 2\sqrt{N-N_p} ( \sqrt{N}  - \sqrt{N-N_p}) \}, J_L-2N_p \} \\
\Gamma_2 &= \{1,\{0,0,0\},\{0,0,0\},0\}\,.
	\end{split}
\end{equation}
Hence the cross-over between $D_a$ and $D_b$ is the cross-over between a
solution describing a BMPV black hole surrounded by a supertube and a
solution describing a black ring, and from now on we will refer to $D_a$
and $D_b$ as $\Dtube$ and $\Dbr$. Thus, the entropy of the two-center
configurations is
\begin{align}
 S_{\text{tube}}&=2\pi \sqrt{D_{\text{tube}}},&
 D_{\text{tube}} &= N^2 \left(1-\sqrt{j-p}\right)^2 p,
 \label{Stube}
\\
 S_{\text{BR}}&=2\pi \sqrt{D_{\text{BR}}},&
 D_{\text{BR}} &= N^2 \left(1-\sqrt{1-p}\right)^2 (1 + p -j).
 \label{SBR}
\end{align}

\subsubsection{New Phases in Supergravity}

We have now established that there exist two maximal entropy
configurations (with cross-over at $j=1$) that have the same quantum
numbers as the new CFT phase. Unfortunately neither of these phases has
the same entropy as the CFT but interestingly they are restricted to the
same regime of validity as the enigmatic CFT phase, namely $2 p \leq j
\leq 1 + p$.  As the bulk entropy is lower than that of the CFT it
seems, as expected, that unprotected states are lifting as we go to
strong coupling.  Surprisingly, however, our results suggest that not
all states lift. The new phases we find in the bulk indicate that many
states that do not contribute to the elliptic genus in fact do not lift
at strong coupling. Furthermore, those are not just a small subset of
the original states: the entropy of the bulk objects has the same growth
with the charges as the entropy of the CFT\@. As mentioned before, this
might be the consequence of some, as yet undiscovered, index that
captures some fraction of the entropy of the enigmatic CFT phase.

\begin{figure}[htbp]
 \begin{center}
 \begin{tabular}{ccc}
   \epsfxsize=8cm \epsfbox{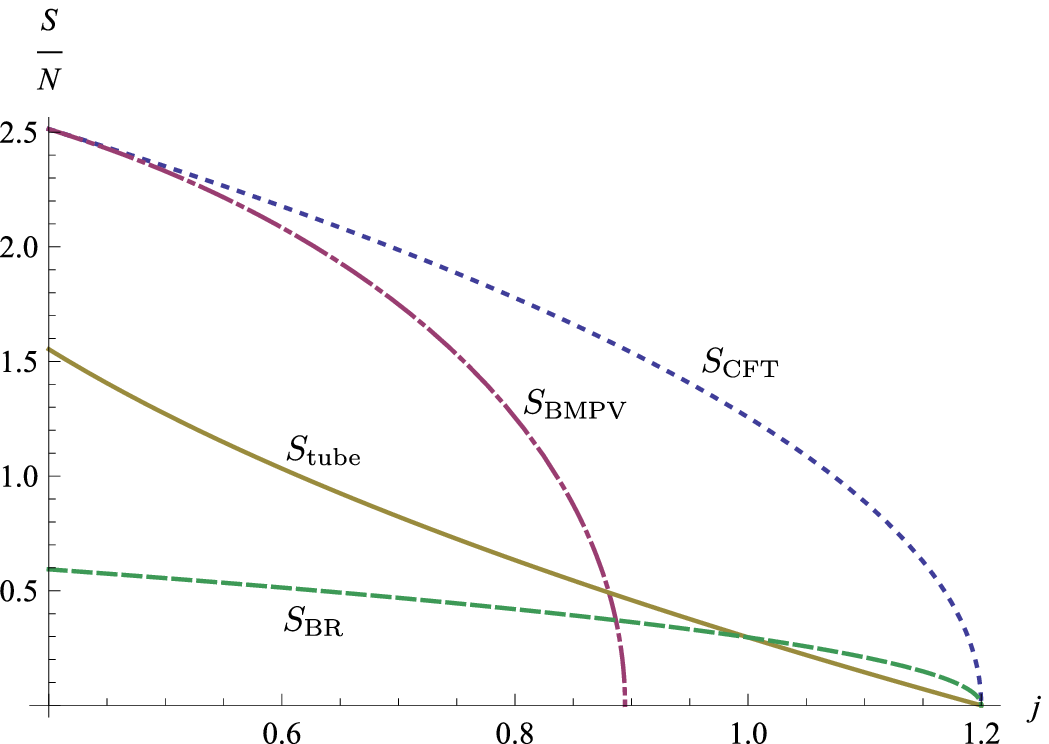} &
   \epsfxsize=8cm \epsfbox{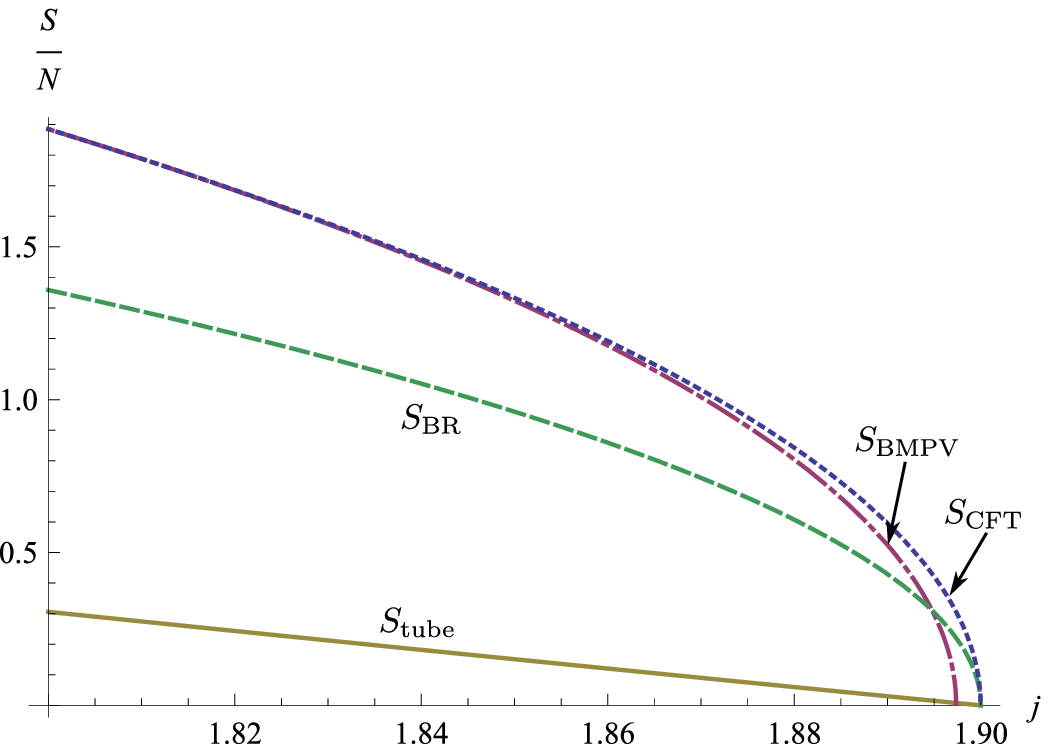} \\
   $p = 0.2$ & $p=0.9$\\
   \epsfxsize=8cm \epsfbox{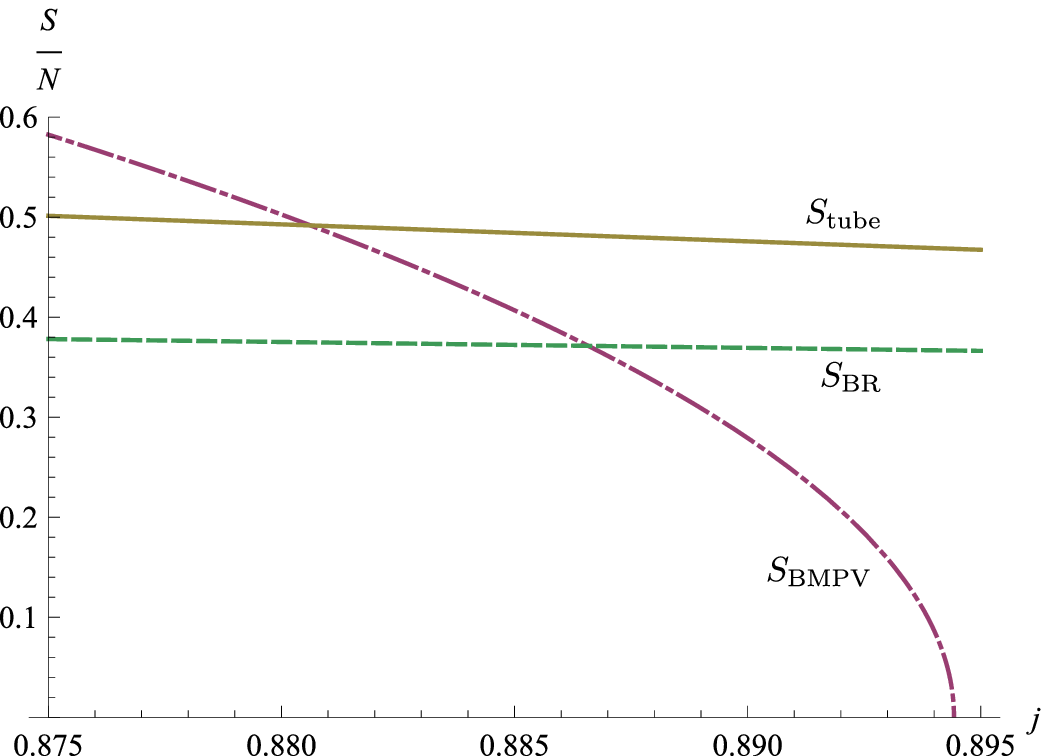} &
   \epsfxsize=8cm \epsfbox{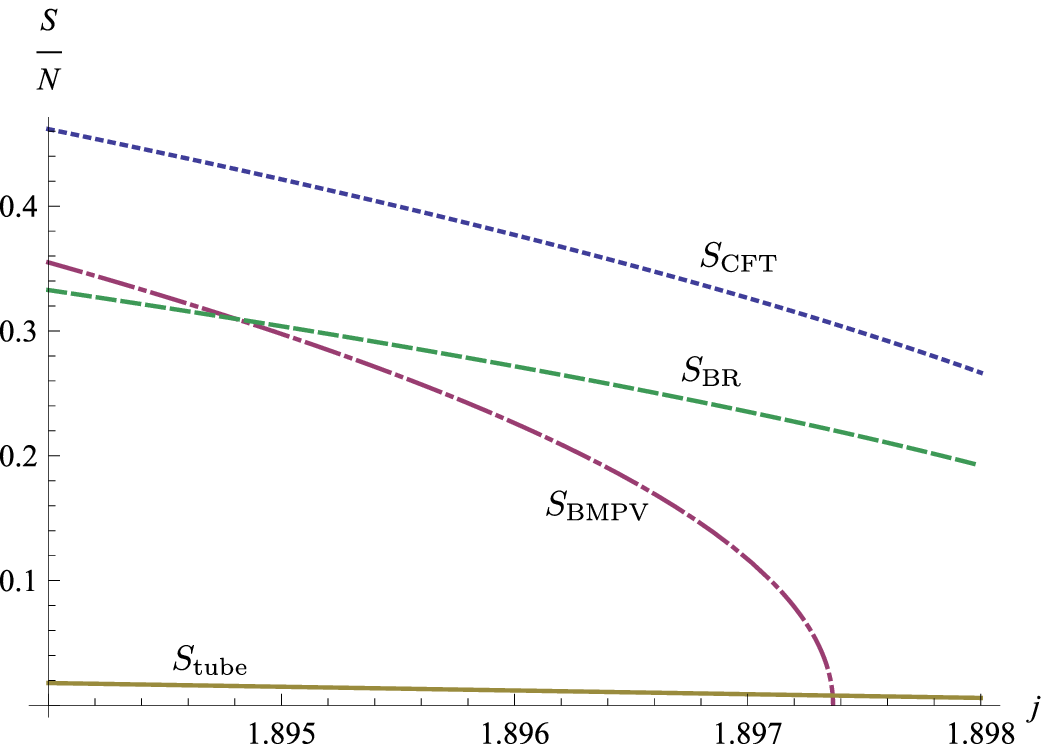} \\
   $p = 0.2$ & $p=0.9$\\
   \epsfxsize=8cm \epsfbox{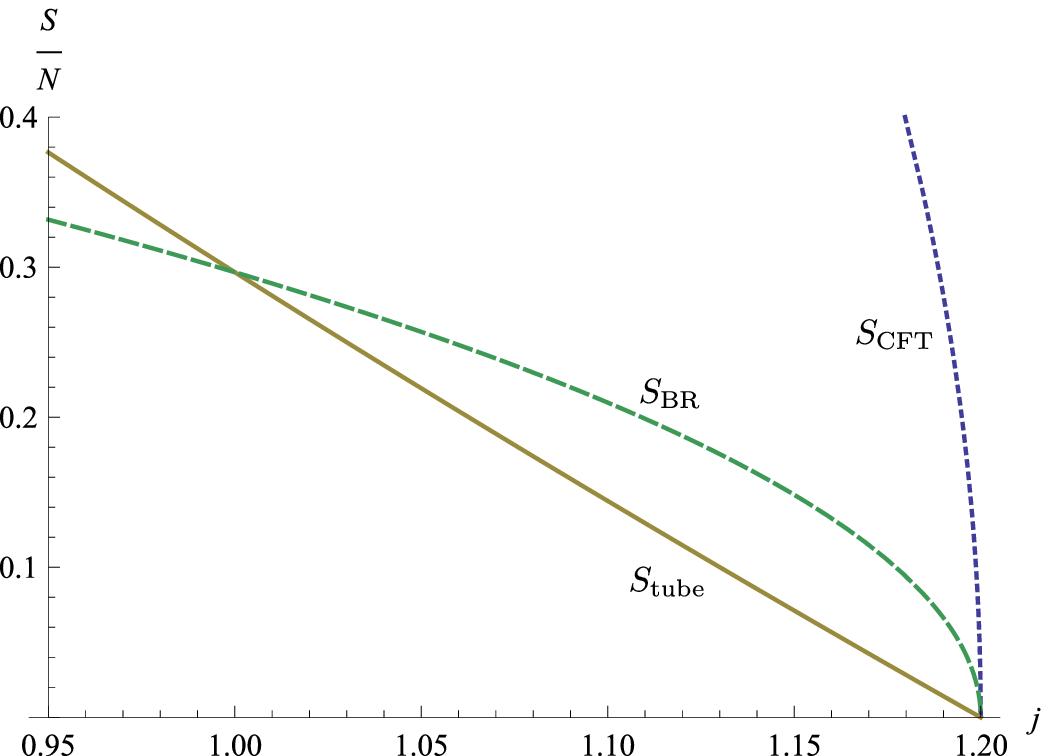} &
   \epsfxsize=8cm \epsfbox{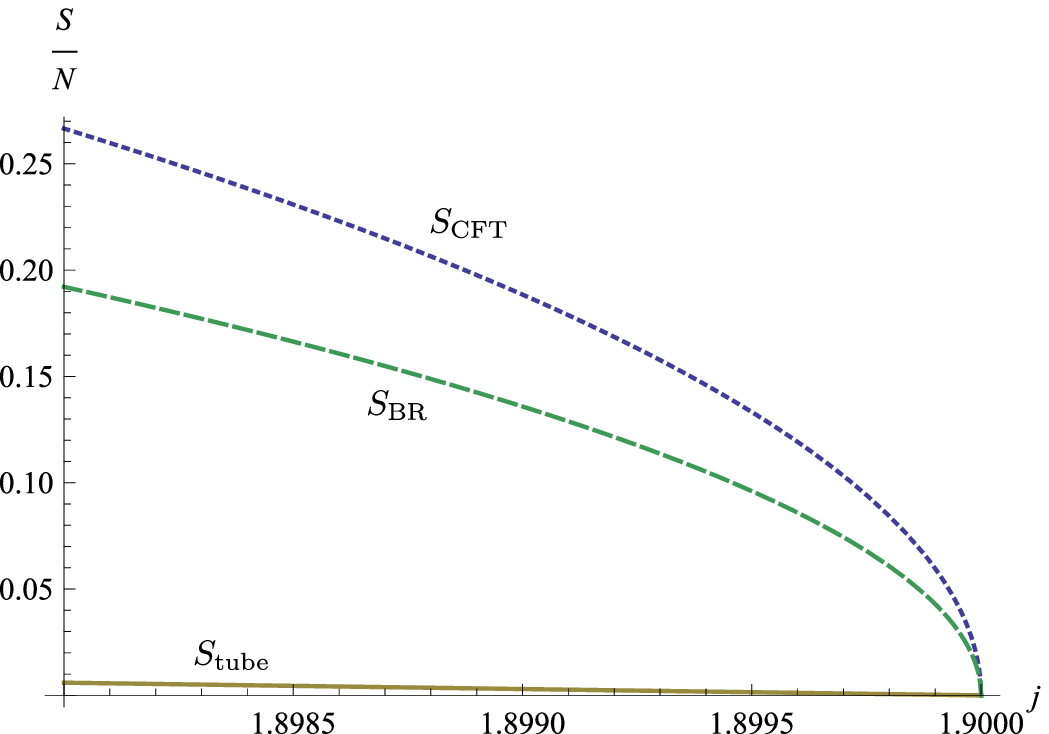}\\
   $p = 0.2$ & $p=0.9$\\
 \end{tabular}\\[2ex]
  \caption{\sl Plots of the various entropies.  The new phases are in green
  (dashed), $\Sbr = 2 \pi N (1- \sqrt{1-p})\sqrt{1+p-j} $, and brown/yellow
  (solid), $\Stube=2\pi N (1- \sqrt{j-p})\sqrt{p}$.  For comparison we plot, in
  blue (dotted), the CFT entropy, $S_{\textrm{CFT}} = 2\pi N \sqrt{p (1 + p
  -j)}$, and, in purple (dot-dashed), the entropy of a single-center BMPV black
  hole with the same charges, $S_{\text{BMPV}} = 2 \pi N \sqrt{ p - j^2/4}$.
  \label{fig:sugraEntropy}}
 \end{center}
\end{figure}

In Fig.\ \ref{fig:sugraEntropy}, we plot the entropy for these
two-center phases, as well as that of the single-center BMPV black hole
and the CFT phase to see how they compare.
We plot the entropies versus $j$, for specific fixed values of $p$,
namely $p=0.2$ (left column) and $p=0.9$ (right column). First, from the
$p=0.2$ graphs, we see that, for $j<1$, $\Dtube$ dominates over $\Dbr$
while, for $j>1$, $\Dbr$ dominates over $\Dtube$.  On the $j=1$ line,
the two entropies are degenerate, $\Dtube=\Dbr$, although the actual
configurations remain different.  However, because these phases exist
only for $j>2p$, if $p$ is too large, including $p=0.9$, the
BMPV+supertube (or ``BMPV+tube'') phase ceases to exist in the
allowed range of $j$ and only the black ring phase exists.

Next, we can ask how do the two-center entropies $\Dtube$ and $\Dbr$ compare with
$D_{\rm BMPV}$, the entropy of the single-center BMPV phase?  For
$j>2\sqrt{p}$, which corresponds to the region below the BMPV bound
$p=j^2/4$, the BMPV phase does not exist and the two-center phases are
the dominant phases (although only one of them is dominant depending on
$j\lessgtr 1$ as we just discussed).  On the other hand, for
$j<2\sqrt{p}$, which corresponds to the region above the BMPV bound, the
two-center phase dominates over the single-center BMPV phase in a
certain small range of $j$ below the bound $j=2\sqrt{p}$.  Again,
depending on the value of $j$, the dominant phase is either the
BMPV+tube phase ($j<1$) or the black ring phase ($j>1$).

\begin{figure}[htb]
 \begin{center}
   \epsfxsize=12.5cm \epsfbox{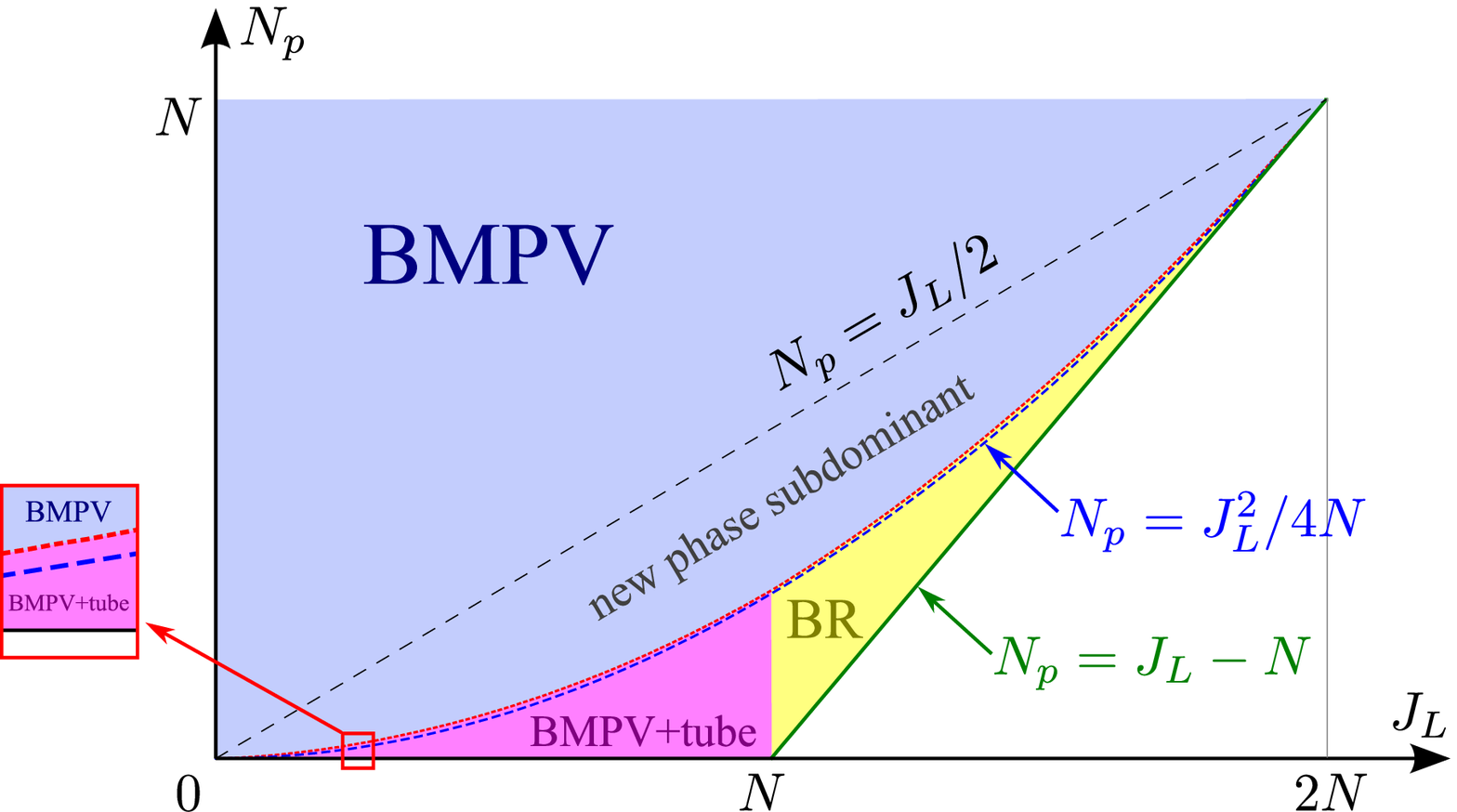} \caption{\sl The bulk
  phase diagram. In the light blue region, the single-center BMPV black
  hole is dominant.  In the pink and yellow regions the
  new phase dominates, either as a BMPV black
  hole surrounded by a supertube for $J_L<N$ (pink), or as a black ring
  for $J_L>N$ (yellow).  Below the thin dashed black line and above the
  dotted red curve, the BMPV phase and the new phase coexist but the BMPV phase is
  dominant.  In the narrow region between the dotted red curve and
  dashed blue curve, the two phases coexist and the new phase is
  dominant.
  \label{fig:phase_diag_grav}}
\end{center}
\end{figure}

All these can be seen much more clearly in Fig.\
\ref{fig:phase_diag_grav}, where we present the phase diagram of the
bulk D1-D5 system on the $J_L$-$N_p$ plane (we have already presented a
schematic version of this in Fig.\ \ref{fig:new_phase_diag}b).  Notably,
even above the BMPV cosmic censorship bound $N_p=J_L^2/4N$, there is a
region in which the new phase dominates over the single-center BMPV
black hole.  Also, the new phase is dominant in the whole region below
the cosmic censorship bound where the phase of a gas of supergravity
particles is subdominant.

\section{Discussion}
\label{sec:discussion}

In this paper, we have carefully investigated the supersymmetric phases of
the D1-D5 system, and found new phases on both sides of the AdS/CFT correspondence.
The new phase in the CFT is always entropically dominant over the BMPV phase
in the whole parameter region where the two phases coexist, whereas the
new phase in supergravity is dominant over the BMPV phase in a much smaller region.  
Below the cosmic censorship bound where the BMPV
phase ceases to exist, the new phases are dominant both in the CFT
as well as in supergravity.

In the CFT we found that the angular momenta of the phase that dominates the entropy satisfy the
relation $J_R=J_L-2N_p$ (eqn.\ \eqref{Eq:CFT_JL_JR_rel}).  We then looked for bulk configurations that satisfy the same relation (eqn.\ \eqref{bulkJLJR}) and obtained the phase diagram shown in
Fig.~\ref{fig:phase_diag_grav}.  

If one relaxes this constraint, and looks instead for bulk configurations that dominate the entropy for fixed charges and $J_L$, one can
find bulk two-center configurations (BMPV+tube and pure black ring)
that have $J_R<J_L-2N_p$ and have slightly \emph{larger} entropy than
the ones having $J_R=J_L-2N_p$.  However, the difference in entropy is
small and the phase diagram is virtually unchanged from
Fig.~\ref{fig:phase_diag_grav}.\footnote{A peculiar thing however is
that, sufficiently inside the BMPV parabola (sufficiently away from the
cosmic censorship bound), the most entropic two-center configuration has
$J_R=0$ and $r_{12}=0$.  This is a collapsing limit of the two-center
solution and is singular.  The entropy in this limit is smaller than
that of the single-center BMPV black hole, and therefore such a
configuration is never realized thermodynamically.  So, this does not
affect the phase diagram at all.}  To avoid this unnecessary
complication, we imposed the constraint $J_R=J_L-2N_p$ in the bulk.

Thus we have found that near the boundary of the region where
single-center black holes exist (the cosmic censorship bound) there
appear new phases with more entropy than the single-center black hole,
which can be thought of as the result of shedding of hair, or moulting,
of the single center black hole.  Moreover, we have seen that in
different regimes of parameter space the BMPV black hole has different
moulting patterns: for small $J_L$ it sheds all its angular momentum in
supertube hair, while for large $J_L$ it sheds a hair of Gibbons-Hawking
or Taub-NUT charge and becomes a black ring.

The phenomenon we find has also been seen for D4-D0
(equivalently M5-P) black holes in ${\cal N}=2$ four-dimensional supergravity
\cite{deBoer:2008fk}. In both situations the new phase dominates only very close
to the cosmic censorship boundary. Note that in an asymptotically-flat setting
one can map the D6-D2-D0 black hole whose moulting we described here to the
D4-D2-D0 black hole whose moulting was described in \cite{deBoer:2008fk} via a
combination of spectral flows, gauge transformations and 4D $S$-duality
(equivalent to six $T$-dualities) \cite{Dall'Agata:2010dy}. This map
however interchanges harmonic functions, and generically may not map
asymptotically $AdS_3\times S^3$ black holes to asymptotically $AdS_3 \times
S^2$ black holes.  Thus it is not immediately obvious whether the $AdS$
moulting pattern we found here maps to the $AdS$ moulting pattern found in
\cite{deBoer:2008fk}.

\subsection{A Supersymmetric Gregory--Laflamme Instability}

By an analysis of the geometry similar to the one done in
\cite{Lunin:2002iz}, one can show that the bulk ``instability'' that
drives the BMPV black hole to a two-center solution can be thought of as
a ``supersymmetric version'' of a Gregory--Laflamme instability
\cite{Gregory:1993vy}. Indeed, all the solutions we study are
supersymmetric and therefore stable.  However, if we make them
infinitesimally non-extremal one naturally expects them to decay into
more entropic configurations; thus a near-extremal BMPV black hole would
decay into a near-extremal black ring or a near-extremal BMPV+supertube
geometry, and would localize on the $S^2$ base as we explain
below. This is very similar to the localization instability found for
the original entropy enigma \cite{Denef:2007vg, deBoer:2008fk}, in which
a supersymmetric black hole localizes in $S^2$.

In the six-dimensional AdS$_3\times S^3$ geometry, the original BMPV
black hole is filling the entire $S^3$ and is pointlike in the
two-dimensional spatial part of AdS$_3$, which can locally be thought of
as $\bbR^2$.  Thus the horizon topology is $S^3\times S^1$, where $S^1$
is coming from fattening a pointlike object in $\bbR^2$.
On the other hand, the new two-center solution made of a BMPV black hole
and a supertube can be thought of as a black hole which wraps the $S^1$
Hopf fiber of the $S^3$ and is pointlike in the $S^2$ Hopf base.  It is
again pointlike in the two-dimensional spatial part of the AdS$_3$. Now
the horizon topology is $S^1\times S^3$, where $S^3$ is coming from
fattening a point in the spatial part of $S^2\times{\rm AdS}_3$, which is
locally $\bbR^{4}$.  Note that the tube uplifts to a smooth point on the
$S^3$ in six dimensions (see \cite{Lunin:2002iz}, page 8).\footnote{Some
more details on the topology of the spacetime, along the lines of
\cite{Lunin:2002iz}, are as follows: the five-dimensional spatial part of
the spacetime can be thought of as $S^1$ fibered over an $\bbR^4$ base.
The supertube worldvolume is a circle in $\bbR^4$, at which the $S^1$ shrinks (because of the KKM dipole charge).
If one considers a disk $D_2$ whose boundary is this circle, the $S^1$
fiber over the $D_2$ gives the $S^3$.  The BMPV black hole sits at the
center of this $D_2$.  Because the $S^1$ fiber does not shrink there,
the BMPV black hole wraps the fiber $S^1$ although it is pointlike in the base.}
So, in this process, the black hole localizes in the base $S^2$.

It is interesting to ask why this localization occurs only in the base
$S^2$ of the $S^3$ but not in the fiber $S^1$, no matter how small the
charges $J_L,N_p$ are. We can argue that a complete localization in
$S^3$ is entropically disfavorable, using an argument similar to that of
\cite{Bhattacharyya:2010yg}:
Consider a small black hole localized in $S^3$.  For this hole to carry
$J_L$, it must be zipping around the equator of the $S^3$.  Let the
velocity and the rest mass of the small black hole be $v$ and $m$,
respectively.  Its angular momentum $J_L$ is\,\footnote{Let the $S^3$ be
given by $\sum_{i=1}^4 (x^i)^2=R^2$. For example, let the hole be
rotating along the circle in the 1-2 plane, {\it i.e.},
$(x^1)^2+(x^2)^2=R^2$, $x^3=x^4=0$.  Then the angular momentum
$J^{ab}=x^a p_b-x^b p_a$ is given by $J^{12}=-J^{21}=Rmv\gamma$ with all
other components vanishing.  If we define $J_L^i,J_R^i$, $i=1,2,3$ by
$J_{L,R}^i=J^{i4}_\pm$, $J^{ij}_{\pm}=(1/2)(\tilde J^{ij}\pm J^{ij})$,
$\tilde J^{ij}=(1/2)\epsilon^{ijkl}J^{kl}$, then we find
$J_L^{3}=-J_R^3=-Rmv\gamma/2$.  According to our definition,
$J_L=2J_L^3=Rmv\gamma$.  } $J_L\sim Rmv\gamma$, while its energy is
$E\sim m\gamma$, where $R$ is the radius of $S^3$ and
$\gamma=(1-v^2)^{-1/2}$. If we assume that this configuration is BPS,
then $N_p=ER\sim Rm\gamma$.
By solving these relations for $m$ and $v$, we find
$m=(N_p^2-J_L^2)^{1/2}/R$, $v=J_L/N_p$. The black hole mass $m$ and
entropy $S_{\text{small}}$ are related to $r_H$ by $m\sim  r_H^3/G_6$ and
$S_{\text{small}}\sim r_H^4/G_6$ where $G_6$ is the six-dimensional
Newton constant.  Therefore, we find $S_{\text{small}}\sim
(N_p^2-J_L^2)^{2/3}G_6^{1/3}R^{-4/3}=N(p^2-j^2)^{2/3}$.  Here, we used
the relation $R\sim G_6^{1/4}N^{1/4}$ which follows from the
AdS$_3$/CFT$_2$ dictionary.
On the other hand, the entropy of the BMPV black hole is
$S_{\text{BMPV}}\sim (NN_p-J_L^2/4)^{1/2}=N(p-j^2/4)^{1/2}$.  
Let us consider the scaling limit $N\to \infty$ with $p=N_p/N$ and
$j=J_L/N$ fixed, as we have been assuming throughout the paper.  We take
$p,j\ll 1$ so that the radius of the BMPV black hole becomes much
smaller than that of $S^3$.\footnote{From $S_{\text{BMPV}}\sim
r_{\text{BMPV}}/G_3$, $R\sim G_6^{1/4}N^{1/4}$ and $G_3\sim G_6/R^3$, it
is easy to see that $r_{\text{BMPV}}/R\sim S_{\text{BMPV}}/N\sim
\sqrt{p-j^2/4}$.  So, $p,j\ll 1$ is sufficient for the radius of the
BMPV black hole, $r_{\text{BMPV}}$, to become much smaller than $R$.}
In order for the small black hole to exist, the reality of
$S_{\text{small}}$ requires that $p\sim j^\alpha$ with $\alpha\ge 1$ as
we consider $p,j\ll 1$.  In this case, $S_{\text{small}}\sim Np^{4/3}\ll
S_{\text{BMPV}}\sim Np^{1/2}$ for $p\ll 1$.
Namely, in the limit in which the radius of the BMPV black hole becomes
much smaller than that of $S^3$, a full localization in $S^3$ is
entropically unfavorable and does not happen. What happens instead is a
partial localization in the $S^2$ base, as we have demonstrated by
constructing the explicit solution.

\subsection{The New Phases in the Canonical Ensemble}

Thus far, we considered the new phases in the microcanonical ensemble,
fixing the conserved charges $N_p$ and $J_L$.  It is interesting to investigate the
role of the new phases in the \emph{canonical} ensemble.\footnote{We
thank S.~Minwalla for suggesting we consider the
canonical ensemble.}  Let us flow to the NS sector 
where the transition from a gas of gravitons to the BMPV phase can be regarded as a Hawking--Page phase transition.  In
the NS sector, the entropy formulas for the BMPV phase and the new phase of the CFT
are
\begin{align}
 S^{\text{NS}}_{\text{BMPV}}(L_0)&=2\pi\sqrt{N\left(L_0-{\f{N}{4}}\right)},\qquad
 S^{\text{NS}}_{\text{new}}(L_0)=2\pi L_0~,
\end{align}
where we have set $J_L=0$ for simplicity and have used the relation
\eqref{DimAndMomentum} to eliminate $N_p$ and write the equations in terms of $L_0$. If we introduce the left-moving temperature $T$, from the
thermodynamical relation ${\partial S/\partial L_0}=1/T\equiv \beta$, we
obtain
\begin{align}
 T_{\text{BMPV}}={1\over \pi}\sqrt{{L_0\over N}-{1\over 4}},\qquad
 T_{\text{new}}={1\over 2\pi}.
\end{align}
Now let us go to the canonical ensemble by defining the free energy\footnote{Because of the shift by $-c/24=-N/4$, the relation between
the partition function and the free energy is
$\Tr_{\text{NS,BPS}}[e^{-\beta (L_0-c/24)}]=e^{-\beta F}$.}
\begin{align}
 F=L_0-{N\over 4}-TS.
\end{align}
We find
\begin{align}
 F_{\text{BMPV}}(T)&=-\pi^2N T^2,\qquad
 F_{\text{new}}=-{N\over 4}.
\end{align}
Note that $F_{\text{new}}$ is defined only for $T={1/2\pi}$.

On the other hand, the thermodynamic quantities for
``thermal''\,\footnote{We have a
non-vanishing left-moving temperature but the right-moving temperature vanishes.
Therefore the physical temperature vanishes.} AdS are given by:
\begin{align}
 F_{\text{tAdS}}=-{N\over 4},\qquad
 S_{\text{tAdS}}=0,\qquad
 (L_0)_{\text{tAdS}}=0.
\end{align}
These simply come from $e^{-\beta F}=\Tr_{\text{tAdS}}[e^{-\beta
(L_0-N/4)}]\sim e^{\beta N/4}$ because only the NS ground state
contributes.

\begin{figure}[htbp]
 \begin{center}
 \begin{tabular}{ccc}
   \epsfxsize=7cm \epsfbox{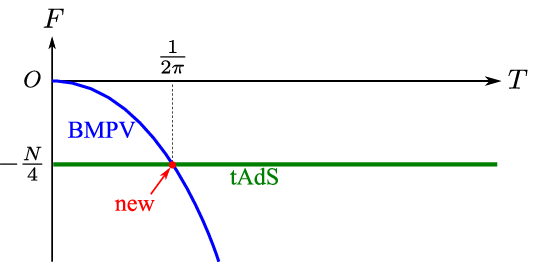}
  &\hspace*{1cm}&
  \epsfxsize=7cm \epsfbox{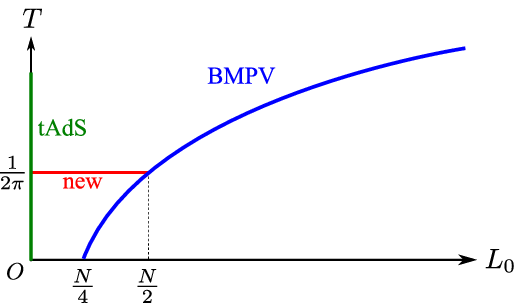}\\[1ex]
  (a) $F$ versus $T$
  &\hspace*{1cm}&
  (b) $T$ versus $L_0$
 \end{tabular}\\[2ex]
  \caption{\sl Thermodynamic quantities for the CFT phases in the canonical ensemble.
  \label{fig:canonicalCFT}}
 \end{center}
\end{figure}

We have plotted $F(T)$ for the three phases in
Fig.~\ref{fig:canonicalCFT}(a). As we increase $T$ from $T=0$, we have a
Hawking--Page transition at $T=T_c=1/2\pi$ where the thermal AdS phase
gives way to the BMPV phase.  Exactly at $T=T_c$, we can have the new
phase as well. The meaning of this is clearer in the graph of $T(L_0)$
shown in Fig.~\ref{fig:canonicalCFT}(b). As we increase $T$ from $T=0$, we
first go along the vertical axis in the thermal AdS phase. Then at
$T=T_c=1/2\pi$, we now move horizontally along the ``new phase'' line,
and then finally reach the BMPV phase.  During this horizontal motion,
the temperature stays constant and the energy put into the system is
used to convert the short strings into the long one.  So, in the
canonical ensemble, the new phase can be interpreted as the coexisting
phase of the thermal AdS (short strings) and BMPV (long string) phases,
much as the coexisting phase of ice and water. The difference is that
ice and water coexist in the real space while the two CFT phases coexist
in the space of effective strings.

We can repeat the same analysis for the bulk configuration. The entropy formulae \eqref{Stube}, spectral-flowed to the NS sector, become
\be
S^{NS}_{new,bulk}(L_0)=2 \pi (\sqrt{N} - \sqrt{N-L_0})\sqrt{L_0}.
\ee
Note that the spectral-flowed expression for the BMPV+supertube and the black ring configurations is the same for $J_L=0$. The temperature for this phase is
\be
T_{new,bulk}=\f{\sqrt{L_0 (N-L_0)}}{2 \pi (L_0 - \f{N}{2}) + \pi \sqrt{N(N-L_0)}}
\ee
and the free energy is
\be
F_{new,bulk}=\f{L_0 \sqrt{N}}{\sqrt{N}+2 \sqrt{N-L_0}} - \f{N}{4}.
\ee
\begin{figure}[htbp]
 \begin{center}
 \begin{tabular}{ccc}
   \epsfxsize=7cm \epsfbox{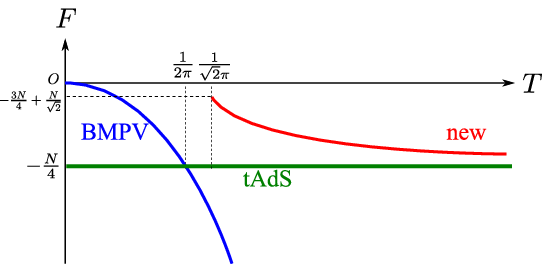}
  &\hspace*{1cm}&
  \epsfxsize=7cm \epsfbox{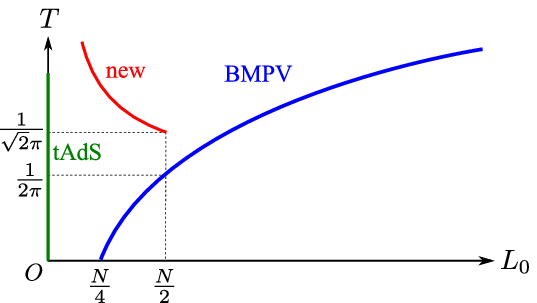}\\[1ex]
  (a) $F$ versus $T$
  &\hspace*{1cm}&
  (b) $T$ versus $L_0$
 \end{tabular}\\[2ex]
  \caption{\sl Thermodynamic quantities for the CFT phases in the canonical ensemble.
  \label{fig:canonicalSUGRA}}
 \end{center}
\end{figure}

We have plotted the $F(T)$ for the BMPV, thermal AdS and the new bulk
phase in Fig.~\ref{fig:canonicalSUGRA}(a). We have a Hawking--Page
transition at $T=T_c=1/2\pi$ where the thermal AdS gives way to the BMPV
phase. The new phase exist for $\f{1}{\sqrt{2} \pi} < T < \infty$. In
Fig.~\ref{fig:canonicalSUGRA}(b) we plot $T(L_0)$ for the three
phases. The temperature of the new phase is infinite for $L_0=0$ and
monotonically goes down to $\f{1}{\sqrt{2} \pi}$ at $L_0=\f{N}{2}$. From
both these graphs we see that the new bulk phase has a negative specific
heat and thus cannot be realized in the canonical ensemble even though
it exists in the microcanonical ensemble.

Including $J_L\neq 0$ does not change the above qualitative picture.

\subsection{Future Directions}

Since our motivation has been mostly AdS/CFT-based we have focused here on a
particular ``moulting'' of the BMPV black hole in an attempt to reproduce the
CFT phase transition.  It is interesting to note however that an
asymptotically-flat BMPV black hole can have more moulting patterns than an
asymptotically-AdS one: in asymptotically-flat solutions the D1, D5, and P charges are on equal footing, and a black hole  can shed
either D1-D5, D1-P or D5-P supertube hair. Nevertheless, since the $AdS_3\times S^3$ near-horizon breaks the interchange symmetry between the three charges, only the D1-D5 supertube hair remains in this limit; the other supertubes are too large and do not fit inside this near-horizon region \cite{Bena:2004wt}.


In \cite{Bena:2004tk} a proposal for a CFT ensemble dual to a bulk
black ring was put forward, which, modulo one phenomenological
assumption about the length of the short strings, reproduces the
seven-parameter entropy of the ring. The phases we discuss in this paper
have short strings that have the smallest-allowed size consistent with
the charges, and hence, almost by construction, have more entropy than
the black ring. However, as one increases the effective coupling to move from the orbifold point to the regime where supergravity is valid, one expects the phase we constructed to lose a finite fraction of its entropy and end up describing the black ring. 

There are three possible scenarios how this might happen: it may be possible that all
states that have short strings of length smaller than that of
\cite{Bena:2004tk} get uplifted, and only the states with short strings
of the length of \cite{Bena:2004tk} or bigger survive. The second
possibility is that the number of short strings stays constant as one
increases the coupling, but their length changes; since the total length
is constant this reduces the entropy carried by the long string, to the
black ring value. The third possibility is that the phenomenological
length of \cite{Bena:2004tk} represents the average of the lengths of
the short string lengths, and that as one increases the coupling, the
kind of small strings that the long string sheds changes, such that the
final average is the phenomenological length.

We would also like to note that our computation of the microscopic
partition function based on \cite{Eguchi:1988vra} can be
straightforwardly generalized to include $J_R$ dependence, and it would
be interesting to see if this can be related to the recent results of
\cite{Manschot:2011xc} (see also \cite{deBoer:2008zn, deBoer:2009un}).

In this paper, we have made a thorough search for the bulk configuration
that maximizes the entropy.  However, it is logically possible, although
we find it unlikely, that there are some bulk configurations that have
larger entropy than the ones we have been able to find. Indeed, we made
an assumption that the relevant two-center configurations have one
smooth center, and the stability argument that we present in Appendix C
indicates that such a configuration is indeed a local maximum of the
entropy; however, we could not establish that this is a global maximum,
and thus it is formally possible that there are some other two-center
configurations with two horizons and more entropy.

Second, we imposed a $U(1)\times U(1)$ symmetry in the bulk because
the entropy-maximizing configuration in the CFT lives in a single $J_R$
multiplet.  However, as we observed above, the value of $J_R$ that
maximizes the bulk entropy is not precisely the same as the one
maximizing the CFT entropy, and it is logically possible that the $J_R$
multiplet matching does not hold; as such the maximum-entropy
configuration in the bulk might break this symmetry and have more than
two centers, or have some inhomogeneities.\footnote{Much like it happens
in some holographic systems where spatially inhomogeneous configurations
can be thermodynamically dominant over homogeneous ones (for an
incomplete list of recent work, see \cite{modulatedPhases}).}

These unlikely possibilities aside, our calculation shows that there
exist many CFT states that are \emph{not} protected by the elliptic
genus, but that nevertheless do \emph{not} lift at strong
coupling. Furthermore, the entropy of these states is not subleading,
but is of the same order of, and sometimes dominant over the entropy of
the black hole. This fact either indicates the existence of a new index,
or hints at a previously unthought-of dynamical mechanism that prevents
the lifting of such a large number of states. We find both possibilities
extremely interesting.

\section*{Acknowledgments}

We thank A. Dabholkar, H. Elvang, R. Emparan, M. Guica, P. Kraus,
G. Mandal, S. Minwalla, S. Murthy, K.  Papadodimas and A. Virmani for
helpful discussions.  MS is very grateful to the ITFA, University of
Amsterdam and the IPhT, CEA-Saclay where part of this work was done for
hospitality. IB and MS are also grateful to the  Aspen Center for Physics for hospitality, and 
support via the NSF grant 1066293. The work of IB and SE was supported in part by the ANR
grant 08-JCJC-0001-0, and by the ERC Starting Independent Researcher
Grant 240210 - String-QCD-BH. The work of BDC is supported by the ERC
Advanced Grant 268088-EMERGRAV. This work is part of the research
programme of the Foundation for Fundamental Research on Matter (FOM),
which is part of the Netherlands Organisation for Scientific Research
(NWO).  The research of SE is also supported by the Netherlands
Organization for Scientific Research (NWO) under a Rubicon grant.

\appendix

\section{The Decoupling Limit}\label{app_decoupling}

In this Appendix we examine the charges and the harmonic functions that give multicenter solutions that in the IIB frame (\ref{IIB_sol}) have AdS$_3\times$S$^3/{\mathbb Z}_n$ asymptotics\footnote{A discussion of this can also be found in \cite{Bena:2004wt} and Appendix B of \cite{Bena:2008dw}.}.
We take the following total charge
\begin{equation}
	\Gamma = \{n, k^I, l_I, m\}
\end{equation}
and set all the constants in the harmonic functions to zero except $m_0$ and
$l_3^0$.  We will generally only be concerned with the asymptotic charges but we
will also need dipole charges to compute $J_R$ asymptotically so we consider a
two-center configuration with the following harmonic functions
\begin{align}
	V &= \frac{n_1}{r_1} + \frac{n_2}{r_2}, \qquad K^I = \frac{k_1^I}{r_1} +
	\frac{k^I_2}{r_2}, \\
	L_I &= \frac{l^1_I}{r_1} +
	\frac{l_I^2}{r_2} + l_3^0 \delta_{I3}, \qquad 
	\tM = \frac{m_1}{r_1} + \frac{m_2}{r_2} + m_0 
\end{align}
with the first center at the origin, $r_1 = |\vec{r}|$, and the second center at
$\vec{a}$, $r_2 = |\vec{r} - \vec{a}|$.  Note that this asymptotic analysis
carries over straightforwardly to more centers.

We first expand the functions appearing in the metric to leading order
\begin{align}
	Z_1 &= \frac{1}{r} \left(l_1 + \frac{k^2 k^3}{n}\right) =: \frac{N_1}{r},
	\qquad 
	Z_2 = \frac{1}{r} \left(l_2 + \frac{k^1 k^3}{n}\right) =: \frac{N_5}{r}
	\label{zident1} \\
	Z_3 &= l_3^0 + \frac{1}{r} \left(L_3 + \frac{k_1 k_2}{n}\right) =: l_3^0 +
	\frac{N_p}{r},  \label{zident2} 
\end{align}
from which we read off the leading terms in the IIB metric
\begin{equation}
	ds_{\rm IIB}^2 \sim 
	- \frac{r}{Z_3 L} (dt + k)^2 + \frac{Z_3 r}{L} (dz + A^3)^2
	+ \frac{n L}{r^2} dr^2 + L \left(n \, d\Omega_2^2 + \frac{\sigma^2}{n}\right)
	\label{2bmetric}
\end{equation}
with $L = \sqrt{N_1 N_5}$ and $\Omega_2$ the standard S$^2$ metric.  To connect
with the standard D1-D5-P metric we consider a total charge (1, 0, $l_I$, 0)
implying that $k=0$ and $A^3 = -Z_3^{-1}(dt+k)$ and then redefine
\begin{equation}
z = x_5 + \tau, \qquad 2 t - l_3^0 z = \tau - x_5, \qquad r = \rho^2
\end{equation}
putting the metric in the form
\begin{equation}
	\frac{ds_{\rm IIB}^2}{4} \sim 
	\frac{\rho^2}{L}\left[ - d\tau^2 + dx_5^2 + \frac{N_p}{\rho^2} (dx_5 +
	d\tau)^2\right]
	+ L \frac{d\rho^2}{\rho^2} +  \frac{L}{4} ( d\Omega_2^2 + \sigma^2)
\end{equation}
where the Hopf metric on S$^3$ now properly normalized.  This justifies our
identification of $N_1, N_5$ and $N_P$ in (\ref{zident1})-(\ref{zident2}).


To determine $J_L$ and $J_R$ we should reduce the metric (\ref{2bmetric}) on
S$^3$ and read off the corresponding v.e.v. from the normalizable mode of the relevant gauge
fields.  A simpler, albeit less direct, way to identify the charges is as
follows.  The relationship between $\mu$ and $J_L$ (see eqn
(\ref{cft_from_asym})) can be fixed by considering a single-center BMPV and
relating its horizon entropy (in terms of harmonic functions) to what we expect
from the CFT\@.  This identification and normalization also follows from the
behavior of $\mu$ under bulk spectral flow.  In the M-theory frame (reduced to
five dimensions) $\mu$ or $J_L$ is related to the angular momentum along the $\psi$ circle
and the other angular momentum comes from the ${\mathbb R}^3$ base of the
solutions (the asymptotic value of $\omega$).  Thus we can identify $J_R$ as the
asymptotic value of $\omega$ and the normalization is fixed with respect to the
normalization of $J_L$ (as we take both charges to be integral rather than
half-integral).


\section{Spectral Flow}\label{app_spectral}

We provide, for reference, the charges of the ``BMPV plus supertube'' solution in terms of the
charges of the original, generic, configuration from whence they were spectral flowed
\begin{equation}
	\begin{split}
\Gamma_1 &= \Biggl\{1,\{0,0,0\},\\
&\qquad\quad\Bigl\{k_2 k_3-l_1 (\alpha -1),\ k_1 k_3-l_2 (\alpha
-1),\\
&\qquad\qquad\quad\frac{p_3 \left(k_2 \left(k_1 p_3+l_2\right)-(\alpha -1) \left(l_3
p_3+m\right)\right)-k_3 \left(l_3 p_3+m\right)+l_1 \left(k_1
p_3+l_2\right)}{\left(k_3+p_3 (\alpha -1)\right){}^2}\Bigr\},\\
&\qquad\quad
{1\over {k_3+p_3 (\alpha -1)}}\Bigl[-k_3 \left(k_1 \left(2 k_2 p_3+l_1\right)+k_2 l_2+(\alpha -1) \left(m-l_3
p_3\right)\right)\\
&\qquad\qquad\qquad\qquad\qquad
	 +(\alpha -1) \left(p_3 \left(k_2 l_2+m (-\alpha )+m\right)+l_1
\left(k_1 p_3+2 l_2\right)\right)+k_3^2 l_3\Bigr]
\Biggr\},\\
\Gamma_2  &= \Biggl\{0,\bigl\{0,0,-\alpha  \left(k_3+p_3 (\alpha
-1)\right)\bigr\},\\
&\qquad\quad\Bigl\{\alpha  \left(p_3 \left(k_2+p_2 (\alpha -1)\right)+k_3
p_2+l_1\right),\alpha  \left(p_3 \left(k_1+p_1 (\alpha -1)\right)+k_3
p_1+l_2\right),0\Bigr\}, \\
&\qquad\quad -\frac{\alpha  \left(p_3 \left(k_1+p_1 (\alpha
-1)\right)+k_3 p_1+l_2\right) \left(p_3 \left(k_2+p_2 (\alpha -1)\right)+k_3
p_2+l_1\right)}{k_3+p_3 (\alpha -1)}\Biggr\},\\
h &=\left\{0,  \{0,0,0\},\{0,0,1\},\alpha  \left(k_3+p_3 (\alpha -1)\right)\right\}.
	\end{split}\label{bmpvtube_charge_detail}
\end{equation}
Note, as emphasized in section \ref{flow_to_bmpv}, the fact that some
entries are fractional poses no physical problem, because it is merely a
result of the fractional spectral flow \eqref{flow_param} that \emph{we}
chose to do.




\section{Stability analysis of two-center solution with one smooth center}\label{app_convexity}

In \cite{deBoer:2008fk} a two-center solution was shown to be entropically
dominant over a single center solution and it was assumed that keeping one
center smooth would maximize the two-center entropy. Here we demonstrate the
validity of this assumption locally in the space of charges for the two-center solutions
considered in this paper, where one center is a BMPV black hole and the other is a
smooth supertube.  As described in the bulk of the paper we can use spectral
flow to map this to a generic configuration with one smooth center, so the analysis
performed here is broadly applicable.

Let us consider a general deformation of the BMPV+supertube system and focus on configurations and variations with equal D1 and D5 charges ($N_1 = N_5$) and equal d1 and d5 dipole charges:
\begin{equation}\label{bmpvtube_def}
	\begin{split}
		\Gamma_{\textrm{bmpv}} &=
\left\{1,\left\{0,0,0\right\},\left\{Q,Q,Q_3\right\}, m\right\},\\
\Gamma_{\textrm{tube}} &= \left\{0,\left\{d,d,d_3\right\},\left\{q, q,
q_3\right\}, m' \right\},\\
h &= \left\{0,\{0,0,0\},\{0,0,1\},-d_3 \right\}.
	\end{split}
\end{equation} 
The equality of the charges is a simplifying assumption but should
not be essential.  Variations of the BMPV D4 charges can be undone by gauge
transformation so the form above captures the most general (continuous)
deformation.  Note also that the D6 charges must remain integer in order for the background to be regular. 

We parameterize the charges as
\begin{eqnarray}
&q = a_0 + a_1 \lambda + a_2 \lambda^2+\dots, \qquad
&q_3 =  b_1 \lambda + b_2 \lambda^2 + \dots\\
&d =  c_1 \lambda + c_2 \lambda^2 + \dots, \qquad
&d_3 = 1
\end{eqnarray}
where we have also imposed the integrality of $d_3$ (which corresponds to a KK dipole charge and must be integer if the background is to be regular).  The other charges can be fixed in terms of the CFT charges $J_L$, $J_R$, $N$ and $N_p$ and the above.  We take the CFT
charges to be fixed but unconstrained (we do impose the unitarity bound $J_L < N +
N_p$ but this should hold for any state).

Note that $a_0$ is related to $m'$ at lowest order via $m' = a_0^2 +
{\mathcal O}(\lambda)$ so that to zeroth order in $\lambda$ the second center is
indeed a supertube.  The no-CTC condition implies the $q_i$ and $Q_i$ must have the
same sign (to leading order) so $b_1 \geq 0$ and $0 \leq a_0 \leq N_1$.  

To get more constraints we consider the entropies of the two centers. To
leading order the square of the entropy, 
$D(\Gamma_{\textrm{tube}})$, (see eqn (\ref{e77inv})), is never positive 
\begin{equation}
	D(\Gamma_{\textrm{tube}}) \sim -\frac{1}{4} \left(b_1-2 a_0 c_1\right){}^2
	\lambda^2
	+\dots
\end{equation}
so we  must take $c_1 = b_1/2 a_0$.  Imposing this allows us to simplify the next
non-vanishing term 
\begin{equation}
	D(\Gamma_{\textrm{tube}}) \sim \frac{b_1^2 \left(2 a_0 \left(a_0
	a_1+b_1\right)-b_1 N_1\right)}{4 a_0^3} \lambda^3 + \dots
\end{equation}
whose positivity requires
\begin{equation}\label{stubepos}
	2 x a_0^2+2 a_0-N_1 \geq 0
\end{equation}
where we have defined $x \equiv a_1/b_1$.

Next we turn to the square of the BMPV entropy, $D(\Gamma_{\textrm{BMPV}})$.  To zeroth order this is a
quartic polynomial in $a_0$
\begin{equation}\label{sbmpv0}
\frac{1}{4} \left( -a_0^4+ 2 a_0^2 J_L+2 a_0^2 N_P-8 a_0 N_1 N_P+2 J_L N_P-J_L^2-N_P^2+4 N_1^2 N_P\right)	
\end{equation}
while its leading deformation is ${\mathcal O}(\lambda)$ and has the following
form
\begin{equation}\label{sbmpvdef}
	\frac{b_1 \left(a_0-N_1\right) \left(a_0 \left(2 x N_P+N_1\right)-J_L+2
	N_P\right)}{a_0} \, \lambda\,.
\end{equation}
In order for the deformation to increase the entropy the expression
(\ref{sbmpvdef}) must be positive (the entropy contribution from the second,
deformed tube, center is subleading) which gives
\begin{equation}
2 a_0 x N_P+a_0 N_1-J_L+2 N_P \leq 0\,.
\end{equation}
Combining this with (\ref{stubepos}) yields upper and lower bounds on $x$ which
are only compatible when
\begin{equation}\label{a0const}
a_0^2 N_1 -a_0 J_L +N_1 N_P \leq 0\,.
\end{equation}
Thus $a_0$ is constrained to lie between the roots of this polynomial.  

On the other hand (\ref{sbmpv0}) is a quartic polynomial in $a_0$ which must be
positive for the leading entropy to be real.  One then checks that positivity of
(\ref{sbmpv0}) is not compatible with (\ref{a0const}).  It then follows that any
deformation that increases the entropy also generates a CTC so the BMPV plus
tube center is (locally) entropically stable.

\section{Why the ``enigmatic states'' do not contribute to the elliptic genus}
\label{sec:newphasedoesntcontributetoindex}

From a numerical analysis in section \ref{subsec:numericalEvaluation} we
saw that the enigmatic phase does not contribute to the elliptic genus
while the BMPV phase does. In this Appendix we will give an explanation
of why the particular states we consider, namely the ones of the form of
a long string with excitations on it plus multiple short strings of
length one, do not contribute to the elliptic genus.

For the enigmatic phase, we determined the number of short strings, $l$,
by maximizing the entropy; this number $l$ is given in Eq.\
\bref{maxmizingl}. However, other states with different number of short
strings, call it $l+\delta l$, also contribute to the elliptic
genus. Here, let us sum up the contributions from the states with
different values of $\delta l$, and show that the sum vanishes, because
of the alternating signs for bosonic and fermionic states.

If we change the number of length-one short strings by $\delta l$, the
total $J_L$ remains the same but $J_R=J_L-2N_p + \delta l$ and it can be
seen that the entropy is \be S_{\delta l} = S_l - \f{\delta l^2}{8
S_l}\,.  \ee This approximation is valid for $\delta l \ll S_l$. For the
enigmatic phase $S_l \sim N$ and this bound is $\delta l \ll N$.  Thus
the elliptic genus \bref{def-EG} is given approximately by \bea
\chi_{EG;enigma} &\approx& e^{S_{\text{enigma}}} \sum_{\delta l =
-\infty}^\infty (-1)^{\delta l}e^{-\f{\delta l^2}{8 S_{\text{enigma}}}}
\nn &=& e^{S_{\text{enigma}}} \vartheta_4\left( 0,e^{-\f{1}{4
S_{\text{enigma}}}}\right) \eea where we have ignored the error in
summing from $-\infty$ to $\infty$ instead of $-N$ to $N$ as it goes to
zero when $N \to \infty$.

We can now use modular transformation properties of theta functions to write this as
\be
\chi_{EG;enigma} \approx e^{S_{\text{enigma}}} \sqrt{8 \pi S_{\text{enigma}}}\, \vartheta_2 \left(0,e^{-16 \pi^2 S_{\text{enigma}}}\right)\,,
\ee
and it is easy to see that this vanishes for $S_{\text{enigma}} \to
\infty$. Thus these states do not contribute to the elliptic genus.

\section{Units and conventions}
\label{app:Conventions}

Newton's constant in $D$ spacetime dimensions is related to the $D$-dimensional Planck length as
\be
G_D = (2 \pi)^{D-3} (\ell_D)^{D-2}\,.
\ee
The tensions of the extended objects in string and M-theory are:
\begin{gather}
T_{F1} = \frac {1}{2\pi l_s^2}, \qquad T_{Dp} = \frac {1} {g_s (2\pi)^p (l_s)^{p+1}},\qquad T_{NS5} = \frac{1} {g_s^2(2\pi)^5 (l_s)^6}, \nn
T_{M2} = \frac{ 1} {(2\pi)^2 (l_{11})^3},\qquad T_{M5} = \frac{1}{(2\pi)^5 (l_{11})^6}\,,
\end{gather}
where $g_s$ is the string coupling constant and $l_s$ is the string length. The eleven-dimensional Planck length is related to these as
\be
l_{11} = g_s^{1/3} l_s\,.
\ee
In a compactification of M-theory along a circle of radius $R_{11}$ we get
\be
R_{11} = g_s l_s\,.
\ee
In a $T^6$ compactification of M-theory, where the radius of each torus circle is $R_5,\ldots ,R_{10}$, the five-dimensional Planck length is related to the eleven-dimensional Planck length as 
\be
G_5 = \frac{G_{11}}{vol(T^6)} = \frac{G_{11}}{(2\pi)^6 R_5 R_6 R_7 R_8 R_9 R_{10}} = \frac \pi 4 \frac{(l_{11})^9}{R_5 R_6 R_7 R_8 R_9 R_{10}} \,. \label{G5}
\ee
The relation between the integer charges counting the number of M2 and M5 branes, $N_I$ and $n^I$, and the physical charges of the five-dimensional solution, $Q_I$ and $q^I$, upon compactification of M-theory on $T^6$ is 
\begin{gather}
Q_1 =\f{(l_{11})^6}{R_7 R_8 R_9 R_{10}} N_1 \,,\qquad Q_2 =\f{(l_{11})^6}{R_5 R_6 R_9 R_{10}} N_2\,, \qquad Q_3 =\f{(l_{11})^6}{R_5 R_6 R_7 R_8} N_3\,, \nn
q^1 =\f{(l_{11})^3}{R_5 R_6} n^1 \,,\qquad q^2 =\f{(l_{11})^3}{R_7 R_8} n^2\,, \qquad q^3 =\f{(l_{11})^3}{R_9 R_{10}} n^3\, . 
\end{gather}
In this paper we choose a system of units where all the three $T^2$ are of equal volume and we have
\be
R_5 R_6 = R_7 R_8 = R_9 R_{10} = \h l_{11}^3 =\h g_s l_s^3
\ee
Note that this is a numerical identity. With this choice we have
\be
G_5 = 2 \pi, \qquad Q_I = 4 N_I, \qquad q^I = 2 n^I.
\ee


\begin{thebibliography}{99}

\bibitem{Gubser:2008px}
  S.~S.~Gubser,
  ``Breaking an Abelian gauge symmetry near a black hole horizon,''
  Phys.\ Rev.\  D {\bf 78}, 065034 (2008)
  [arXiv:0801.2977 [hep-th]].



\bibitem{Hartnoll:2008vx}
  S.~A.~Hartnoll, C.~P.~Herzog and G.~T.~Horowitz,
  ``Building a Holographic Superconductor,''
  Phys.\ Rev.\ Lett.\  {\bf 101}, 031601 (2008)
  [arXiv:0803.3295 [hep-th]].

\bibitem{Hartnoll:2008kx}
  S.~A.~Hartnoll, C.~P.~Herzog and G.~T.~Horowitz,
  ``Holographic Superconductors,''
  JHEP {\bf 0812}, 015 (2008)
  [arXiv:0810.1563 [hep-th]].

\bibitem{Denef:2009tp}
  F.~Denef and S.~A.~Hartnoll,
  ``Landscape of superconducting membranes,''
  Phys.\ Rev.\  D {\bf 79}, 126008 (2009)
  [arXiv:0901.1160 [hep-th]].

\bibitem{Gubser:2009qm}
  S.~S.~Gubser, C.~P.~Herzog, S.~S.~Pufu and T.~Tesileanu,
  ``Superconductors from Superstrings,''
  Phys.\ Rev.\ Lett.\  {\bf 103}, 141601 (2009)
  [arXiv:0907.3510 [hep-th]].


\bibitem{Gauntlett:2009dn}
  J.~P.~Gauntlett, J.~Sonner and T.~Wiseman,
  ``Holographic superconductivity in M-Theory,''
  Phys.\ Rev.\ Lett.\  {\bf 103}, 151601 (2009)
  [arXiv:0907.3796 [hep-th]];
  J.~P.~Gauntlett, J.~Sonner and T.~Wiseman,
  ``Quantum Criticality and Holographic Superconductors in M-theory,''
  JHEP {\bf 1002}, 060 (2010)
  [arXiv:0912.0512 [hep-th]].

\bibitem{Bhattacharyya:2010yg}
  S.~Bhattacharyya, S.~Minwalla and K.~Papadodimas,
  ``Small Hairy Black Holes in $AdS_5 x S^5$,''
  JHEP {\bf 1111}, 035 (2011)
  [arXiv:1005.1287 [hep-th]].

\bibitem{Gauntlett:2004wh}
  J.~P.~Gauntlett and J.~B.~Gutowski,
  ``Concentric black rings,''
  Phys.\ Rev.\  D {\bf 71}, 025013 (2005)
  [arXiv:hep-th/0408010].
 
\bibitem{Denef:2007vg}
  F.~Denef and G.~W.~Moore,
  ``Split states, entropy enigmas, holes and halos,''
  JHEP {\bf 1111}, 129 (2011)
  [arXiv:hep-th/0702146].

\bibitem{deBoer:2008fk}
  J.~de Boer, F.~Denef, S.~El-Showk, I.~Messamah and D.~Van den Bleeken,
  ``Black hole bound states in $AdS_3 \times S^2$,''
  JHEP {\bf 0811}, 050 (2008)
  [arXiv:0802.2257 [hep-th]].

\bibitem{Maldacena:1997de}
  J.~M.~Maldacena, A.~Strominger and E.~Witten,
  ``Black hole entropy in M theory,''
  JHEP {\bf 9712}, 002 (1997)
  [arXiv:hep-th/9711053].

\bibitem{Minasian:1999qn}
  R.~Minasian, G.~W.~Moore and D.~Tsimpis,
  ``Calabi-Yau black holes and (0,4) sigma models,''
  Commun.\ Math.\ Phys.\  {\bf 209}, 325 (2000)
  [arXiv:hep-th/9904217].

\bibitem{Breckenridge:1996is}
  J.~C.~Breckenridge, R.~C.~Myers, A.~W.~Peet and C.~Vafa,
  ``D-branes and spinning black holes,''
  Phys.\ Lett.\ B {\bf 391}, 93 (1997)
  [arXiv:hep-th/9602065].

\bibitem{hep-th/0005003} 
  R.~Dijkgraaf, J.~M.~Maldacena, G.~W.~Moore and E.~P.~Verlinde,
  ``A Black hole Farey tail,''
  hep-th/0005003.

\bibitem{deBoer:1998us}
  J.~de Boer,
  ``Large N Elliptic Genus and AdS/CFT Correspondence,''
  JHEP {\bf 9905}, 017 (1999)
  [arXiv:hep-th/9812240].

\bibitem{Maldacena:1999bp}
  J.~M.~Maldacena, G.~W.~Moore and A.~Strominger,
  ``Counting BPS black holes in toroidal type II string theory,''
  arXiv:hep-th/9903163.

\bibitem{Dabholkar:2009dq}
  A.~Dabholkar, M.~Guica, S.~Murthy and S.~Nampuri,
  ``No entropy enigmas for N=4 dyons,''
  JHEP {\bf 1006}, 007 (2010)
  [arXiv:0903.2481 [hep-th]].

\bibitem{Eguchi:2010ej}
  T.~Eguchi, H.~Ooguri and Y.~Tachikawa,
  ``Notes on the K3 Surface and the Mathieu group $M_{24}$,''
  Exper.\ Math.\  {\bf 20}, 91 (2011)
  [arXiv:1004.0956 [hep-th]].

\bibitem{Elvang:2004rt}
  H.~Elvang, R.~Emparan, D.~Mateos and H.~S.~Reall,
  ``A supersymmetric black ring,''
  Phys.\ Rev.\ Lett.\  {\bf 93}, 211302 (2004)
  [arXiv:hep-th/0407065].

\bibitem{Bena:2004de}
  I.~Bena and N.~P.~Warner,
  ``One ring to rule them all ... and in the darkness bind them?,''
  Adv.\ Theor.\ Math.\ Phys.\  {\bf 9}, 667 (2005)
  [arXiv:hep-th/0408106].

\bibitem{Elvang:2004ds}
  H.~Elvang, R.~Emparan, D.~Mateos and H.~S.~Reall,
  ``Supersymmetric black rings and three-charge supertubes,''
  Phys.\ Rev.\ D {\bf 71}, 024033 (2005)
  [arXiv:hep-th/0408120].

\bibitem{Gauntlett:2004qy}
  J.~P.~Gauntlett and J.~B.~Gutowski,
  ``General concentric black rings,''
  Phys.\ Rev.\  D {\bf 71}, 045002 (2005)
  [arXiv:hep-th/0408122].

\bibitem{Bena:2004tk}
  I.~Bena and P.~Kraus,
  ``Microscopic description of black rings in AdS/CFT,''
  JHEP {\bf 0412}, 070 (2004)
  [arXiv:hep-th/0408186].

\bibitem{Cyrier:2004hj}
  M.~Cyrier, M.~Guica, D.~Mateos and A.~Strominger,
  ``Microscopic entropy of the black ring,''
  Phys.\ Rev.\ Lett.\  {\bf 94}, 191601 (2005)
  [arXiv:hep-th/0411187].

\bibitem{Bena:2005va}
  I.~Bena and N.~P.~Warner,
  ``Bubbling supertubes and foaming black holes,''
  Phys.\ Rev.\  D {\bf 74}, 066001 (2006)
  [arXiv:hep-th/0505166].

\bibitem{Berglund:2005vb}
  P.~Berglund, E.~G.~Gimon and T.~S.~Levi,
  ``Supergravity microstates for BPS black holes and black rings,''
  JHEP {\bf 0606}, 007 (2006)
  [arXiv:hep-th/0505167].
  
\bibitem{Iizuka:2005uv}
  N.~Iizuka and M.~Shigemori,
  ``A Note on D1-D5-J system and 5-D small black ring,''
  JHEP {\bf 0508}, 100 (2005)
  [arXiv:hep-th/0506215].

\bibitem{Dabholkar:2005qs}
  A.~Dabholkar, N.~Iizuka, A.~Iqubal and M.~Shigemori,
  ``Precision microstate counting of small black rings,''
  Phys.\ Rev.\ Lett.\  {\bf 96}, 071601 (2006)
  [arXiv:hep-th/0511120].

\bibitem{Alday:2005xj}
  L.~F.~Alday, J.~de Boer and I.~Messamah,
  ``What is the dual of a dipole?,''
  Nucl.\ Phys.\ B {\bf 746}, 29 (2006)
  [arXiv:hep-th/0511246].

\bibitem{Dabholkar:2006za}
  A.~Dabholkar, N.~Iizuka, A.~Iqubal, A.~Sen and M.~Shigemori,
  ``Spinning strings as small black rings,''
  JHEP {\bf 0704}, 017 (2007)
  [arXiv:hep-th/0611166].

\bibitem{MasakiMITTalk}
M.~Shigemori, ``The Phases of D1-D5 CFT ---
	Towards Understanding Black Ring Microscopics,'' talk given at
	Massachusetts Institute of Technology, Oct.\ 10,
	2006.


\bibitem{David:2002wn}
  J.~R.~David, G.~Mandal and S.~R.~Wadia,
  ``Microscopic formulation of black holes in string theory,''
  Phys.\ Rept.\  {\bf 369}, 549 (2002)
  [arXiv:hep-th/0203048].


\bibitem{Avery:2010er}
  S.~G.~Avery, B.~D.~Chowdhury and S.~D.~Mathur,
  ``Deforming the D1D5 CFT away from the orbifold point,''
  JHEP {\bf 1006}, 031 (2010)
  [arXiv:1002.3132 [hep-th]].

\bibitem{Avery:2010hs}
  S.~G.~Avery, B.~D.~Chowdhury and S.~D.~Mathur,
  ``Excitations in the deformed D1D5 CFT,''
  JHEP {\bf 1006}, 032 (2010)
  [arXiv:1003.2746 [hep-th]].

\bibitem{Avery:2010vk}
  S.~G.~Avery and B.~D.~Chowdhury,
  ``Intertwining Relations for the Deformed D1D5 CFT,''
  JHEP {\bf 1105}, 025 (2011)
  [arXiv:1007.2202 [hep-th]].

\bibitem{Schwimmer:1986mf}
  A.~Schwimmer and N.~Seiberg,
  ``Comments on the N=2, N=3, N=4 Superconformal Algebras in Two-Dimensions,''
  Phys.\ Lett.\  {\bf B184}, 191 (1987).

\bibitem{Balasubramanian:2000rt}
  V.~Balasubramanian, J.~de Boer, E.~Keski-Vakkuri and S.~F.~Ross,
  ``Supersymmetric conical defects: Towards a string theoretic description  of
  black hole formation,''
  Phys.\ Rev.\  D {\bf 64}, 064011 (2001)
  [arXiv:hep-th/0011217].

\bibitem{Maldacena:2000dr}
  J.~M.~Maldacena and L.~Maoz,
  ``De-singularization by rotation,''
  JHEP {\bf 0212}, 055 (2002)
  [arXiv:hep-th/0012025].
  

\bibitem{Mathur:2005ai}
  S.~D.~Mathur,
  ``The Quantum structure of black holes,''
  Class.\ Quant.\ Grav.\  {\bf 23}, R115 (2006).
  [hep-th/0510180].



\bibitem{Eguchi:1988vra}
  T.~Eguchi, H.~Ooguri, A.~Taormina and S.~K.~Yang,
  ``Superconformal Algebras and String Compactification on Manifolds with SU(N)
  Holonomy,''
  Nucl.\ Phys.\  B {\bf 315}, 193 (1989).

\bibitem{Kawai:1993jk}
  T.~Kawai, Y.~Yamada and S.~K.~Yang,
  ``Elliptic Genera And N=2 Superconformal Field Theory,''
  Nucl.\ Phys.\  B {\bf 414}, 191 (1994)
  [arXiv:hep-th/9306096].

\bibitem{Dijkgraaf:1996xw}
  R.~Dijkgraaf, G.~W.~Moore, E.~P.~Verlinde and H.~L.~Verlinde,
  ``Elliptic genera of symmetric products and second quantized strings,''
  Commun.\ Math.\ Phys.\  {\bf 185}, 197 (1997)
  [arXiv:hep-th/9608096].

\bibitem{Castro:2008ys}
  A.~Castro and S.~Murthy,
  ``Corrections to the statistical entropy of five dimensional black holes,''
  JHEP {\bf 0906}, 024 (2009)
  [arXiv:0807.0237 [hep-th]].

\bibitem{Mateos:2001qs}
  D.~Mateos, P.~K.~Townsend,
  ``Supertubes,''
  Phys.\ Rev.\ Lett.\  {\bf 87}, 011602 (2001).
  [hep-th/0103030].
  
\bibitem{Denef:2000nb}
  F.~Denef,
  ``Supergravity flows and D-brane stability,''
  JHEP {\bf 0008}, 050 (2000).
  [hep-th/0005049].   

\bibitem{Gutowski:2004yv}
  J.~B.~Gutowski, H.~S.~Reall,
  ``General supersymmetric AdS(5) black holes,''
  JHEP {\bf 0404}, 048 (2004).
  [hep-th/0401129].

\bibitem{Bena:2005ni}
  I.~Bena, P.~Kraus and N.~P.~Warner,
  ``Black rings in Taub-NUT,''
  Phys.\ Rev.\  D {\bf 72}, 084019 (2005)
  [arXiv:hep-th/0504142].

\bibitem{Bena:2008dw}
  I.~Bena, N.~Bobev, C.~Ruef, N.~P.~Warner,
  ``Supertubes in Bubbling Backgrounds: Born-Infeld Meets Supergravity,''
  JHEP {\bf 0907}, 106 (2009).
  [arXiv:0812.2942 [hep-th]].
  
\bibitem{Gibbons:1987sp}
  G.~W.~Gibbons, P.~J.~Ruback,
  ``The Hidden Symmetries of Multicenter Metrics,''
  Commun.\ Math.\ Phys.\  {\bf 115}, 267 (1988).

\bibitem{Bena:2008wt}
  I.~Bena, N.~Bobev and N.~P.~Warner,
  ``Spectral Flow, and the Spectrum of Multi-Center Solutions,''
  Phys.\ Rev.\  D {\bf 77}, 125025 (2008)
  [arXiv:0803.1203 [hep-th]].

  
\bibitem{Balasubramanian:2006gi}
  V.~Balasubramanian, E.~G.~Gimon and T.~S.~Levi,
  ``Four Dimensional Black Hole Microstates: From D-branes to Spacetime Foam,''
  JHEP {\bf 0801}, 056 (2008)
  [arXiv:hep-th/0606118].

\bibitem{Bena:2004wt}
  I.~Bena, P.~Kraus,
  ``Three charge supertubes and black hole hair,''
  Phys.\ Rev.\  {\bf D70}, 046003 (2004).
  [hep-th/0402144].
  
\bibitem{Marolf:2005cx}
  D.~Marolf, A.~Virmani,
  ``A Black hole instability in five dimensions?,''
  JHEP {\bf 0511}, 026 (2005).
  [hep-th/0505044].
  
\bibitem{Bena:2005zy}
  I.~Bena, C.-W.~Wang, N.~P.~Warner,
  ``Sliding rings and spinning holes,''
  JHEP {\bf 0605}, 075 (2006).
  [hep-th/0512157].

\bibitem{Dall'Agata:2010dy}
  G.~Dall'Agata, S.~Giusto, C.~Ruef,
  ``U-duality and non-BPS solutions,''
  JHEP {\bf 1102}, 074 (2011).
  [arXiv:1012.4803 [hep-th]].

 \bibitem{Lunin:2002iz}
  O.~Lunin, J.~M.~Maldacena and L.~Maoz,
  ``Gravity solutions for the D1-D5 system with angular momentum,''
  arXiv:hep-th/0212210.

\bibitem{Gregory:1993vy}
  R.~Gregory, R.~Laflamme,
  ``Black strings and p-branes are unstable,''
  Phys.\ Rev.\ Lett.\  {\bf 70}, 2837-2840 (1993).
  [hep-th/9301052].


\bibitem{Manschot:2011xc}
  J.~Manschot, B.~Pioline and A.~Sen,
  ``A Fixed point formula for the index of multi-centered N=2 black holes,''
  JHEP {\bf 1105}, 057 (2011)
  [arXiv:1103.1887 [hep-th]].

\bibitem{deBoer:2008zn}
  J.~de Boer, S.~El-Showk, I.~Messamah, D.~Van den Bleeken,
  ``Quantizing N=2 Multicenter Solutions,''
  JHEP {\bf 0905}, 002 (2009).
  [arXiv:0807.4556 [hep-th]].


\bibitem{deBoer:2009un}
  J.~de Boer, S.~El-Showk, I.~Messamah, D.~Van den Bleeken,
  ``A Bound on the entropy of supergravity?,''
  JHEP {\bf 1002}, 062 (2010).
  [arXiv:0906.0011 [hep-th]].

\bibitem{modulatedPhases}
  S.~Nakamura, H.~Ooguri and C.~S.~Park,
  ``Gravity Dual of Spatially Modulated Phase,''
  Phys.\ Rev.\  D {\bf 81}, 044018 (2010)
  [arXiv:0911.0679 [hep-th]];
%
  H.~Ooguri and C.~S.~Park,
  ``Holographic End-Point of Spatially Modulated Phase Transition,''
  Phys.\ Rev.\  D {\bf 82}, 126001 (2010)
  [arXiv:1007.3737 [hep-th]];
%
  A.~Aperis, P.~Kotetes, E.~Papantonopoulos, G.~Siopsis, P.~Skamagoulis and G.~Varelogiannis,
  ``Holographic Charge Density Waves,''
  Phys.\ Lett.\  B {\bf 702}, 181 (2011)
  [arXiv:1009.6179 [hep-th]];
%
  R.~Flauger, E.~Pajer and S.~Papanikolaou,
  ``A Striped Holographic Superconductor,''
  Phys.\ Rev.\  D {\bf 83}, 064009 (2011)
  [arXiv:1010.1775 [hep-th]];
%
  H.~Ooguri and C.~S.~Park,
  ``Spatially Modulated Phase in Holographic Quark-Gluon Plasma,''
  Phys.\ Rev.\ Lett.\  {\bf 106}, 061601 (2011)
  [arXiv:1011.4144 [hep-th]];
%
  C.~A.~B.~Bayona, K.~Peeters and M.~Zamaklar,
  ``A Non-homogeneous ground state of the low-temperature Sakai-Sugimoto
  model,''
  JHEP {\bf 1106}, 092 (2011)
  [arXiv:1104.2291 [hep-th]];
%
  A.~Donos and J.~P.~Gauntlett,
  ``Holographic striped phases,''
  JHEP {\bf 1108}, 140 (2011)
  [arXiv:1106.2004 [hep-th]];
  S.~Takeuchi,
  ``Modulated Instability in Five-Dimensional U(1) Charged AdS Black Hole with
  R**2-term,''
  JHEP {\bf 1201}, 160 (2012)
  [arXiv:1108.2064 [hep-th]].

\end{thebibliography}
\end{document}